\documentclass{aa}
\usepackage{graphicx}
\usepackage{txfonts}
\usepackage{natbib}
\usepackage[utf8]{inputenc}
\usepackage{amsmath}
\usepackage{amssymb}
\usepackage{mathtools}
\newcommand{\fMDM}{{f_{M_\mathrm{DM}}}}
\newcommand{\Qual}{q}

\usepackage[colorlinks,allcolors=blue]{hyperref}

\begin{document}

   \title{Galactic mass-to-light ratios with superfluid dark matter}

   \author{T. Mistele
          \inst{1}
          \and
          S. McGaugh\inst{2}
          \and
          S. Hossenfelder \inst{1}
          }

   \institute{Frankfurt Institute for Advanced Studies,
              Ruth-Moufang-Str. 1, D 60438 Frankfurt am Main, Germany\\
              \email{mistele@fias.uni-frankfurt.de}
         \and
             Department of Astronomy, Case Western Reserve University, 10900 Euclid Avenue,
             Cleveland, OH 44106, USA\\
             }

   \date{\today}

  \abstract
   {We make rotation curve fits to test the superfluid dark matter model.}
   {
      In addition to
      verifying that the resulting fits match the rotation curve data reasonably well,
            we aim to evaluate how satisfactory they are with respect to two criteria, namely, how reasonable the resulting stellar mass-to-light ratios are
            and
            whether the fits end up in the regime of superfluid dark matter where the model resembles modified Newtonian dynamics (\textsc{MOND}).
    }
   {We fitted the superfluid dark matter model to the rotation curves of 169 galaxies in the \textsc{SPARC} sample.}
   {We found that the mass-to-light ratios obtained with superfluid dark matter are generally acceptable in terms of stellar populations. However, the best-fit mass-to-light ratios have an unnatural dependence on the size of the galaxy in that giant galaxies have systematically lower mass-to-light ratios than dwarf galaxies.
    A second finding is that the superfluid often fits the rotation curves best in the regime where the superfluid's force cannot resemble that of MOND without adjusting a boundary condition separately for each galaxy.
    In that case, we can no longer expect superfluid dark matter to reproduce the phenomenologically observed scaling relations that make \textsc{MOND} appealing. If, on the other hand, we consider only solutions whose force approximates  \textsc{MOND} well, then the total mass of the superfluid is in tension with gravitational lensing data.}
   {
    We conclude that even the best fits with superfluid dark matter are still unsatisfactory for two reasons.
    First, the resulting stellar mass-to-light ratios show an unnatural trend with galaxy size.
    Second, the fits do not end up in the regime that automatically resembles \textsc{MOND,} and if we force the fits to do so, the total dark matter mass is in tension with strong lensing data.
   }


   \maketitle

\section{Introduction}

 In 2015, \citeauthor{Berezhiani2015} proposed a new hypothesis that combines features of cold dark matter (\textsc{CDM}) and modified Newtonian dynamics \citep[\textsc{MOND};][]{Milgrom1983a,Milgrom1983b,Milgrom1983c,Bekenstein1984}: superfluid dark matter (\textsc{SFDM}). In \textsc{SFDM}, dark matter is composed of a light (on the order of eV) scalar field that can condense to a superfluid. In the superfluid phase, phonons mediate a force that is similar to the force of \textsc{MOND}. This hypothesis has since passed several observational tests \citep{Berezhiani2018, Hossenfelder2019, Hossenfelder2020}.

However, it was recently found that \textsc{SFDM} needs about 20\% less baryonic mass than \textsc{MOND} to fit the Milky Way rotation curve at $R \lesssim 25$ kpc \citep{Hossenfelder2020}.
Though a modest effect, this underestimates the stellar mass required by microlensing \citep{Wegg2017}. It also underestimates the amplitude of the spiral structure required to reconcile the Galactic rotation curve measured independently by stars and gas \citep{McGaugh2019b}. This offset is similar to that found for emergent gravity \citep{Verlinde2017} by \citet{Lelli2017EG}, which shares some properties of \textsc{SFDM}, thus raising the prospect that it might be a general trend.
Consequently, \textsc{SFDM} may require a systematically smaller stellar mass-to-light ratio ($M/L_*$) than \textsc{MOND}.
Since \textsc{MOND} generally agrees with the $M/L_*$ expected from stellar population synthesis (SPS) models \citep{McGaugh2004}, such a systematic trend can be problematic for \textsc{SFDM}.
To investigate this, we fitted \textsc{SFDM} to the \textit{Spitzer} Photometry and Accurate Rotation Curves (\textsc{SPARC}) data \citep{Lelli2016} with $M/L_*$ as a fitting parameter.

\section{Models}
\label{sec:models}

  Four parameters are required for \textsc{SFDM}; we used the
 fiducial values from \citet{Berezhiani2018}, $m = 1\,\mathrm{eV}$, $\Lambda = 0.05\,\mathrm{meV}$, $\alpha = 5.7$, and $\beta = 2$.
We kept those parameters fixed during our analysis.
In Appendix~\ref{sec:sfdm:params}, we argue that our conclusions are generally robust against variations in these parameters.

The total acceleration inside the superfluid core of a galaxy is
$
 \vec{a}_{\mathrm{tot}} = \vec{a}_\theta + \vec{a}_b + \vec{a}_{\mathrm{SF}} \,$,
where  $\vec a_{\theta}$ is the acceleration created by the phonon force, $\vec a_{\rm SF}$ the acceleration stemming from the normal gravitational attraction of the superfluid, and $\vec a_b$ that stemming from the mass of the baryons.
The position dependence of those accelerations is determined by the \textsc{SFDM} equations of motion and the distribution of baryonic mass.
At a transition radius where the superfluid condensate is estimated to break down, one matches the superfluid core to a Navarro-Frenk-White (\textsc{NFW}) halo \citep{Berezhiani2018}.

From integrating the standard Poisson equation including the superfluid's energy density $\rho_{\mathrm{SF}}$ as a source term, one obtains $\hat{\mu}(\vec{x}) = \mu_{\mathrm{nr}} - m \phi_N(\vec{x})$, where $\mu_{\mathrm{nr}}$ is the chemical potential and $\phi_N(\vec{x})$ is the Newtonian gravitational potential.
The gradient of $\phi_N(\vec{x})$ gives $\vec{a}_b + \vec{a}_{\mathrm{SF}}$.
In the so-called no-curl approximation, one obtains the phonon force, $\vec{a}_\theta$, as an algebraic function of $\vec{a}_b$ and $\varepsilon_*(\vec{x})$ (see Appendix~\ref{sec:appendix:estar}),
\begin{align}
 \label{eq:estar}
 \varepsilon_*(\vec x) := \frac{2 m^2}{\alpha M_{\mathrm{Pl}} |\vec{a}_b(\vec x)|} \frac{\hat{\mu} (\vec{x})}{m} \,,
\end{align}
where $M_{\rm Pl}$ is the Planck mass (it enters through Newton's constant).
The quantity $\varepsilon_*(\vec{x})$ controls how closely \textsc{SFDM} resembles \textsc{MOND}.
We refer to $|\varepsilon_*| \ll 1$ as the \textsc{MOND} limit and to $|\varepsilon_*| = {\cal O}(1)$ as the pseudo-\textsc{MOND} limit.
In the \textsc{MOND} limit and assuming the no-curl approximation, the phonon force points into the same direction as $\vec{a}_b$ with magnitude $a_\theta = \sqrt{a_0 a_b}$.
Here, $a_0 = \alpha^3 \Lambda^2/M_{\mathrm{Pl}}$ is the acceleration scale below which the phonon force becomes important compared to $a_b$.
At larger accelerations, it is subdominant.
This gives a typical \textsc{MOND}-like total acceleration $a_{\mathrm{tot}}$, at least as long as $a_{\mathrm{SF}}$ stays negligible.
Usually, $a_{\mathrm{SF}}$ is indeed negligible in the proper \textsc{MOND} limit $|\varepsilon_*| \ll 1$ but less so in the pseudo-MOND limit $|\varepsilon_*| = \mathcal{O}(1)$.
More details on the definition and rationale behind these limits are in Appendix~\ref{sec:appendix:models} (see also \citealt{Mistele2020}).
The actual value of $\varepsilon_*$ depends on the baryonic mass distribution and a boundary condition needed to solve the equations of motion.

A big advantage of MOND is that galactic scaling relations such as the radial acceleration relation \citep[RAR;][]{Lelli2017b} arise automatically with no intrinsic scatter.
The same goes for SFDM in the MOND limit $|\varepsilon_*| \ll 1$.
In this limit, SFDM predicts a tight RAR irrespective of the precise value of the boundary condition.
This is different outside the MOND limit where the total acceleration $a_{\mathrm{tot}}$ depends sensitively on the choice of the boundary condition.
Thus, outside the MOND limit, scaling relations such as the RAR can arise only by adjusting this boundary condition separately for each galaxy.
Otherwise, increased scatter and systematic deviations are likely.

In principle, it might be possible that galaxy formation selects the right boundary condition for each galaxy to produce a tight RAR even outside the MOND limit.
However, then SFDM loses one of its main advantages over $\Lambda$CDM and one might as well stick with $\Lambda$CDM.

We compared \textsc{SFDM} to \textsc{MOND} using one of the standard interpolation functions \citep{Lelli2017b},
\begin{align}
 \nu_e(y) = \frac{1}{1 - e^{-\sqrt{y}}} \,,
\end{align}
where $y = |{\vec a}_b|/a_0$ and $a_0$ is the one free parameter in \textsc{MOND}.
In \textsc{SFDM} the interpolation function is slower to reach its limits for large and small $y$.
Also, usually $a_0$ is chosen smaller in SFDM compared to MOND to account for the presence of $a_{\mathrm{SF}}$ \citep{Berezhiani2018}.
For MOND, we adopt $a_0^{\mathrm{MOND}} \approx 1.2 \cdot 10^{-10}\,\mathrm{m}/\mathrm{s}^2$ from \citet{Lelli2017b}.
For SFDM, the fiducial parameters from \citet{Berezhiani2018} give
$ a_0^{\mathrm{SFDM}} \approx 0.87 \cdot 10^{-10}\,\mathrm{m}/\mathrm{s}^2$.

To check how sensitive our results are to the particular theoretical realization of \textsc{SFDM,} we included the two-field model from \cite{Mistele2020}.
In this two-field model, the phenomenology on galactic scales is similar to standard \textsc{SFDM}, but it has the advantages that (a) it does not require ad-hoc finite-temperature corrections for stability, (b) its phonon force is always close to its \textsc{MOND}-limit, and (c) the superfluid can remain in equilibrium much longer than galactic timescales.
Both models are described in more detail in Appendix~\ref{sec:appendix:models}.

\section{Data}
\label{ssec:data}

We took the observed rotation velocity $V_{\mathrm{obs}}$ directly from \textsc{SPARC} \citep{Lelli2016}.
To find the best \textsc{SFDM} fit, we then needed the baryonic energy density $\rho_b(R,z)$ because it is a source for the equation of motion of the superfluid.
For this, we used updated high-resolution mass models including resolved gas surface density profiles for 169 of the 175 SPARC galaxies (Lelli 2021, private communication).
We excluded the six galaxies that lack radial profiles for the gas distribution.

These mass models provide surface densities $\Sigma$ for the bulge, the stellar disk, and the HI disk of each galaxy for a discrete set of positions starting at $R=0$.
We linearly interpolated the data points and assumed zero surface density outside the outermost surface density data point. This gives a simple, data-compatible approximation for the density distribution at all radii.

For the bulge, we assumed spherical symmetry and extracted its energy density from its surface density using an Abel transform,\begin{align}
 \rho_{\mathrm{bulge}}(r) = -\frac{1}{\pi} \int_0^{\infty} d\bar{r} \frac{\Sigma_{\mathrm{bulge}}'(\sqrt{\bar{r}^2+r^2})}{\sqrt{\bar{r}^2+r^2}} \,.
\end{align}

For the stellar disk, we assumed a scale height, $h_*$, of \citep{Lelli2016}
\begin{align}
 h_* = 0.196 \cdot (R_{\mathrm{disk}}[\mathrm{kpc}])^{0.633} \,\mathrm{kpc} \,,
\end{align}
where $R_{\mathrm{disk}}$ is the disk scale length from \textsc{SPARC}. Again, we used a linear interpolation of the \textsc{SPARC} surface brightness data points.

For the gas disk we did the same as for the stellar disk, except that in this case we assumed a fixed scale height, $h_g = 0.130\,\mathrm{kpc}$.
This is the same scale height used in \citet{Hossenfelder2020}.
We do not expect this choice of scale height to significantly affect the results.
To account for the non-HI gas,
    we multiplied the HI surface density by $1.4$ \citep{McGaugh2020}.

\section{Method}

Solutions of the equations of motion can be parameterized by one boundary condition, $\varepsilon:= \varepsilon_*(R_{\mathrm{ mid}})$, where $R_{\mathrm{mid}}:= (R_{\mathrm{min}} + R_{\mathrm{max}})/2$, and $R_{\mathrm{min}}$ ($R_{\mathrm{max}}$) is the smallest (largest) radius with a rotation curve data point.
The value of $\varepsilon$ quantifies how closely the phonon force resembles a \textsc{MOND} force in the middle of the observed rotation curve.

In our fitting procedure, we kept $V_{\mathrm{obs}}$ and the fiducial model parameters of \textsc{SFDM} fixed,
    but we allowed a common factor, $Q_*$, to adjust the stellar disk and bulge $M/L_*$ relative to
    the nominal stellar population values in the Spitzer [3.6] band
    $(M/L_*)_{\mathrm{disk}} = 0.5$ and $(M/L_*)_{\mathrm{bulge}} = 0.7$ \citep{Lelli2017b},
\begin{eqnarray}
 \rho_b(R,z) &=& \rho_{\mathrm{gas}}(R,z) \nonumber + 0.5 \cdot Q_* \cdot \rho_*(R,z) \nonumber \\
 &+& 0.7 \cdot Q_* \cdot \rho_{\mathrm{bulge}}(\sqrt{R^2+z^2}) \,.
\end{eqnarray}

Using this total baryonic energy density, we solved the SFDM equations of motion for different boundary conditions. From that we then obtained the expected rotation curve.

We assume that all rotation curve data points are within the superfluid core; otherwise, rotation curves cannot be automatically \textsc{MOND}-like since the MOND-like phonon force is active only inside the superfluid.
In our fits, we took the superfluid phase to end only when its energy density reaches zero.
That is, we only required that
 $\varepsilon_*(\vec x)$ (see Eq.\ (\ref{eq:estar})) is  larger than an algebraic minimum value $\varepsilon_{\mathrm{min}}$ everywhere within the superfluid.
This minimum value is reached when $\rho_{\mathrm{SF}}$ vanishes and (for the case of $\beta=2$) is given by $\varepsilon_{\mathrm{min}} = - \sqrt{3/32} \approx - 0.31$.

For the best fits, we then checked whether all data points lie within the superfluid core according to a different criterion based on thermal equilibrium.
It turns out that for 31 of the 169 galaxies this is not the case. However, this criterion for the value of the transition radius to the \textsc{NFW} halo is quite ad hoc. We therefore do not discard these solutions, though we checked that they do not alter the main conclusions (see also Appendix~\ref{sec:results:sfdm:thermal}).

Then we compared how well this rotation curve matches with the observed velocities, $V_{\mathrm{obs}}$, from \textsc{SPARC}. For this, we defined the best fit for each galaxy as that with the smallest $\chi^2$,
\begin{align}
 \chi^2 =  \frac{1}{N-f} \sum_{R} \frac{(V_{\mathrm{obs}}(R) - V_c(R))^2}{\sigma_{V_{\mathrm{obs}}}^2(R)} \,.
\end{align}
Here, $N$ is the number of data points in the galaxy, $f = 2$ is the number of fit parameters ($Q_*$ and $\varepsilon$), $\sigma_{V_{\mathrm{obs}}}$ is the uncertainty on the velocity $V_{\mathrm{obs}}$ from \textsc{SPARC}, $V_c(R)$ is the calculated rotation curve in \textsc{SFDM}, and the sum is over the data points at radius $R$.

We minimized $\chi^2$ for
\begin{align}
\label{eq:rangeQ}
 10^{-2} \leq \, & Q_* \leq 15 \,, \\
\label{eq:rangeestar}
 10^{-2} \leq \, & \left(\varepsilon - \varepsilon_{\mathrm{min}}\right) \leq 10^4 \,.
\end{align}
In our fit code, we scanned values of $\log_{10}(Q_*)$ and $\log_{10}(\varepsilon - \varepsilon_{\mathrm{min}})$.

In the \textsc{SPARC} data, the Newtonian acceleration due to gas sometimes points outward from the galactic center, not toward it, because of a hole in the HI data, possibly due to a transition from atomic to molecular gas.
Usually, such a negative gas contribution is countered by the positive contributions from the stellar disk and the bulge and does not pose a problem. When this is not the case, there is technically no stable circular orbit so we cannot calculate a rotation curve.
When this happened, we omitted those data points when calculating $\chi^2$.

As a cross-check and as a comparison for \textsc{SFDM},
    we also fitted the \textsc{RAR} to the \textsc{SPARC} data, that is, we fitted the \textsc{SPARC} data with \textsc{MOND} assuming no curl term and the exponential interpolation function $\nu_e$ \citep{Lelli2017b}.
In this case, we have only one free fit parameter, $Q_*$, and consequently, when calculating $\chi^2$, we set $f=1$.
We describe our fitting and calculation methods in more detail in Appendix~\ref{sec:method}.

\section{Results}
\label{sec:results}

The result of our \textsc{MOND} fit is similar to that of \citet{Li2018},
    which also fitted the \textsc{RAR} to \textsc{SPARC} galaxies.
The major difference is that \citet{Li2018} used a Markov chain Monte Carlo (MCMC) procedure with Gaussian priors, while we used a simple parameter scan to minimize $\chi^2$.
We also did not vary distance and inclination and did not separately vary the mass-to-light ratio of the stellar disk and the bulge. As a consequence of this simplified fitting procedure,
    our distribution of best-fit $M/L_*$ has more outliers and looks less Gaussian than that of \citet{Li2018}.

\begin{figure}
 \centering
 \includegraphics[width=.48\textwidth]{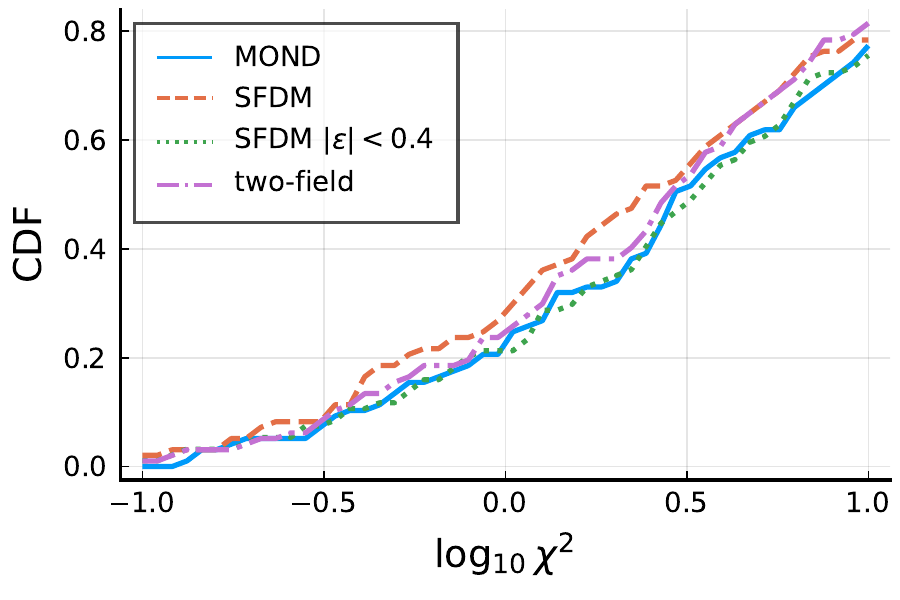}
 \caption{
     Best-fit $\chi^2$ cumulative distribution functions for the $\Qual=1$ galaxies for different models.
    }
 \label{fig:chi2cdfs}
\end{figure}

Still, our median best-fit stellar mass-to-light ratios and the best-fit $\chi^2$ values are similar to those from \citet{Li2018}.
The median stellar disk $M/L_*$ is $0.39$.
When we restrict ourselves to galaxies with high quality data ($\Qual=1$), this becomes $0.47$, very similar to the $0.50$ from \citet{Li2018}.
We show the $\chi^2$ cumulative distribution function (CDF) in Fig.~\ref{fig:chi2cdfs}, which is also in reasonable agreement with \citet{Li2018}.

\begin{figure}
 \centering
 \includegraphics[width=.48\textwidth]{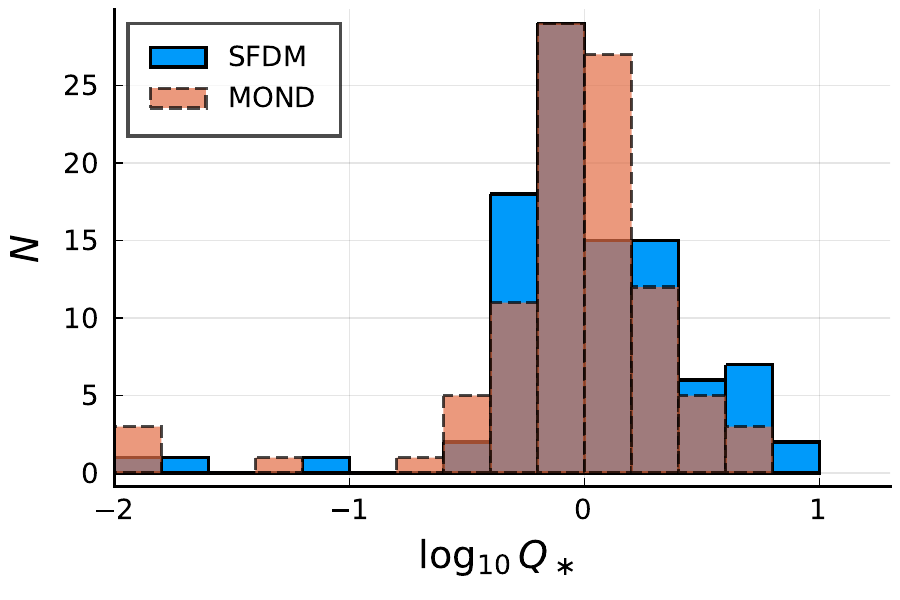}
 \caption{
     Histograms of the best-fit $Q_\ast$ values for the SFDM and MOND fits restricted to the $\Qual=1$ galaxies.
    }
 \label{fig:fYhists}
\end{figure}

In Fig.~\ref{fig:fYhists} one sees that some galaxies end up at the minimum stellar mass-to-light ratio allowed in our fitting method, corresponding to $Q_* \approx 0.01$.
If we do not restrict ourselves to $\Qual=1$, this peak at $Q_* \approx 0.01$ is even more pronounced.
As discussed in Appendix~\ref{sec:rar}, this is an artifact of our fitting procedure and can be ignored in what follows.

\subsection{MOND versus SFDM}
\label{ssec:rar}

Figure~\ref{fig:fYhists} shows the best-fit $Q_*$ for the 97 galaxies with $\Qual=1$.
Contrary to what one might naively expect from the Milky Way result \citep{Hossenfelder2019},
    the \textsc{SFDM} fits do not have significantly smaller $Q_*$ than the \textsc{MOND} fits.
Indeed, the median $Q_*$ for the $\Qual=1$ galaxies is about $4\%$ larger than for \textsc{MOND}.

\begin{figure}[t]
 \centering
 \includegraphics[width=.48\textwidth]{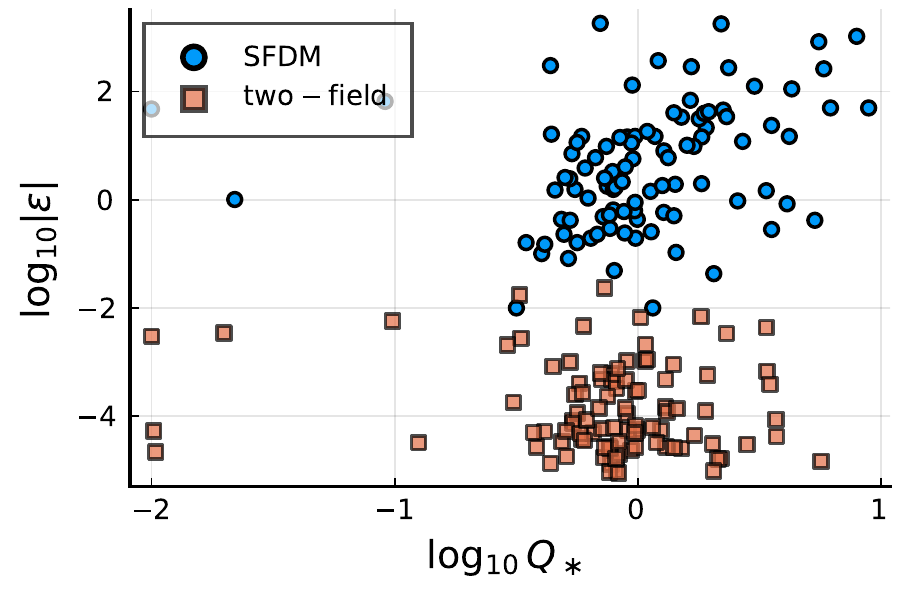}
 \caption{
     Best-fit $\varepsilon$ values versus the best-fit $Q_*$ values for the $\Qual=1$ galaxies.
     We show $\log_{10} |\varepsilon|$ rather than $\log_{10} |\varepsilon - \varepsilon_{\mathrm{min}}|$ to show how many galaxies end up in the MOND limit (corresponding to $|\varepsilon_*| \ll 1$) rather than how many galaxies end up close to $\rho_{\mathrm{SF}} = 0$ (corresponding to $|\varepsilon_* - \varepsilon_{\mathrm{min}}| \ll 1$).
     For standard SFDM, the correlation coefficient is $r = 0.28$.
    }
 \label{fig:estarscatter}
\end{figure}

\begin{figure}
 \centering
 \includegraphics[width=.48\textwidth]{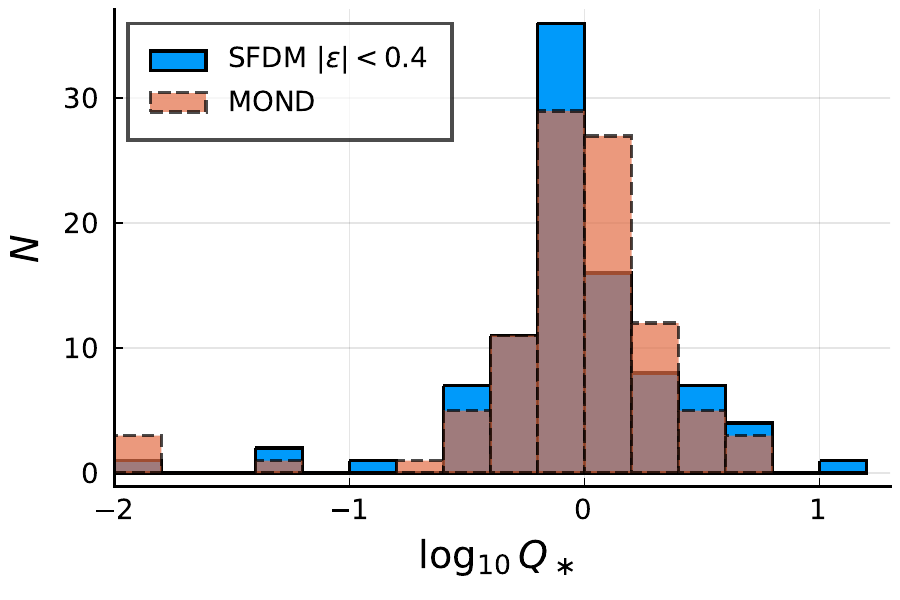}
 \caption{
     Same as Fig.~\ref{fig:fYhists} but for SFDM restricted to $|\varepsilon| < 0.4$.
    }
 \label{fig:fYhist04}
\end{figure}

One reason for this is that for many galaxies the superfluid is not in the \textsc{MOND} limit $|\varepsilon| \ll 1$, as one sees from Fig. \ref{fig:estarscatter}.
We theoretically explain why going outside the MOND limit allows for larger $M/L_*$ in Appendix~\ref{sec:sfdm:mond}.
To confirm this, we did the fits again but required that the galaxies are in the \textsc{MOND} limit, $|\varepsilon| < 0.4$. For the rationale behind the precise value 0.4, please refer to Appendix~\ref{sec:nomondrequired}.
The resulting $Q_*$ values are shown in Fig.~\ref{fig:fYhist04}.

As one can see from Fig.~\ref{fig:chi2cdfs}, the fits with the requirement  $|\varepsilon| < 0.4$ are not much worse than those without.
The averaged $Q_*$ is now smaller than in \textsc{MOND}; for the $\Qual=1$ galaxies, the median stellar disk $M/L_*$ is about $10\%$ smaller than for \textsc{MOND}.
This confirms superfluids outside the MOND limit as one reason for the large $Q_*$ values in \textsc{SFDM} (see also Appendix~\ref{sec:largeML}).

\begin{figure*}
    \centering
    \includegraphics[width=.49\textwidth]{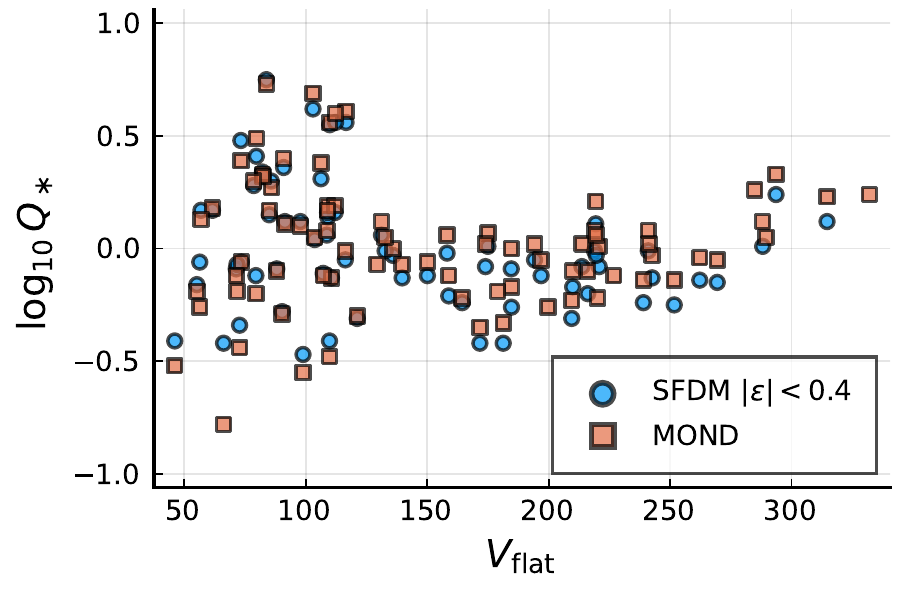}
    \includegraphics[width=.49\textwidth]{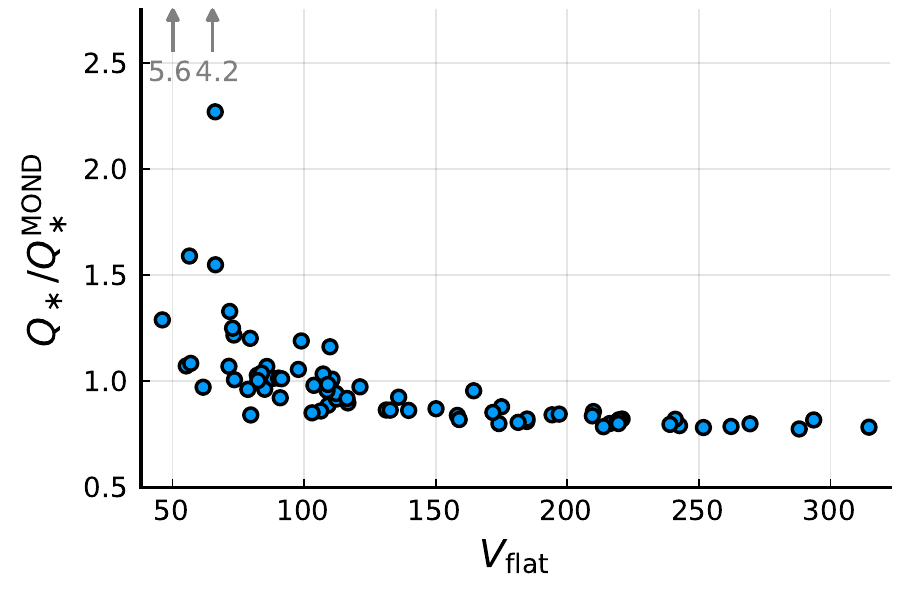}
    \caption{
        Best-fit $Q_*$ for SFDM restricted to the MOND limit ($|\varepsilon| < 0.4$) and MOND as a function of the observed flat rotation velocity, $V_{\mathrm{flat}}$, for the $\Qual=1$ galaxies.
        For SFDM, some galaxies can barely satisfy the condition $|\varepsilon| < 0.4$ and therefore give a bad fit to the data.
        Their best-fit $Q_*$ is meaningless, and they are excluded from the SFDM fit results.
        Specifically, we exclude galaxies that have both $\varepsilon > 0.38$ and $\chi^2 > 100$.
        Left:
        Best-fit $Q_*$ for SFDM and MOND.
        As discussed at the beginning of Sect.~\ref{sec:results}, a few
        galaxies fall below the lower boundary of the plot.
        Their $(V_{\mathrm{flat}} \cdot (\mathrm{km}\,\mathrm{s}^{-1})^{-1},\, \log_{10} Q_*,\, \log_{10} Q_*^{\mathrm{MOND}})$ values are
            $(50.1, -1.25, -2.00)$, $(65.2, -1.38, -2.00)$, and $(66.3, -1.81, -2.00)$.
        Right:
        Best-fit $Q_*$ values for SFDM relative to those for MOND.
        Gray arrows indicate two outliers with relatively large $Q_*/Q_*^{\mathrm{MOND}}$.
    }
    \label{fig:this}
\end{figure*}

Another reason why \textsc{SFDM} does not universally give smaller $Q_*$ than \textsc{MOND} is that the best-fit $Q_*$ depends on the type of galaxy.
In \textsc{SFDM}, $Q_*$ is systematically smaller for galaxies with relatively large accelerations $a_b$, but
    not for those with small accelerations.
This can be seen, for example, in the right panel of Fig.~\ref{fig:this}, which shows the best-fit $Q_*$ of each galaxy in \textsc{SFDM} relative to the best-fit $Q_*$ for \textsc{MOND} as a function of the observed asymptotic rotation velocity, $V_{\mathrm{flat}}$.
A larger $V_{\mathrm{flat}}$ is associated with larger accelerations -- this is, where \textsc{SFDM} systematically gives smaller $Q_*$ than MOND.
There are similar trends for surface brightness and the gas fraction;
    both also correlate with the accelerations $a_b$ (see Appendix~\ref{sec:results:sfdm:MLtrends} for more details).

The reason for this trend is that the smaller $a_0$ value of SFDM makes the acceleration $\sqrt{a_0 a_b}$  smaller than in \textsc{MOND}.
This acceleration $\sqrt{a_0 a_b}$ is dominant at small $a_b$, so that \textsc{SFDM} needs more baryonic mass than MOND to get the same total acceleration  (at least if $a_{\mathrm{SF}}$ is negligible).
This is explained in more detail in Appendix~\ref{sec:sfdm:less}.

This trend in $Q_* / Q_*^{\mathrm{MOND}}$ shows not only how SFDM is different from MOND but also how SFDM does not comply with expectations from SPS models.
The general idea is that MOND is known to
be in good agreement with the expectations of SPS
\citep{Sanders1996,Sanders1998,McGaugh2004,GalRev2020}, so any systematic trend in $Q_*/Q_*^{\mathrm{MOND}}$ is potentially problematic.

In our case, both the MOND and the SFDM fits show increased scatter in $Q_*$ for small galaxies.
This is shown in the left panel of Fig.~\ref{fig:this}.
One reason for the increased scatter is that the data for smaller galaxies is generally of lower quality.
Another reason is that these galaxies tend to be gas-dominated, in which case adjusting $Q_*$ has only a small effect on the overall fit. Consequently, larger changes in $Q_*$ are needed to
impact the fit quality. Indeed, the scatter increases dramatically at precisely the scale where gas typically begins to dominate the mass budget \citep{McGaugh2011,Lelli2022}.

The SFDM fits show an upward trend in $Q_*$ for small galaxies.
This corresponds to the upward trend in $Q_*/Q_*^{\mathrm{MOND}}$ shown in the right panel of Fig.~\ref{fig:this}.
The problem is that such trends of the stellar $M/L_*$ with galaxy properties are not expected from SPS models
for the late type galaxies that compose the SPARC sample.
If anything, we expect the stellar $M/L_*$ to increase with mass \citep[e.g.,][]{BdJ2001}, opposite the sense of the trend found here. Indeed, we utilize the near-infrared Spitzer [3.6] band specifically to minimize variations in the mass-to-light ratio.
In the most recent stellar population models of late type galaxies
\citep[e.g.,][]{Schombert2019,Schombert2022}, accounting for the shape of the stellar metallicity distribution tends to counteract the modest effect of stellar age in the near-infrared, leading to the expectation of a nearly constant $M/L_*$.

Given our simplistic fitting procedure, Fig.~\ref{fig:this}, left, alone may or may not be convincing evidence for a systematic trend in $Q_*$.
Still, our fitting procedure is well suited to identify relative differences between MOND and SFDM, and we have a good theoretical understanding of this difference.
Thus, we expect the trends in our best-fit $Q_*/Q_*^{\mathrm{MOND}}$ to be robust.
Since $Q_*^{\mathrm{MOND}}$ is known to be in good agreement with SPS expectations, we interpret the systematic trend seen in the right panel of Fig.\ \ref{fig:this} as a good indicator of trends in absolute $Q_*$ revealing a tension between SFDM and SPS.

In Appendix~\ref{sec:RARplots} we show how our fit results illustrate that only the MOND limit of SFDM can reproduce MOND-like galactic scaling relations such as the RAR without having to adjust the boundary condition $\varepsilon$ separately for each galaxy.

\subsection{Tension with strong lensing}

Irrespective of the resulting $M/L_*$ values,
    there is a price to pay for enforcing the \textsc{MOND} limit in \textsc{SFDM}.
A \textsc{MOND}-like rotation curve in the \textsc{MOND} limit $|\varepsilon_*| \ll 1$ can only be achieved by reducing the acceleration created by the gravitational pull of the superfluid.
As a result, the total dark matter mass in those galaxies, $M_{200}^{\mathrm{DM}}$, comes out to be quite small.
Here, $M_{200}^{\mathrm{DM}}$ is the dark matter mass within the radius $r_{200}$ where the mean dark matter density drops below $\rho_{200} = 200 \times 3 H^2/(8 \pi G)$ with the Hubble constant $H$.
We adopt $H = 67.3\,\mathrm{km}\,\mathrm{s}^{-1}\,\mathrm{Mpc}^{-1}$.

\begin{figure}[t]
 \centering
\includegraphics[width=0.48\textwidth]{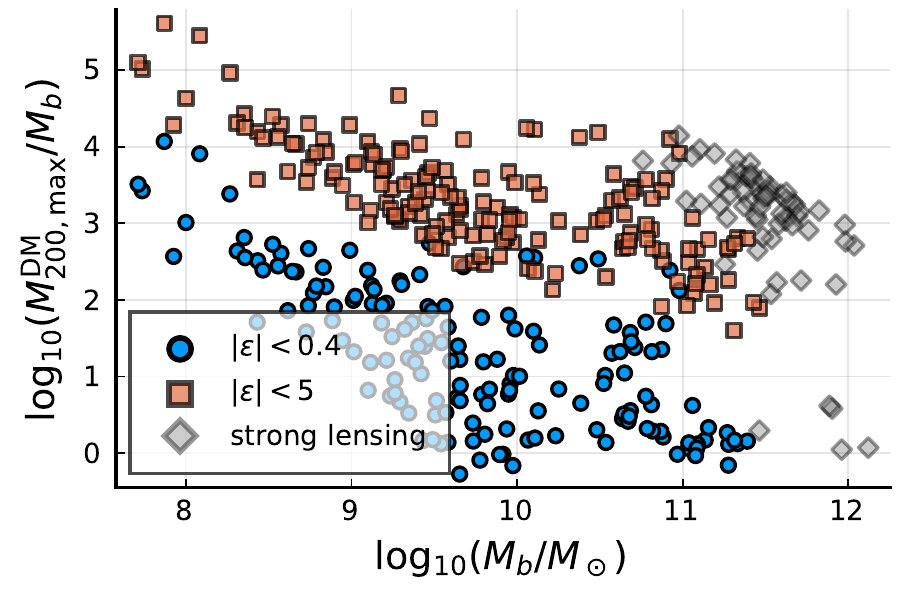}
 \caption{
     Total baryonic mass, $M_b$, versus the upper bound, $M_{200,\mathrm{max}}^{\mathrm{DM}}/M_b$, of the ratio of the total dark matter mass, $M_{200}^{\mathrm{DM}}$, and the baryonic mass for the SPARC galaxies.
     This is for $(M/L_*)_{\mathrm{disk}} = 0.5$ and $(M/L_*)_{\mathrm{bulge}} = 0.7$.
     The upper bound comes from the condition that the rotation curve is in the proper MOND limit ($|\varepsilon| < 0.4$, blue circles) or at least the pseudo-MOND limit ($|\varepsilon| < 5$, red squares).
     Also shown are the best-fit results from the strong lensing analysis of \cite{Hossenfelder2019},
        where we use their best-fit $M_{200}^{\mathrm{DM}}$ for the vertical axis.
    }
 \label{fig:M200max}
\end{figure}

A small $M_{200}^{\mathrm{DM}}$ is not a problem for fitting \textsc{SFDM} to the observed rotation curves, but it is a problem if one \emph{also} wants to fit strong lensing data.
Indeed, \citet{Hossenfelder2019} find that \textsc{SFDM} requires ratios  $M_{200}^{\mathrm{DM}}/M_b \gtrsim 1000$ to fit strong lensing constraints, where $M_b$ is the total baryonic mass.
Requiring a rotation curve in the \textsc{MOND} limit $|\varepsilon_*| \ll 1$ for the \textsc{SPARC} galaxies produces average masses at least an order of magnitude smaller.

To illustrate the problem with strong lensing, we have in Fig.~\ref{fig:M200max} plotted the (logarithm of) the total baryonic and the maximum possible total dark matter mass given our requirement $|\varepsilon| < 0.4$ in comparison to the values found in \citet{Hossenfelder2019}.
For this, we used $Q_* = 1$ for all \textsc{SPARC} galaxies because the precise stellar mass-to-light ratio is irrelevant here.
``Maximum possible'' here refers not only to the requirement $|\varepsilon| < 0.4$ but also to uncertainties in how to determine the radius where the superfluid core is matched to an \textsc{NFW} halo:
We used the transition radius that gives the largest total dark matter mass (see Appendix~\ref{sec:M200estar04} for details).

The best \textsc{SFDM} fits to strong lensing data tend to have $M_b \gtrsim 10^{11}\,M_\odot$ and $M_{200}^{\mathrm{DM}}/M_b \gtrsim 1000$.
In contrast, despite our generous NFW matching procedure, the \textsc{SPARC} galaxies with $M_b > 10^{11}\,M_\odot$ have $M_{200}^{\mathrm{DM}}/M_b < 10$ when restricted to have rotation curves in the \textsc{MOND} limit $|\varepsilon| < 0.4$.
This is a stark contrast.

The \textsc{SPARC} galaxies do not reach baryonic masses quite as large as the lensing galaxies from \citet{Hossenfelder2019}.
But from Fig.~\ref{fig:M200max} it seems clear that the trend goes into the wrong direction:
The larger the galaxy, the smaller the maximum possible $M_{200}^{\mathrm{DM}}/M_b$ (given $|\varepsilon| < 0.4$).

The quoted values of $M^{\mathrm{DM}}_{200}/M_b$ for the strong lensing fits may seem high.
But at least from a $\Lambda$CDM abundance matching perspective, these are actually expected due to the large baryonic masses of the lensing galaxies \citep{Hossenfelder2019}.
Nevertheless, somewhat smaller ratios may be possible.
The fitting procedure of \citet{Hossenfelder2019} did not aim to produce small $M^{\mathrm{DM}}_{200}/M_b$ values.
It only aimed to simultaneously fit the observed Einstein radii and velocity dispersions of the lensing galaxies.
Probably somewhat smaller $M^{\mathrm{DM}}_{200}/M_b$ ratios are possible at the cost of somewhat worse fits of the Einstein radii and velocity dispersions.
However, given the size of the discrepancy in Fig.~\ref{fig:M200max},
    we do not expect that the proper MOND limit of SFDM can reasonably fit these data.

\begin{figure}[t]
 \centering
\includegraphics[width=0.48\textwidth]{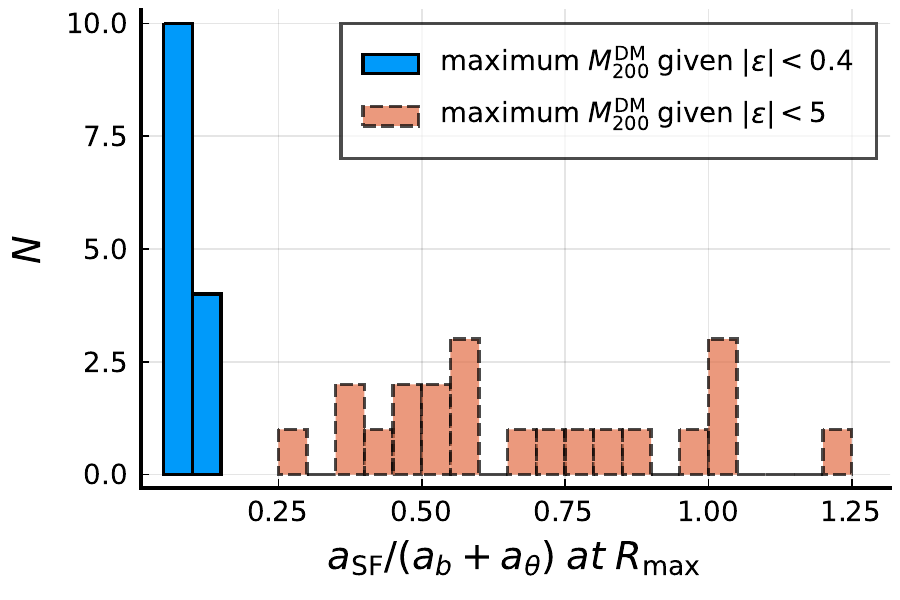}
 \caption{
        Histogram of $a_{\mathrm{SF}}$ relative to $a_b + a_\theta$ at the last rotation curve data point at $R = R_{\mathrm{max}}$.
        This is for the maximum possible total dark matter mass, $M_{200}^{\mathrm{DM}}$, given the condition $|\varepsilon| < 0.4$ (blue) and $|\varepsilon| < 5$ (red).
        We take $Q_* = 1$ for all galaxies and show only the galaxies with $M_b > 10^{11}\,M_\odot$, relevant for strong lensing.
    }
 \label{fig:M200maxVccorrections}
\end{figure}

To study this closer, we did another calculation in which we allowed galaxies into the pseudo-\textsc{MOND} limit.
Concretely, we redid the maximum $M_{200}^{\mathrm{DM}}/M_b$ calculation with the requirement   $|\varepsilon| < 5$.
The precise value 5 is again somewhat arbitrary.
We explain why this is a pragmatic choice in Appendix~\ref{sec:nomondrequired}.
We see from Fig.~\ref{fig:M200max} that in the pseudo-\textsc{MOND}-limit galaxies with $M_b > 10^{11}\,M_\odot$ still have smaller total dark matter masses than what is required for strong lensing, although the problem is less severe than in the proper \textsc{MOND} limit.
Somewhat worse but still acceptable fits to the strong lensing data might be able to ameliorate this.

The pseudo-\textsc{MOND} limit, however, is unsatisfactory for two reasons.
First, it relies sensitively on ad hoc finite-temperature corrections of \textsc{SFDM} that may be unphysical.
Second, the pseudo-MOND limit has the disadvantage that the acceleration from the superfluid, $a_{\mathrm{SF}}$, can be significant.
If $a_{\mathrm{SF}}$ is significant, we do not automatically get the \textsc{MOND}-type galactic scaling relations,
    since then the superfluid boundary condition must be adjusted for each galaxy to get the correct total acceleration.
In this case, \textsc{SFDM} loses its advantage over \textsc{CDM} despite the phonon force being close to $\sqrt{a_0 a_b}$.

Figure~\ref{fig:M200maxVccorrections} shows the size of $a_{\mathrm{SF}}$ relative to $a_b + a_\theta$ at the last rotation curve data point at $R = R_{\mathrm{max}}$,
    assuming the maximum total dark matter masses from Fig.~\ref{fig:M200max}.
Indeed, for the pseudo-\textsc{MOND} limit, $a_{\mathrm{SF}}$ is significant for the galaxies with $M_b > 10^{11}\,M_\odot$ relevant for strong lensing.
This is despite SFDM having a very cored dark matter profile.
Thus, also with the pseudo-MOND limit, we cannot get MOND-like rotation curves and strong lensing at the same time.

\subsection{Two-field SFDM}
\label{sec:twofield}

\begin{figure}[t]
 \centering
 \includegraphics[width=.48\textwidth]{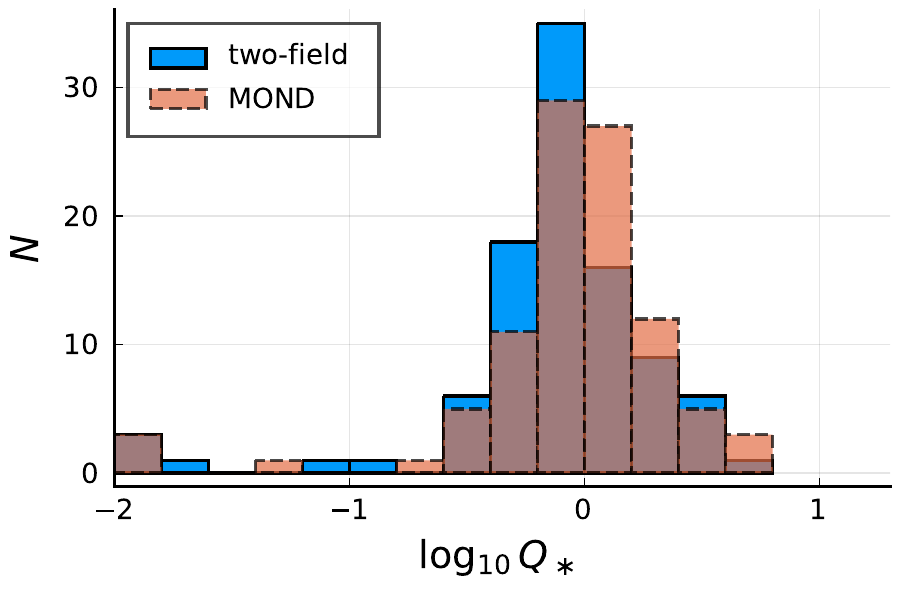}
 \caption{
     Same as Fig.~\ref{fig:fYhists} but for two-field SFDM.
    }
 \label{fig:fYhisttwofield}
\end{figure}

For two-field \textsc{SFDM}, the $Q_*$-distribution (Fig.\ \ref{fig:fYhisttwofield}) and the corresponding \textsc{CDF} (Fig.\ \ref{fig:chi2cdfs}) are similar to those of standard \textsc{SFDM}.
However, the two-field model is constructed so that it is easier for the phonon force to be close to the \textsc{MOND}-like value $\sqrt{a_0 a_b}$.
For this reason, the best fits for two-field \textsc{SFDM} all have $|\varepsilon_*| \ll 1$, as expected (Fig.~\ref{fig:estarscatter}).
Only for two galaxies (NGC6789, UGC0732) does $\varepsilon_*$ become larger than $0.1$.
Its largest value is $0.36$ for NGC6789.
That is, the acceleration $a_b + a_\theta$ is almost always close to the MOND-like value $a_b + \sqrt{a_0 a_b}$ (see Appendix~\ref{sec:twofield:ML} for more details).

Thus, two-field \textsc{SFDM} can easily have large dark matter masses and $|\varepsilon_*| \ll 1$ at the same time.
It does not have the same problem with strong lensing as the proper MOND limit $|\varepsilon_*| \ll 1$ of standard SFDM.
Two-field \textsc{SFDM} does, however, still have a problem with strong lensing similar to the pseudo-\textsc{MOND} limit of standard \textsc{SFDM}.
Large total dark matter masses imply that the rotation curve receives significant corrections from the superfluid's gravitational pull $a_{\mathrm{SF}}$.
This is despite two-field \textsc{SFDM} having, like standard \textsc{SFDM}, a very cored density profile.
For this reason, large total dark matter masses imply systematically higher rotation curve velocities than \textsc{MOND}.

\begin{figure}[t]
 \centering
 \includegraphics[width=0.48\textwidth]{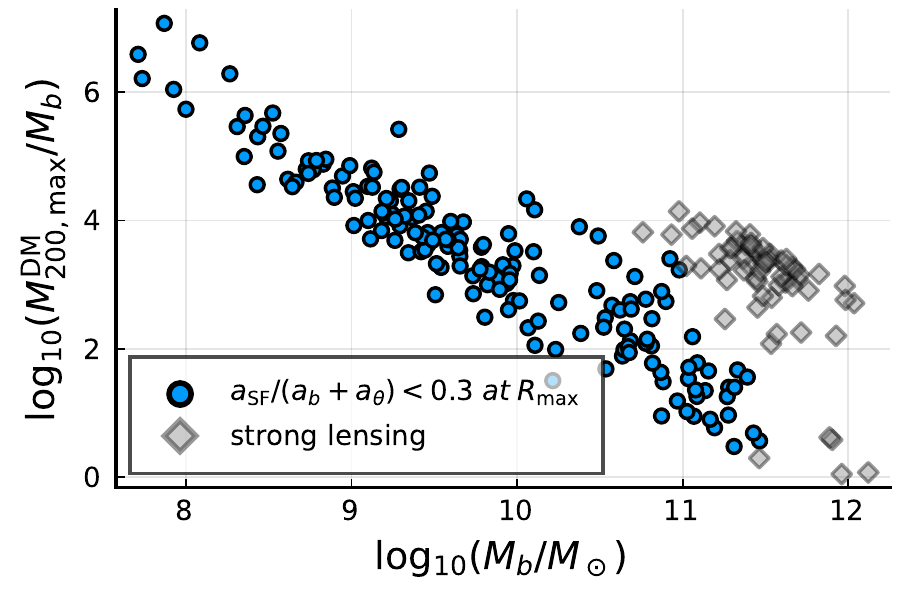}
 \caption{
    Same as Fig.~\ref{fig:M200max} but for the two-field model and with the requirement that $a_{\mathrm{SF}}$ is at most $30\%$ as large as $a_b + a_\theta$ at the last rotation curve data point, $R_{\mathrm{max}}$.
    }
 \label{fig:M200maxtf}
\end{figure}

To illustrate this problem we depict in Fig.~\ref{fig:M200maxtf} the maximum possible total dark matter mass for the two-field model,
    given the requirement that $a_{\mathrm{SF}}$ is at most $30\%$ as large as $a_b + a_\theta$ at the last rotation curve data point at $R=R_{\mathrm{max}}$ (see Appendix~\ref{sec:appendix:twofieldlensing}).
The scatter in the distribution is smaller in the two-field model because it depends less on the details of the baryonic matter distribution (see Appendix~\ref{sec:appendix:twofieldlensing}).
As one can see, in the two-field model the discrepancy with the lensing data is weaker than for the proper \textsc{MOND} limit $|\varepsilon_*| \ll 1$ of standard \textsc{SFDM}, but still present.
Avoiding this tension with the lensing data would require rotation curves that are even less \textsc{MOND}-like.
Whether or not somewhat worse but still acceptable fits to the strong lensing data could ameliorate this problem requires further investigation.

\section{Conclusion}

We have found that it is difficult to reproduce the achievements of \textsc{MOND} with the models that have so far been proposed for SFDM.

\begin{acknowledgements}

This work was supported by the DFG (German Research Foundation) under grant number HO 2601/8-1 together with the joint NSF grant PHY-1911909.

\end{acknowledgements}

\bibliographystyle{aa}
\bibliography{sfdm-sparc}

\begin{appendix}

\section{The models}
\label{sec:appendix:models}

Here, we introduce both the original SFDM model from \cite{Berezhiani2015} and the two-field model from \cite{Mistele2020} in more detail.

\subsection{Standard SFDM}
\label{sec:appendix:estar}

In standard SFDM, in an equilibrium superfluid core of a galaxy, the phonon field, $\theta$, is determined by the equation \citep{Berezhiani2018}
\begin{align}
 \label{eq:phononFull}
 \vec{\nabla} \left( \frac{(\vec{\nabla} \theta)^2 + 2m (\frac{2 \beta}{3} -1) \hat{\mu} }{\sqrt{(\vec{\nabla} \theta)^2 + 2m(\beta-1)\hat{\mu}}}  \vec{\nabla} \theta \right) =  \frac{\alpha}{2 M_{\mathrm{Pl}}} \, \rho_b \,,
\end{align}
and the field $\hat{\mu}$ is determined by the Poisson equation
\begin{align}
 \label{eq:poissonFull}
 \Delta \left( - \frac{\hat{\mu}}{m} \right) = 4 \pi G \left( \rho_b + \rho_{\mathrm{SF}}\left[\hat{\mu}, \vec{\nabla} \theta\right] \right) \,,
\end{align}
with the superfluid energy density, $\rho_{\mathrm{SF}}$,
\begin{align}
 \rho_{\mathrm{SF}}\left[\hat{\mu}, \vec{\nabla} \theta\right] = \frac{2\sqrt{2}}{3} m^{5/2} \Lambda \, \frac{3(\beta-1) \hat{\mu} + (3-\beta) \frac{(\vec{\nabla} \theta)^2}{2m}}{\sqrt{ (\beta-1) \hat{\mu}  + \frac{(\vec{\nabla} \theta)^2}{2m} }} \,.
\end{align}
Here, $m$, $\Lambda$, and $\alpha$ are model parameters.
The quantity $\hat{\mu}(\vec{x}) = \mu_{\mathrm{nr}} - m \phi_N(\vec{x})$ is a combination of the (constant) non-relativistic chemical potential $\mu_{\mathrm{nr}}$ and the Newtonian gravitational potential $\phi_N(\vec{x})$.
It controls how much the superfluid weighs, depending on a boundary condition (see Appendix~\ref{sec:sfdm:calc}).
The parameter $\beta$ parametrizes finite-temperature corrections, which are needed to avoid an instability \citep{Berezhiani2015}.
The phonon force $\vec{a}_\theta$ is given by
\begin{align}
 \vec{a}_\theta = - \frac{\alpha \Lambda}{M_{\mathrm{Pl}}} \vec{\nabla} \theta \,.
\end{align}

We mainly used the no-curl approximation for the $\theta$ equation of motion.
That is, for the solution of this equation,
    which is of the form $\vec{\nabla} ( g \cdot \vec{\nabla} \theta) = \vec{a}_b$ for some $g$,
    we assumed $g \vec{\nabla} \theta = \vec{a}_b$.
This is a standard approximation in MOND and it works well also for SFDM \citep{Hossenfelder2020}.

In the no-curl approximation, the quantity $\varepsilon_*(\vec{x})$ (see Eq.~\eqref{eq:estar}) is useful.
As we will see, it controls how closely SFDM resembles MOND.
As discussed in \citet{Mistele2020}, we have
\begin{align}
  |\vec{a}_\theta| = \sqrt{a_0 |\vec{a}_b|} \cdot \sqrt{x_\beta(\varepsilon_*)} \,,
\end{align}
where
\begin{align}
 a_0 = \frac{\alpha^3 \Lambda^2}{M_{\mathrm{Pl}}} \,
\end{align}
and where $x_\beta(\varepsilon_*)$ is determined by the cubic equation
\begin{align}
 \label{eq:xeq}
  0 &= x_\beta^3 + 2 \left(\frac{2\beta}{3} - 1\right) \varepsilon_* \cdot x_\beta^2 + \left( \left(\frac{2\beta}{3}-1\right)^2 (\varepsilon_*)^2 - 1\right) x_\beta - \left(\beta-1\right) \varepsilon_* \,.
\end{align}
That is, $\vec{a}_\theta$ is an algebraic function of $\vec{a}_b$ and $\varepsilon_*$.
This also allows us to write $\rho_{\mathrm{SF}}$ as a function of $|\vec{a}_b|$ and $\varepsilon_*$,
\begin{align}
 \label{eq:rhoSFnocurl}
 \rho_{\mathrm{SF}}(\varepsilon_*, |\vec{a}_b|) = \frac23 \frac{m^2}{\alpha} \sqrt{a_0 |\vec{a}_b|} M_{\mathrm{Pl}} \cdot f_\beta(\varepsilon_*) \,,
\end{align}
where
\begin{align}
 f_\beta(\varepsilon_*) = \frac{
     x_\beta(\varepsilon_*) (3-\beta) + 3(\beta-1) \varepsilon_*
}{\sqrt{
    x_\beta(\varepsilon_*) + (\beta-1) \varepsilon_*
}} \,.
\end{align}

\begin{figure*}
  \centering
  \includegraphics[width=\textwidth]{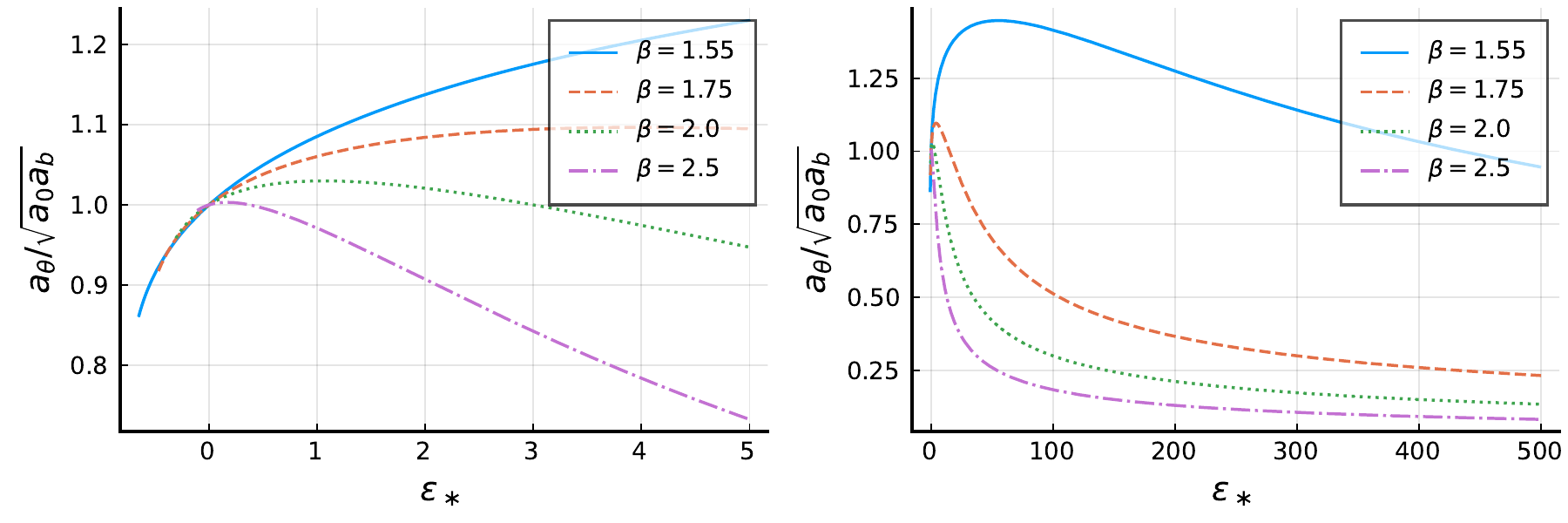}
  \caption{
     How close the phonon force, $a_\theta$, is to its MOND limit value $\sqrt{a_0 a_b}$ as a function of $\varepsilon_*$.
     Left: For $\varepsilon_* < 5$ and for various values of the parameter, $\beta$, that parametrizes finite-temperature corrections.
     Right: Same but up to $\varepsilon_* = 500$.
  }
  \label{fig:atheta-estar}
\end{figure*}

\begin{figure}
 \centering
 \includegraphics[width=.48\textwidth]{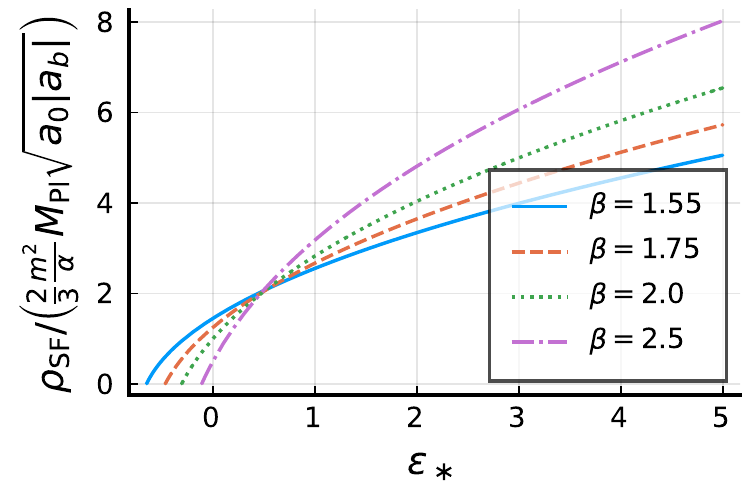}
 \caption{
     Function $f_\beta(\varepsilon_*) = \rho_{\mathrm{SF}} \cdot \left(\frac23 \frac{m^2}{\alpha} M_{\mathrm{Pl}} \sqrt{a_0 |\vec{a}_b|} \right)^{-1}$ as a function of $\varepsilon_*$ for different values of $\beta$.
     This is a concave, monotonically increasing function.
     It does not depend on any model parameters except $\beta$.
    }
 \label{fig:rhoSF-estar}
\end{figure}

For both $a_\theta$ and $\rho_{\mathrm{SF}}$, we have a prefactor proportional to $\sqrt{a_0 a_b}$ multiplied by a function that depends on $\varepsilon_*$ and $\beta$ only.
For later use, we record the expansion of this second function for small and large values of $\varepsilon_*$.
For $|\varepsilon_*| \ll 1$, we have
\begin{subequations}
\begin{align}
 \label{eq:estarexpansion:small:atheta}
 \sqrt{x_\beta(\varepsilon_*)} &= 1 + \mathcal{O}(\varepsilon_*) \,, \\
 \label{eq:estarexpansion:small:rhoSF}
 f_\beta(\varepsilon_*) &= (3-\beta) + \mathcal{O}(\varepsilon_*) \,.
\end{align}
\end{subequations}
For $\varepsilon_* \gg 1$,
\begin{subequations}
\begin{align}
  \sqrt{x_\beta(\varepsilon_*)} &= \frac{\sqrt{\beta-1}}{2\beta-3} \frac{3}{\sqrt{\varepsilon_*}} + \mathcal{O}(\varepsilon_*^{-3/2}) \,, \\
 f_\beta(\varepsilon_*) &= 3 \sqrt{\varepsilon_*} \sqrt{\beta-1} + \mathcal{O}(\varepsilon_*^{-3/2}) \,.
\end{align}
\end{subequations}
In general, $f_\beta$ is a monotonically increasing, concave function of $\varepsilon_*$ (see Fig.~\ref{fig:rhoSF-estar}).
The function $\sqrt{x_\beta}$ is not monotonic (see Fig.~\ref{fig:atheta-estar}).

To avoid a negative or imaginary $\rho_{\mathrm{SF}}$ as well as an instability,
    $\varepsilon_*$ must be larger than some minimum value $\varepsilon_{*\mathrm{min}}$.
\cite{Berezhiani2015} assumed $\hat{\mu} > 0$, corresponding to $\varepsilon_* > 0$,
    but this is not required from their Lagrangian.
It is an assumption with unclear justification.
Here, we were more generous to the model and allowed $\hat{\mu}$ to become negative as long as $\rho_{\mathrm{SF}}$ stays positive.
The corresponding minimum value of $\varepsilon_*$ is determined by $\rho_{\mathrm{SF}} = 0,$ which is equivalent to $f_\beta(\varepsilon_{*\mathrm{min}}) = 0$.
For example,
\begin{align}
 \label{eq:estarmin}
 \varepsilon_{*\mathrm{min}} = \begin{cases}
        - \frac{29}{31} \sqrt{\frac{15}{31}} \approx -0.65\,, \quad &\beta=1.55 \\
        - \frac14 \sqrt{\frac32} \approx -0.31\,, \quad &\beta=2 \\
        0\,, \quad &\beta=3
                               \end{cases} \,.
\end{align}
With Eq.~\eqref{eq:estar}, this translates into a minimum value for $\hat{\mu}/m$,
\begin{align}
 \label{eq:muhatmin}
 \frac{\hat{\mu}(\vec{x})}{m} > \varepsilon_{*\mathrm{min}} \frac{\alpha M_{\mathrm{Pl}} |\vec{a}_b(\vec{x})|}{2m^2} \,.
\end{align}

\subsubsection{MOND limit}
\label{sec:appendix:sfdm:mond}

In SFDM, the total acceleration inside the superfluid core of a galaxy can be written as
\begin{align}
 \vec{a}_{\mathrm{tot}} = \vec{a}_\theta + \vec{a}_b + \vec{a}_{\mathrm{SF}} \,,
\end{align}
where  $\vec a_{\theta}$ is the acceleration created by the phonon force, $\vec a_{\rm SF}$ the acceleration stemming from the normal gravitational attraction of the superfluid, and $\vec a_b$ that stemming from the mass of the baryons.

For \textsc{SFDM} to make sense, one needs the superfluid to at least approximately reproduce \textsc{MOND} rotation curves without being sensitive to the choice of the boundary condition $\varepsilon$.
Otherwise, one does not get the observed MOND-like scaling relations without carefully adjusting the boundary condition separately for each galaxy.
That is to say, without the \textsc{MOND} limit of \textsc{SFDM}, one might as well use \textsc{CDM}.

Rotation curves in SFDM approximate those in MOND when the total acceleration $a_{\mathrm{tot}}$ approximately has the form $a_b + \sqrt{a_0 a_b}$.
This corresponds to two conditions.
First, the phonon force $a_\theta$ must be close to $\sqrt{a_0 a_b}$.
Second, the superfluid's gravitational pull $a_{\mathrm{SF}}$ must be negligible.

The numerical values of the model parameters and the boundary condition of the Poisson equation for $\hat{\mu}$ determine in which coordinate-range \textsc{SFDM} approximates \textsc{MOND} for a given baryonic mass distribution.
Specifically, the \textsc{MOND} limit corresponds to $|\varepsilon_*(\vec{x})| \ll 1$.
In this $|\varepsilon_*| \ll 1$ limit, both conditions to reproduce MOND rotation curves are automatically fulfilled:
The phonon force is close to $\sqrt{a_0 a_b}$ and the superfluid's gravitational pull $a_{\mathrm{SF}}$ is negligible.
The phonon force is close to $\sqrt{a_0 a_b}$ because, for $|\varepsilon_*| \ll 1$, the (no-curl version of) the phonon field equation Eq.~\eqref{eq:phononFull} has the MOND-like form $ |\vec{a}_\theta| \vec{a}_\theta = a_0 \vec{a}_b$.
This corresponds to the small-$|\varepsilon_*|$ expansion $\sqrt{x_\beta} = 1$ from Eq.~\eqref{eq:estarexpansion:small:atheta}.
We explicitly show that $a_{\mathrm{SF}}$ is negligible (i.e., that the second condition is fulfilled) at the end of this subsection.

However, even when $\varepsilon_*$ is of order one, deviations of the phonon force from the \textsc{MOND} form $\sqrt{a_0 a_b}$ remain within the percent range,
    at least for $\beta = 2$ (see Fig.~\ref{fig:atheta-estar}).
It will therefore in the following be handy to define a ``pseudo-MOND limit,'' $\varepsilon_* = \mathcal{O}(1)$.
If this condition is fulfilled, the phonon field no longer satisfies a \textsc{MOND}-like equation, but the acceleration $a_\theta$ of an isolated\footnote{If a galaxy is not isolated, the phonon force may be different than in MOND because the external field effect will be different since $\theta$ does not satisfy a MOND-like equation.}  galaxy is numerically still relatively close to $\sqrt{a_0 a_b}$.
One difference to the proper MOND limit $|\varepsilon_*| \ll 1$ is that now the second condition for having MOND-like rotation curves is not automatically fulfilled.
The superfluid's gravitational pull $a_{\mathrm{SF}}$ can be significant.
So the observed scaling relations are fulfilled automatically only if $a_{\mathrm{SF}}$ stays sufficiently small,
    which needs to be checked separately for each solution.

A different problem with the pseudo-MOND limit is that it depends sensitively on the details of the ad hoc finite-temperature corrections introduced in \cite{Berezhiani2015} to avoid an instability.
For example, the pseudo-MOND limit works only for $\beta$ close to $2$, as can be seen from Fig.~\ref{fig:atheta-estar}, left.
Just as these ad hoc finite-temperature corrections, the pseudo-MOND limit may turn out to be unphysical.

It now remains to show that the superfluid's gravitational pull $a_{\mathrm{SF}}$ is negligible in the proper MOND limit $|\varepsilon_*| \ll 1$.
For simplicity, we assume a point mass baryonic energy density, $\rho_b(\vec{x}) = M_b \delta(\vec{x})$, which gives $a_b = GM_b/r^2$.
Then, for $|\varepsilon_*| \ll 1$,
    we have $\rho_{\mathrm{SF}} \propto 1/r$ (see Eq.~\eqref{eq:estarexpansion:small:rhoSF}).
The superfluid's mass is then
\begin{align}
\label{eq:MSFMOND}
M_{\mathrm{SF}}(r) = M_b \cdot \left(\frac{r}{r_c}\right)^2 \,,
\end{align}
where
\begin{align}
\label{eq:rc}
r_c^{-2} = \sqrt{2\pi} \frac{m^2}{\alpha}  \sqrt{\frac{a_0}{M_b}} \left(1 - \frac{\beta}{3}\right) \,.
\end{align}
We can now estimate the superfluid's gravitational pull $a_{\mathrm{SF}}$ compared to $a_b + a_\theta$.
Roughly,
\begin{align}
 \frac{a_{\mathrm{SF}}}{a_b + a_\theta} \simeq \frac{(r/r_c)^2}{1 + r/ r_{\mathrm{MOND}}} \,,
\end{align}
with $r_{\mathrm{MOND}} = \sqrt{G M_b/a_0}$.
This ratio $a_{\mathrm{SF}}/(a_b + a_\theta)$ can be larger than a fraction $\delta$ only at a radius $r_\delta$ that satisfies
\begin{align}
 r_\delta \gtrsim \delta \frac{r_c^2}{r_{\mathrm{MOND}}} = \delta \cdot \frac{1}{\sqrt{2\pi G}} \frac{\alpha}{m^2} \frac{1}{1 - \frac{\beta}{3}} = 53\,\mathrm{kpc} \cdot \left(\frac{\delta}{10\%}\right) \,,
\end{align}
where we used the fiducial numerical parameters from \cite{Berezhiani2018} for the last equality.
That is, assuming the proper MOND limit $|\varepsilon_*| \ll 1$,
    the superfluid's mass becomes important only at radii larger than where rotation curves are measured.

\subsubsection{Reaching the proper MOND limit}
\label{sec:sfdm:reachingmond}

As already mentioned in \cite{Mistele2020}, reaching the proper MOND limit $|\varepsilon_*| \ll 1$ is not always possible.
To avoid a negative $\rho_{\mathrm{SF}}$ there is a minimum value for $\hat{\mu}$ (see Eq.~\eqref{eq:muhatmin}).
Typically, $\hat{\mu}$ is a decreasing function of galactocentric radius and the Poisson equation Eq.~\eqref{eq:poissonFull} tells us that $\hat{\mu}/m$ has a derivative of about $-G M/r^2$ where $M$ includes both the baryonic and superfluid mass.
Using the baryonic mass $M_b$ as a lower bound on $M$ then gives a lower bound on $\hat{\mu}/m$.
Roughly, $\hat{\mu}/m \gtrsim \hat{\mu}_{\mathrm{min}}(r)/m + G M_b/r$.
This translates into a rough lower bound on $\varepsilon_*$,
\begin{align}
 \label{eq:estarlowerbound}
  \varepsilon_*
  \gtrsim \frac{2m^2 r}{\alpha M_{\mathrm{Pl}}} + \varepsilon_{*\mathrm{min}}
  = 0.10 + 0.41 \cdot \left(\frac{r}{18\,\mathrm{kpc}}-1\right)\,,
\end{align}
where we used $a_b = GM_b/r^2$ and the fiducial parameter values from \cite{Berezhiani2018}.

Thus, small galaxies can easily reach the proper MOND limit $|\varepsilon_*| \ll 1$ over the whole range where their rotation curve is measured.
One just needs to ensure that the superfluid mass is not too large, which is usually possible.

In contrast, larger galaxies sometimes struggle to satisfy the MOND limit condition $|\varepsilon_*| \ll 1$,
    even when the superfluid mass is as small as possible.

\subsubsection{Naive upper bound on MOND limit dark matter mass}
\label{sec:appendix:M200}

In Appendix~\ref{sec:appendix:sfdm:mond}, we saw that being in the proper MOND limit $|\varepsilon_*| \ll 1$ restricts the superfluid's gravitational pull $a_{\mathrm{SF}}$ to be relatively small.
Similarly, the MOND limit restricts the total dark matter mass to be relatively small,
    even if we include the non-condensed phase outside the superfluid core.

To see this, consider a galaxy with a superfluid core in the MOND limit $|\varepsilon_*| \ll 1$ and,
    for simplicity, assume a point mass baryonic mass distribution $\rho_b = M_b \delta(\vec{x})$.
Then, the superfluid's mass is $M_{\mathrm{SF}} = M_b \cdot (r/r_c)^2$ (see Eq.~\eqref{eq:MSFMOND}).
In SFDM one usually assumes that the superfluid ends at some finite radius $r_{\mathrm{NFW}}$ where the superfluid's density is matched to that of an NFW halo.
The total dark matter mass $M_{200}^{\mathrm{DM}}$ can be calculated from
\begin{align}
 M_{200}^{\mathrm{DM}} = \frac{4\pi}{3} r_{200}^3 \rho_{200} = M_{\mathrm{SF}}(r_{\mathrm{NFW}}) + M_{\mathrm{NFW}}(r_{\mathrm{NFW}}, r_{200}) \,.
\end{align}
Here, $M_{\mathrm{NFW}}$ denotes the mass of the NFW halo between the radii $r_{\mathrm{NFW}}$ and $r_{200}$.
The NFW halo energy density falls off faster than the superfluid energy density (i.e., faster than $1/r$).
Thus, $M_{\mathrm{NFW}}$ grows slower than quadratically in $r$ and we have the inequality
\begin{align}
 \frac{4\pi}{3} r_{200}^3 \rho_{200} < M_b \left(\frac{r_{200}}{r_c}\right)^2 \,.
\end{align}
That is,
\begin{align}
 \frac{r_{200}}{r_c} < \frac{M_b}{\frac{4\pi}{3} \rho_{200} r_c^3} \,,
\end{align}
which is equivalent to
\begin{align}
 \frac{M_{200}^{\mathrm{DM}}}{M_b} < \left(\frac{M_b}{\frac{4\pi}{3} \rho_{200} r_c^3}\right)^2
 = \frac{1}{\sqrt{2\pi}} \frac94 \frac{1}{\rho_{200}^2} \left(\frac{m^2}{\alpha}\right)^3 a_0^{3/2} \sqrt{M_b} \left(1 - \frac{\beta}{3}\right)^3 \,.
\end{align}
Numerically, with the fiducial parameter values from \cite{Berezhiani2018} and $H = 67.3\,\mathrm{km}/(\mathrm{s} \cdot \mathrm{Mpc})$, this is
\begin{align}
 \frac{M_{200}^{\mathrm{DM}}}{M_b} < 0.9 \cdot \left(\frac{M_b}{10^{10}\,M_\odot}\right)^{1/2} \,.
\end{align}
This is too little for strong lensing even for very massive galaxies (see Appendix~\ref{sec:M200estar04}).
This upper bound is independent of the matching procedure to the NFW halo.

\subsection{Two-field SFDM}

Two-field SFDM contains two fields $\theta_+$ and $\theta_-$ instead of just one field $\theta$ like standard SFDM \citep{Mistele2020}.
Still, in equilibrium only two nontrivial equations must be solved.
One for $\theta_+$ that carries the MOND-like force and one for the Newtonian gravitational potential $\phi_N$.
As in standard SFDM, we write the equations in terms of $\hat{\mu}/m$ where $\hat{\mu} = \mu_{\mathrm{nr}} - m \phi_N$, $\mu_{\mathrm{nr}}$ is the non-relativistic chemical potential, and $m$ is the mass of the superfluid's constituent particles.
Also, as in standard SFDM, the MOND limit of the phonon force is controlled by a quantity $\varepsilon_* = (2 m^2/ \alpha M_{\mathrm{Pl}} a_b) (\hat{\mu}/m) $.
Thus, we use the same notation $\varepsilon_*$ in both models.

This model has two contributions to the superfluid energy density, $\rho_{\mathrm{SF}+}$ and $\rho_{\mathrm{SF}-}$.
As discussed in \cite{Mistele2020}, usually $\rho_{\mathrm{SF}-}$ dominates.
For our calculation below, we assumed that this is the case and neglected $\rho_{\mathrm{SF}+}$.
We verified that $\rho_{\mathrm{SF}+}$ is always at most $5\%$ as large as $\rho_{\mathrm{SF}-}$ at $R = R_{\mathrm{mid}}$ for the best fits.

As in standard SFDM, we can get the phonon force from a no-curl approximation as a function of $\vec{a}_b$ and $\varepsilon_*$.
We use the same notation $\vec{a}_\theta$ as in standard SFDM.
The quantity $\hat{\mu}/m$ is determined as in standard SFDM just with a different $\rho_{\mathrm{SF}}$.
Concretely, we have in the no-curl approximation for an equilibrium superfluid
\begin{align}
 a_\theta &= \sqrt{a_0 a_b} \sqrt{\sqrt{1 + \frac{\varepsilon_*^2}{4}} + \frac{\varepsilon_*}{2}} \,
\end{align}
and
\begin{align}
 \rho_{\mathrm{SF}} = \rho_{\mathrm{SF}-} = \frac{2 M_{\mathrm{Pl}}^2}{r_0^2} \frac{\hat{\mu}}{m} \,,
\end{align}
where $r_0$ is a parameter of the model.

We used the numerical parameter values from \cite{Mistele2020}.
That is, $r_0 = 50\,\mathrm{kpc}$, $a_0 = 0.87\cdot 10^{-10}\,\mathrm{m}/\mathrm{s}^2$, and, unless stated otherwise, $\bar{a} = 10^{-12}\,\mathrm{m}/\mathrm{s}^2$.
The quantity $m^2/\alpha$ that enters $\varepsilon_*$ is a combination of these, namely $m^2/\alpha =  10^{-7/2} \, (\bar{a}/a_0)^{1/4} (M_{\mathrm{Pl}}/r_0)$.

One difference to standard SFDM is that $\varepsilon_*$ is almost always small so that the phonon force $a_\theta$ almost always has the MOND-like form $\sqrt{a_0 a_b}$.
But, in contrast to standard SFDM, a small $|\varepsilon_*|$ implies only that $a_\theta \approx \sqrt{a_0 a_b}$, not that $a_{\mathrm{SF}}$ is small.
Thus, even for $|\varepsilon_*| \ll 1$ one must check that $a_{\mathrm{SF}}$ is small in order to get MOND-like rotation curves.
Otherwise, the total acceleration will be systematically larger than in MOND.

The energy density $\rho_{\mathrm{SF}}$ reaches zero for $\hat{\mu} = 0$ in two-field SFDM (i.e. $\varepsilon_{*\mathrm{min}} = 0$).

\section{Comparison to MOND}
\label{sec:appendix:MONDvsSFDM}

\begin{figure}
  \centering
  \includegraphics[width=.48\textwidth]{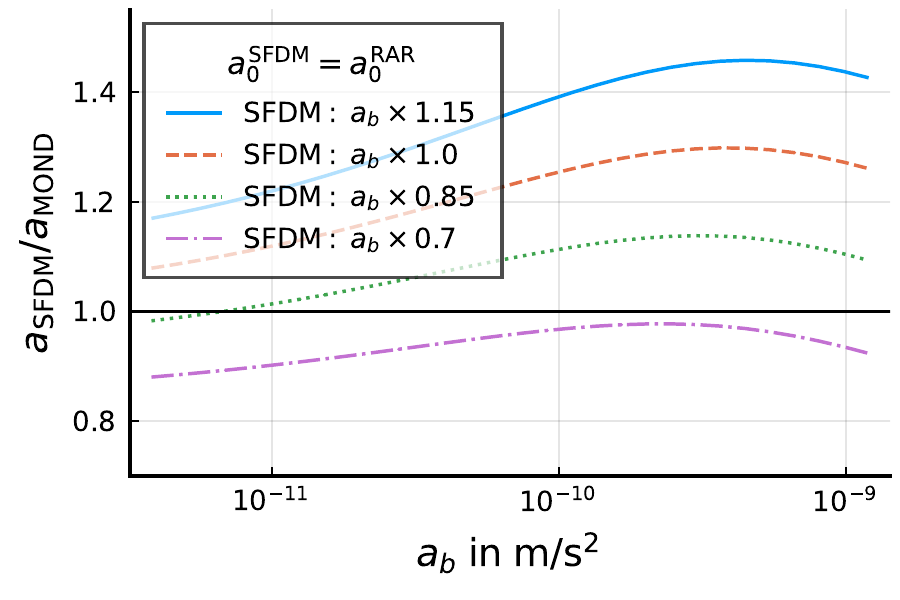}
  \includegraphics[width=.48\textwidth]{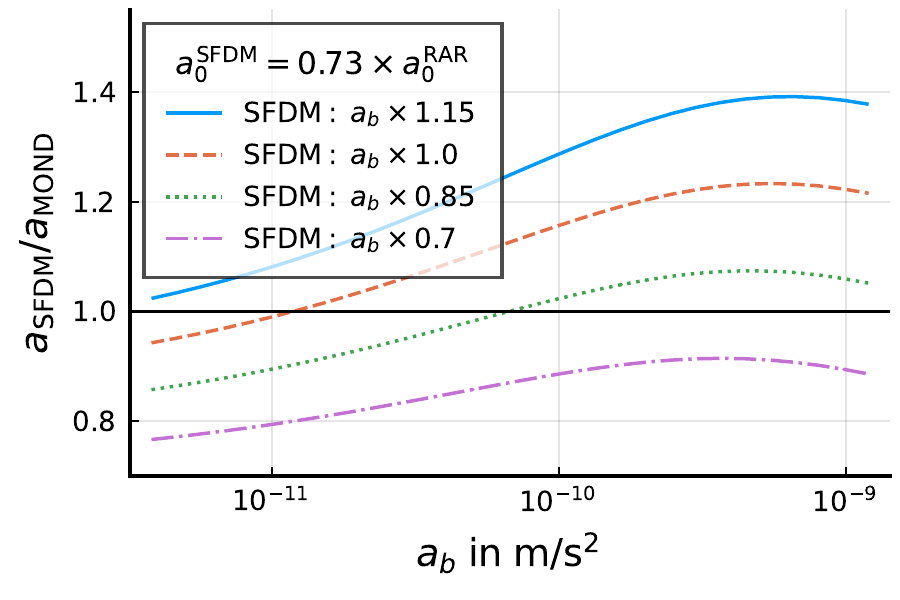}
  \caption{
     Ratio of the accelerations $a_{\mathrm{SFDM}} = a_b \, \nu_\theta(a_b/a_0)$ and $a_{\mathrm{MOND}} = a_b \, \nu_e(a_b/a_0)$ as a function of $a_b$.
     Top:
     With $a_{\mathrm{SFDM}}$ and $a_{\mathrm{MOND}}$ both using the same value for $a_0$, namely $a_0 = 1.2 \cdot10^{-10}\,\mathrm{m}/\mathrm{s}^2$,
        but with $a_{\mathrm{SFDM}}$ using a baryonic acceleration $a_b$ that is multiplied by an overall factor relative to $a_b$ in $a_{\mathrm{MOND}}$.
Bottom:
    Same as top, but now $a_{\mathrm{SFDM}}$ and $a_{\mathrm{MOND}}$ use different values for $a_0$, namely $a_0^{\mathrm{SFDM}} = 0.87\cdot 10^{-10}\,\mathrm{m}/\mathrm{s}^2$ and $a_0^{\mathrm{RAR}} = 1.2\cdot 10^{-10}\,\mathrm{m}/\mathrm{s}^2$, respectively.
  }
  \label{fig:nue-nutheta}
\end{figure}

\subsection{Assuming the MOND limit of SFDM}
\label{sec:sfdm:less}

For \textsc{SFDM} in the \textsc{MOND} limit we approximately have
$
 a_{\mathrm{tot}} \approx \sqrt{a_0 a_b} + a_b
$,
which, in \textsc{MOND}, would correspond to the interpolation function
\begin{align}
 \nu_\theta(y) = 1 + \frac{1}{\sqrt{y}} \,,
\end{align}
with $y = a_b/a_0$.

At baryonic accelerations not much smaller or much larger than $a_0$ (i.e., $y = \mathcal{O}(1)$), the additional acceleration from \textsc{SFDM} is significantly larger than what one obtains from standard \textsc{MOND} interpolation functions such as \citep{Lelli2017b}
\begin{align}
 \nu_e(y) = \frac{1}{1 - e^{-\sqrt{y}}} \,.
\end{align}
It is because of this difference in the interpolation functions that one may naively expect \textsc{SFDM} to require less baryonic mass than standard \textsc{MOND} models, at least in the \textsc{MOND} limit.

This is illustrated in Fig.~\ref{fig:nue-nutheta}, top.
The total acceleration in SFDM is always larger than in MOND,
    if both use the same baryonic $a_b$.
At intermediate accelerations ($a_b \sim a_0 \sim 10^{-10}\,\mathrm{m}/\mathrm{s}^2$) the difference between MOND and SFDM is significant.
This can be countered by making $a_b$ in SFDM smaller,
    that is to say, by choosing a smaller mass-to-light ratio in SFDM than in MOND.

This discussion so far assumes the same $a_0$ for both SFDM and MOND.
However, in practice, one usually chooses a somewhat smaller value for $a_0$ in SFDM.
Indeed, \cite{Berezhiani2018} chose $a_0^{\mathrm{SFDM}} \approx 0.87 \cdot 10^{-10}\,\mathrm{m}/\mathrm{s}^2$, while MOND typically requires $a_0^{\mathrm{RAR}} \approx 1.2 \cdot 10^{-10}\,\mathrm{m}/\mathrm{s}^2$ \citep{Lelli2017b}.
The motivation of \cite{Berezhiani2018} to choose a lower value is to take into account a possible effect of the superfluid's gravitational pull $a_{\mathrm{SF}}$.
Indeed, at small accelerations $a_b$, the total acceleration in MOND is close to
\begin{align}
 \label{eq:a0RAR}
 \sqrt{a_0^{\mathrm{RAR}} a_b} \,,
\end{align}
while in SFDM we have
\begin{align}
 \label{eq:a0SFDM}
 \sqrt{a_0^{\mathrm{SFDM}} a_b} + a_{\mathrm{SF}} \,.
\end{align}
The smaller value of $a_0$ in SFDM allows us to get the same total acceleration even with a nonzero $a_{\mathrm{SF}}$.
Numerically, the smaller $a_0$ value is compensated for when $a_{\mathrm{SF}}$ is about $0.15 \sqrt{a_0^{\mathrm{RAR}} a_b}$.

Neglecting $a_{\mathrm{SF}}$, this smaller value of $a_0$ makes the total acceleration smaller,
    so it counters the need for less baryonic mass in SFDM.
This is illustrated in Fig.~\ref{fig:nue-nutheta}, bottom.
The smaller $a_0$ value has the biggest impact at small accelerations $a_b$.
At small accelerations, $a_b \ll a_0$, SFDM may even require more baryonic mass than MOND,
    at least if we neglect $a_{\mathrm{SF}}$.
Indeed, $a_{\mathrm{SF}}$ is usually negligible in the proper MOND limit $|\varepsilon_*| \ll 1$, as discussed above.
Thus, assuming the proper MOND limit, we expect to find systematically smaller $M/L_*$ in SFDM than in MOND for galaxies with large $a_b$ but not for galaxies with small $a_b$.
This is roughly what we find in our fits below (see Appendix~\ref{sec:results:sfdm:MLtrends}).

\subsection{Caveat: MOND limit}
\label{sec:sfdm:mond}

The above discussion applies in the MOND limit of SFDM.
Outside this MOND limit, the phonon force does not necessarily have the form $a_\theta = \sqrt{a_0 a_b}$ and the superfluid's gravitational pull may not be negligible.
For example, at $\varepsilon_* \to \infty$, we find that $a_\theta \propto 1/\sqrt{\varepsilon_*}$ (see Fig.~\ref{fig:atheta-estar}, right).
That is, having a large $\varepsilon_*$ makes $a_\theta$ small.
A smaller acceleration may allow for larger baryonic masses.
Thus, having galaxies end up at $\varepsilon_* \gg 1$ is a way to allow for relatively large mass-to-light ratios in our fits.

One might be skeptical of this argument for the following reason.
The argument relies on the total acceleration $a_{\mathrm{tot}}$ becoming smaller for $\varepsilon_* \gg 1$.
But this is not necessarily the case.
A large $\varepsilon_*$ does make $a_\theta$ smaller.
But it is possible that the decrease in $a_\theta$ is compensated for by an increase in $a_{\mathrm{SF}}$.
Indeed, at large $\varepsilon_*$, the superfluid's energy density scales as
\begin{align}
 \rho_{\mathrm{SF}} \propto \sqrt{\hat{\mu}} \propto \sqrt{\varepsilon_* a_b} \,.
\end{align}
Thus, at fixed $a_b$, a larger $\varepsilon_*$ makes the superfluid heavier and thus $a_{\mathrm{SF}}$ larger.
For $\varepsilon_* \to \infty$, the acceleration $a_{\mathrm{SF}}$ can become arbitrarily large.
Thus, the total acceleration does not become smaller for $\varepsilon_* \to \infty$,
    despite the smaller phonon acceleration $a_\theta$.

Still, in practice there is a significant window of large values of $\varepsilon_*$ where the total acceleration does become smaller.
To see this explicitly, expand for large large $\varepsilon_*$.
Then, roughly, $a_{\mathrm{tot}}$ scales with $\varepsilon_*$ as
\begin{align}
 a_b + 3 \sqrt{a_0 a_b} \frac{\sqrt{\beta-1}}{2 \beta - 3} \left(\frac{1}{\sqrt{\varepsilon_*}}  + \sqrt{\varepsilon_*} \cdot \frac{2\beta-3}{3-\beta} \cdot \frac{\left.a_{\mathrm{SF}}\right|_{\varepsilon_*=0}}{\sqrt{a_0 a_b}} \right) \,,
\end{align}
where we treated $\varepsilon_*$ as a constant that we can pull out of $a_{\mathrm{SF}}$ (see Appendix~\ref{sec:appendix:estar}).
Thus, at fixed $a_b$, the total acceleration $a_{\mathrm{tot}}$ decreases as a function of large $\varepsilon_*$ as long as
\begin{align}
 \varepsilon_* < \frac{\sqrt{a_0 a_b}}{\left.a_{\mathrm{SF}}\right|_{\varepsilon_*=0}} \cdot \frac{3-\beta}{2\beta-3} \simeq \frac{r_c^2}{r_{\mathrm{MOND}} \cdot r} \cdot \frac{3-\beta}{2\beta-3} = \frac{6}{2\beta-3} \frac{M_{\mathrm{Pl}}}{r} \frac{\alpha}{m^2} \,,
\end{align}
with the MOND radius $r_{\mathrm{MOND}} = \sqrt{G M_b/a_0}$ and $r_c$ as defined in Appendix~\ref{sec:appendix:sfdm:mond}.
Numerically, for the fiducial parameter values from \cite{Berezhiani2018},
\begin{align}
 \varepsilon_* \lesssim 107 \cdot \left(\frac{5\,\mathrm{kpc}}{r}\right) \,.
\end{align}
Thus, the total acceleration is a decreasing function of $\varepsilon_*$ for a significant range of large $\varepsilon_*$ values so that going to large $\varepsilon_*$ is one way to allow for relatively large baryonic masses.

\section{Method}
\label{sec:method}

\subsection{Data}
\label{sec:method:data}

As already mentioned in Sect.~\ref{ssec:data}, we used the observed rotation velocity $V_{\mathrm{obs}}$ directly from SPARC.
We did not allow distance or inclination as a fit parameter, so we did not vary $V_{\mathrm{obs}}$ in our fitting procedure.
As also described in Sect.~\ref{ssec:data}, we obtained the baryonic energy density $\rho_b(R,z)$ from the surface densities provided by SPARC,
\begin{eqnarray}
 \label{eq:Qstar}
 \rho_b(R,z) &=& \rho_{\mathrm{gas}}(R,z) \nonumber + 0.5 \cdot Q_* \cdot \rho_*(R,z) \nonumber \\
 &+& 0.7 \cdot Q_* \cdot \rho_{\mathrm{bulge}}(\sqrt{R^2+z^2}) \,.
\end{eqnarray}
The fit parameter $Q_*$ parametrizes the stellar mass-to-light ratio.
For later use, we numerically solved the Poisson equation
\begin{align}
 \Delta \left(- \frac{\hat{\mu}_x}{m}\right) &= 4 \pi G \rho_x \,,
\end{align}
separately for $\rho_x \in \{\rho_{\mathrm{gas}}, \rho_*, \rho_{\mathrm{bulge}}\}$ using the Mathematica code used in \cite{Hossenfelder2020}.
This allows us to quickly get a solution to the Poisson equation sourced by the full $\rho_b$ with arbitrary $Q_*$ by the rescaling
\begin{align}
 \frac{\hat{\mu}_b}{m} = \frac{\hat{\mu}_{\mathrm{gas}}}{m} + 0.5 \cdot Q_* \cdot \frac{\hat{\mu}_*}{m} + 0.7 \cdot Q_* \cdot \frac{\hat{\mu}_{\mathrm{bulge}}}{m} \,,
\end{align}
where the quantity $\hat{\mu}_b/m$ is minus the standard Newtonian gravitational potential up to an additive constant (see also the next subsection).
The numerical procedure solves the Poisson equation within a sphere with radius $r_\infty$ assuming a $z \to -z$ symmetry.
We assumed spherically symmetric boundary conditions for $\hat{\mu}_b$.
Specifically,
\begin{align}
 \left. \frac{\hat{\mu}_x}{m}\right|_{\sqrt{R^2+z^2} = r_\infty} = 0 \,.
\end{align}
This is reasonable for sufficiently large $r_\infty$.
We used $r_\infty = 100\,\mathrm{kpc}$ except when the SPARC $V_{\mathrm{obs}}$ data extend to radii larger than $100\,\mathrm{kpc}$.
Then, we increased $r_\infty$ in steps of $5\,\mathrm{kpc}$ until $r_\infty$ was larger than the maximum radius of the $V_{\mathrm{obs}}$ data points.

\subsection{SFDM calculation}
\label{sec:sfdm:calc}

We assume that each galaxy's $V_{\mathrm{obs}}$ data points lie within its superfluid core.
This is discussed in more detail in Appendix~\ref{sec:results:sfdm:thermal}.
Then, in SFDM, there are two equations for a galaxy in equilibrium inside the superfluid core.
One for the phonon acceleration, $\vec{a}_\theta$, and one for the quantity $\hat{\mu}$, which contains the Newtonian gravitational potential (see Appendix~\ref{sec:appendix:estar}).

Even in a fully axisymmetric calculation,
    one can impose spherically symmetric boundary conditions for the fields $\hat{\mu}$ and $\theta$ at some large radius $r_\infty$ \citep{Hossenfelder2020}.
The value of $\theta$ at $r_\infty$ is inconsequential, so one can choose $\theta(r_\infty) = 0$.
For $\hat{\mu}$, its value $\mu_\infty$ at $r_\infty$ is important.
It determines the size of the superfluid halo and is a free parameter in the boundary conditions.
We used a parameter similar to $\mu_\infty$ as a free fit parameter in our fitting procedure.

It is useful to split $\hat{\mu}$ into a part called $\hat{\mu}_b$ sourced only by $\rho_b$ and the rest called $\hat{\mu}_{\mathrm{SF}}$.
That is, $\hat{\mu} = \hat{\mu}_b + \hat{\mu}_{\mathrm{SF}}$ with
\begin{align}
 \label{eq:mub}
 \Delta \left(- \frac{\hat{\mu}_b}{m}\right) &= 4 \pi G \rho_b \,, \\
 \label{eq:muSF}
 \Delta \left(- \frac{\hat{\mu}_{\mathrm{SF}}}{m}\right) &= 4 \pi G \rho_{\mathrm{SF}}\left[\hat{\mu}_b + \hat{\mu}_{\mathrm{SF}}, \vec{\nabla} \theta\right]\,.
\end{align}
We used boundary conditions $\hat{\mu}_b(r_\infty) = 0$ and $\hat{\mu}_{\mathrm{SF}}(r_\infty) = \mu_\infty$.
We calculated $\hat{\mu}_b$ as described in the previous subsection.

\subsubsection{A simple approximation}
\label{sec:simple}

For our fits, we did not do a fully axisymmetric calculation.
Instead, we used an approximation that is much faster to compute.
Our approximation mainly consists of using a no-curl approximation for $\vec{a}_\theta$ and assuming spherical symmetry for $\hat{\mu}_{\mathrm{SF}}$.
As discussed in Appendix~\ref{sec:appendix:estar}, the no-curl approximation means that we get $\vec{a}_\theta$ as an algebraic function of $\vec{a}_b$ and $\hat{\mu}$.

The second part of our approximation is assuming spherical symmetry for $\hat{\mu}_{\mathrm{SF}}$ in $\hat{\mu} = \hat{\mu}_b + \hat{\mu}_{\mathrm{SF}}$.
That is, the part of $\hat{\mu}$ due to the superfluid's self-gravity is spherically symmetric.
Only the baryonic part produces an axisymmetric $\hat{\mu}$.
This is a reasonable approximation for the following reason.
A fully axisymmetric calculation gives a $\hat{\mu}$ that is not spherically symmetric only at relatively small radii.
At these radii, $\hat{\mu}_b$ dominates.
At larger radii, even a fully axisymmetric calculation gives a spherically symmetric $\hat{\mu}$ \citep{Hossenfelder2020}.
Indeed, we imposed spherically symmetric boundary conditions at larger radii.
Only at these larger radii does $\hat{\mu}_{\mathrm{SF}}$ usually become important.

We calculated $\hat{\mu}_{\mathrm{SF}}(r)$ from Eq.~\eqref{eq:muSF}, which contains the function $\hat{\mu}_b(R,z)$.
To solve this equation in spherical symmetry,
    we need to make a choice of which $R$ and $z$ to use in evaluating $\hat{\mu}_b(R,z)$ for each $r$.
The same applies to $a_b(R,z)$, which enters indirectly through $a_\theta$.
We chose $R=r$ and $z=0$.
Different choices may give slightly different results.

For $\rho_{\mathrm{SF}}$, we used the expression Eq.~\eqref{eq:rhoSFnocurl} valid in the no-curl approximation.
The function $f_\beta(\varepsilon_*)$ in this expression for $\rho_{\mathrm{SF}}$ is known analytically but is relatively slow to evaluate numerically.
To speed up the calculation, for a given $\beta$, we evaluated $f_\beta$ as a function of $\log_{10} (\varepsilon_* - \varepsilon_{*\mathrm{min}})$ on an evenly spaced grid with grid spacing $0.01$ and linearly interpolated between the grid points.
We used the resulting linear interpolation in our calculation since it is faster to evaluate numerically than the analytical form of $f_\beta$.

Below, we refer to this approximation as the ``simple'' approximation.
In Appendix~\ref{sec:method:validatesimple}, we explicitly demonstrate that it works well using a few example galaxies.

\subsubsection{Calculation using the ``simple'' approximation}
\label{sec:sfdm:calcSF}

We calculated $\hat{\mu}_b$ as described above in Appendix~\ref{sec:method:data}.
From this, we got $\vec{a}_b$ as $-\vec{\nabla} \hat{\mu}_b/m$.
In accordance with our simple approximation, we used the no-curl approximation for $\vec{a}_\theta$ so that we got $\vec{a}_\theta$ as an algebraic function of $\vec{a}_b$ and $\hat{\mu}_b + \hat{\mu}_{\mathrm{SF}}$ (see Appendix~\ref{sec:appendix:estar}).
The remaining part was to calculate $\hat{\mu}_{\mathrm{SF}}$.
This then also gave $\vec{a}_{\mathrm{SF}}$ as $- \vec{\nabla} \hat{\mu}_{\mathrm{SF}}/m$.

For $\hat{\mu}_{\mathrm{SF}}$ we assumed spherical symmetry and we used the form Eq.~\eqref{eq:rhoSFnocurl} for $\rho_{\mathrm{SF}}$ valid in the no-curl approximation.
Then, Eq.~\eqref{eq:muSF} becomes
\begin{align}
\frac{1}{r^2} \partial_r \left(r^2 \partial_r \left(- \frac{\hat{\mu}_{\mathrm{SF}}(r)}{m}\right) \right) = 4 \pi G \, \rho_{\mathrm{SF}}\left(\varepsilon_*, |\vec{a}_b| \right) \,,
\end{align}
where
\begin{align}
 \varepsilon_* = \frac{2 m^2}{\alpha M_{\mathrm{Pl}} |\vec{a}_b|} \frac{\hat{\mu}_b + \hat{\mu}_{\mathrm{SF}}}{m} \,.
\end{align}
The non-spherically symmetric functions $a_b(R,z)$ and $\hat{\mu}_b(R,z)$ are evaluated at $R =r$, $z=0$.

This is a second-order ODE for $\hat{\mu}_{\mathrm{SF}}$.
As boundary conditions we chose
\begin{align}
 \hat{\mu}_{\mathrm{SF}}'(dr) &= 0\,, \\
 \hat{\mu}_{\mathrm{SF}}(R_{\mathrm{mid}}) + \hat{\mu}_b(R_{\mathrm{mid}}, 0) &= \varepsilon \frac{\alpha M_{\mathrm{Pl}} |\vec{a}_b(R_{\mathrm{mid}}, 0)|}{2 m}  \,.
\end{align}
With $dr=0$, the first boundary condition is a standard regularity condition at the origin.
To avoid numerical issues, we chose a small nonzero value for $dr$, usually $dr = 10^{-8}\,\mathrm{kpc}$.
The second condition corresponds to a choice of $\varepsilon = \varepsilon_*(R_{\mathrm{mid}})$ (see Eq.~\eqref{eq:estar}).
It parametrizes how close to the MOND limit $|\varepsilon_*| \ll 1$ a galaxy is in the middle of the $V_{\mathrm{obs}}$ data points.

Solutions for $\hat{\mu}$ are such that they typically reach their minimum allowed value $\hat{\mu}_{\mathrm{min}}$ (see Eq.~\eqref{eq:muhatmin}) at some finite radius.
Beyond this radius, assuming a superfluid core makes no sense.
Thus, whenever solutions ended up with $\hat{\mu} < \hat{\mu}_{\mathrm{min}}$ at some radius $dr \leq r \leq R_{\mathrm{max}}$, we discarded them,
    since we assumed all data points lie within the superfluid core.

Sometimes, Mathematica fails to solve the equations for numerical reasons.
This is indicated by its ``NDSolve'' producing a ``FindRoot::sszero,'' a ``NDSolveValue::berr,'' or a ``NDSolveValue::evcvmit'' error that we can check for.
In this case, we automatically decreased $dr$ by factor of $100$ and retried.

\subsubsection{Validating the simple approximation}
\label{sec:method:validatesimple}

Here, we explicitly compare the simple approximation described in Appendix~\ref{sec:simple} against a fully axisymmetric calculation.
For definiteness, we used the best-fit $Q_*$ and $f_{\varepsilon_*}$ values for SFDM (see Appendix~\ref{sec:fit}).

For the fully axisymmetric calculation we used the Mathematica code from \cite{Hossenfelder2020}.
This code expects boundary conditions in the form $\hat{\mu}(r_\infty) = \mu_\infty$ for some $r_\infty$ and $\mu_\infty$.
We chose $r_\infty = 100\,\mathrm{kpc}$ unless stated otherwise.
Our simple approximation instead uses a value of $\hat{\mu}_{\mathrm{SF}}(R_{\mathrm{mid}})$ as a boundary condition.
To compare our simple calculation and the fully axisymmetric calculation for the same physical boundary conditions,
    we first did the simple calculation with the best-fit values for $Q_*$ and $f_{\varepsilon_*}$.
We then evaluated the solution $\hat{\mu}$ from this simple calculation at $r=r_\infty$ and used the resulting value as the boundary condition $\mu_\infty$ for the fully axisymmetric calculation.

Our simple calculation makes two main approximations.
First, we used the no-curl approximation for the phonon force.
Second, we assumed spherical symmetry for $\hat{\mu}_{\mathrm{SF}}$.
When our simple calculation disagrees with the fully axisymmetric calculation we want to know which of these two parts is responsible for the deviation.
To this end,
    we did a third calculation where we used the fully axisymmetric calculation for $\hat{\mu}$,
    but then used the no-curl approximation when calculating $a_\theta$ for the rotation curve.
We refer to this as the ``full+nocurl'' calculation.

\begin{figure*}
  \centering
  \includegraphics[width=\textwidth]{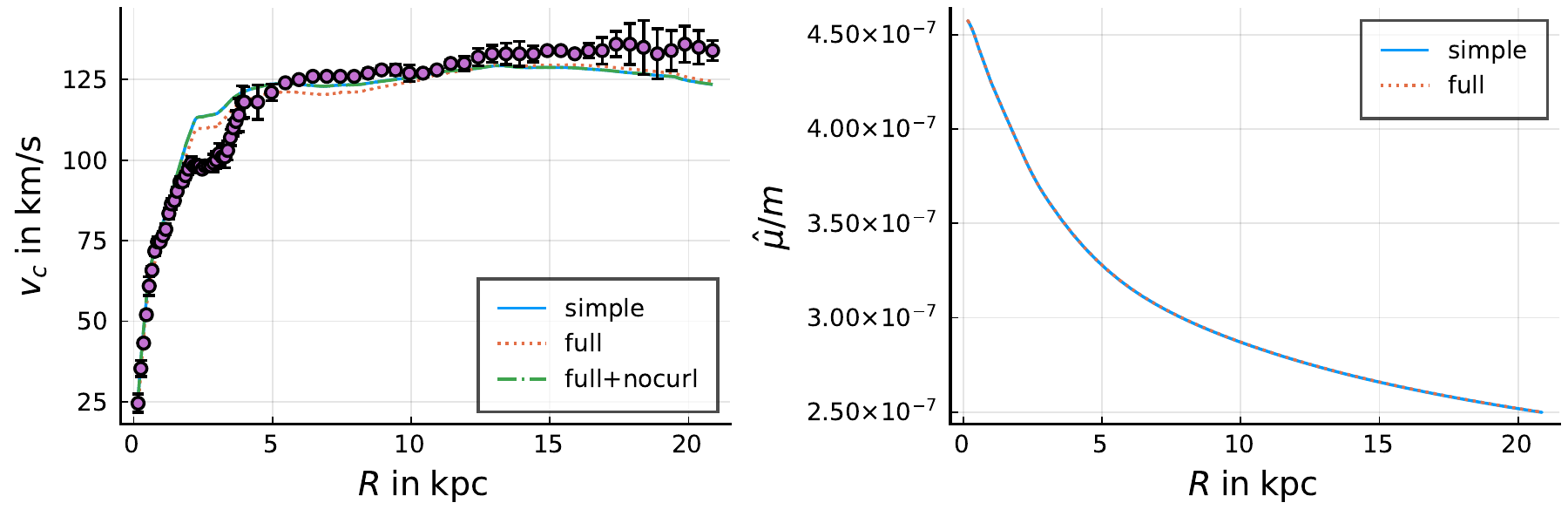}
  \caption{
      Results of different types of calculations in SFDM for NGC2403.
      The simple calculation is the approximation described in Appendix~\ref{sec:simple}.
      The full calculation is the fully axisymmetric SFDM calculation.
      The full+nocurl calculation uses the same $\hat{\mu}(R, z)/m$ as the full calculation but uses a no-curl approximation for the phonon acceleration, $\vec{a}_\theta$.
      This is for the best-fit parameters obtained in Appendix~\ref{sec:results:sfdm}.
      Left: Rotation curve for the different types of calculations (lines) and the observed rotation curve from the SPARC data (circles with error bars).
      Right: Field $\hat{\mu}/m$ for the same types of calculations,
        except for full+nocurl, which has the same $\hat{\mu}/m$ as the full calculation.
  }
  \label{fig:fullsimple-NGC2403}
\end{figure*}

In Fig.~\ref{fig:fullsimple-NGC2403}, left, we show the rotation curve and $\hat{\mu}/m$ for NGC2403 for the different types of calculation described above.
The calculations differ by a few percent at intermediate radii.
The full+nocurl rotation curve lies pretty much on top of the simple rotation curve,
    while the ``full'' rotation curve differs from the two others at intermediate radii.
Thus, the no-curl approximation is the source of the this difference between the full and the simple calculations.
For $\hat{\mu}$, all calculations agree almost perfectly with each other (see Fig.~\ref{fig:fullsimple-NGC2403}, right).

\begin{figure*}
  \centering
  \includegraphics[width=\textwidth]{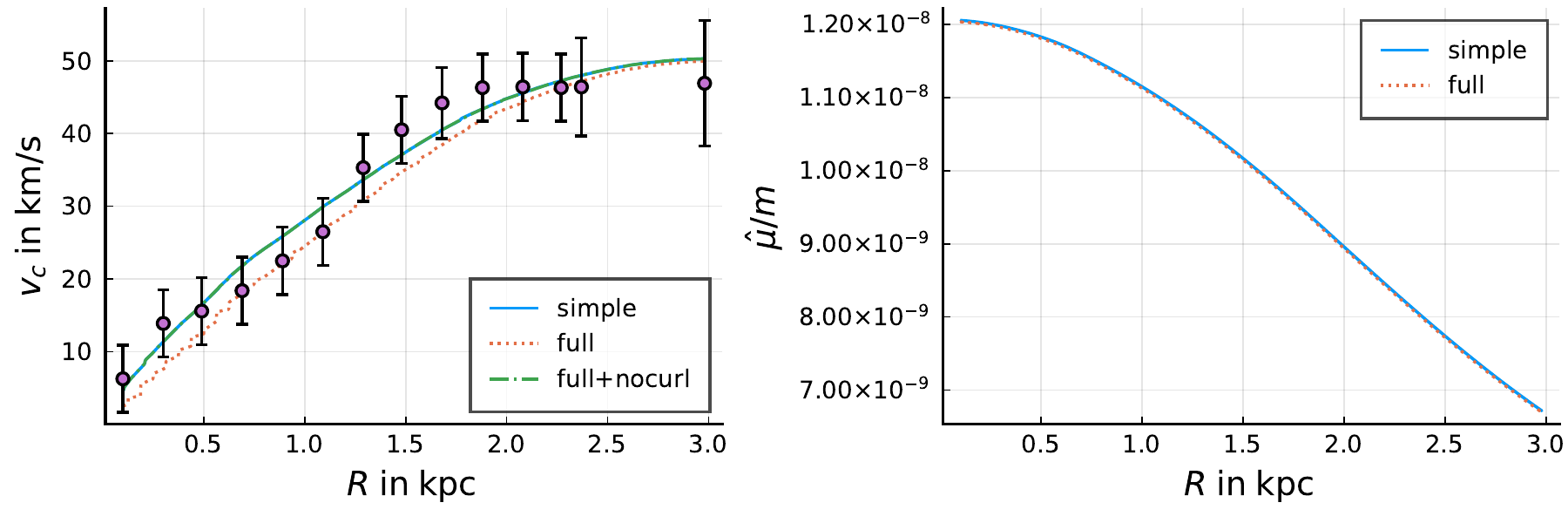}
  \caption{
     Same as Fig.~\ref{fig:fullsimple-NGC2403} but for DDO064.
     This is an example of a galaxy in the MOND limit $|\varepsilon_*| \ll 1$.
  }
  \label{fig:fullsimple-DDO064}
\end{figure*}

\begin{figure*}
  \centering
  \includegraphics[width=\textwidth]{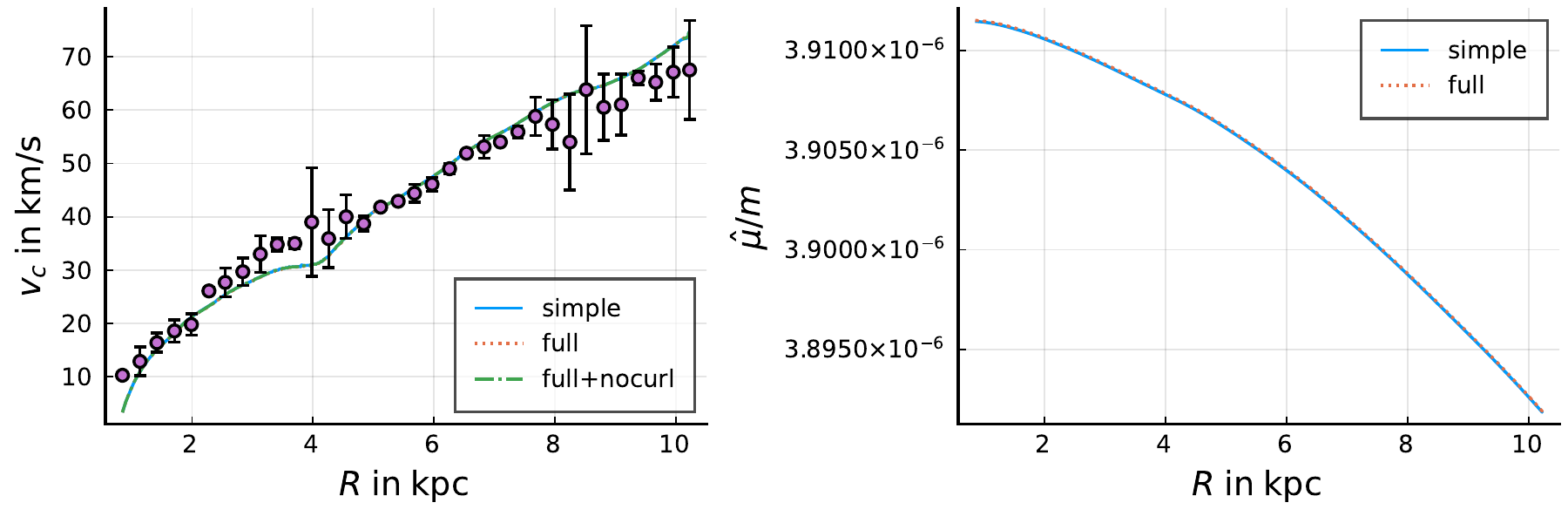}
  \caption{
     Same as Fig.~\ref{fig:fullsimple-NGC2403} but for IC2574.
  }
  \label{fig:fullsimple-IC2574}
\end{figure*}

This same qualitative result holds for DDO064 shown in Fig.~\ref{fig:fullsimple-DDO064}.
This is an example of a galaxy that is in the proper MOND limit $|\varepsilon_*| \ll 1$ almost everywhere at $R_{\mathrm{min}} \leq R \leq R_{\mathrm{max}}$.
The no-curl approximation does not always lead to visible deviations between the full and the simple calculations.
An example is IC2574 where the full and simple calculations agree almost perfectly with each other (see Fig.~\ref{fig:fullsimple-IC2574}).

Thus, our simple approximation works well, with the main error being due to the no-curl approximation.

\subsection{Fitting method}
\label{sec:fit}

For SFDM, we used the two parameters
\begin{align}
f_Y \equiv \log_{10}(Q_*), \,
\end{align}
for the stellar mass-to-light ratio (see Eq.~\eqref{eq:Qstar}), and
\begin{align}
 \label{eq:festar}
 f_{\varepsilon_*} \equiv \log_{10}\left( \varepsilon - \varepsilon_{*\mathrm{min}}\right),  \,
\end{align}
for the superfluid halo  (see Eq.~\eqref{eq:festar}), as fit parameters.
Here, $\varepsilon_{*\mathrm{min}}$ is the minimum possible value of $\varepsilon_*$ where $\rho_{\mathrm{SF}}$ vanishes (see Eq.~\eqref{eq:estarmin}).
We did not vary the model parameters $m$, $\alpha$, $\Lambda$, and $\beta$.
We used Mathematica's ``NMinimize'' with the ``NelderMead'' method to find the smallest $\chi^2$ for each galaxy,
\begin{align}
 \chi^2 =  \frac{1}{N-f} \sum_{R} \frac{(V_{\mathrm{obs}}(R) - V_c(R))^2}{\sigma_{V_{\mathrm{obs}}}^2(R)} \,.
\end{align}
Here, $N$ is the number of data points in the galaxy, $f = 2$ is the number of fit parameters, $\sigma_{V_{\mathrm{obs}}}$ is the uncertainty on the velocity $V_{\mathrm{obs}}$ from SPARC, $V_c(R)$ is the calculated rotation curve in SFDM, and the sum is over the data points at radii $R$.

We minimized $\chi^2$ for $f_Y$ and $f_{\varepsilon_*}$ in the range Eq.~\eqref{eq:rangeQ} and Eq.~\eqref{eq:rangeestar}.
When $f_{\varepsilon_*}$ is too small,
    it can happen that $\hat{\mu}$ does not exist with the desired parameters, as discussed in Appendix~\ref{sec:sfdm:calcSF}.
In this case, we artificially set $\chi^2 = 10^{10}$.
Then, NMinimize continued searching elsewhere.

The NelderMead search method is faster than a simple grid search but can get stuck in local minima.
To avoid this, we ran NMinimize three times with different starting points.
Of the three results, we used that with the smallest $\chi^2$.
The first run is with the ``RandomSeed'' option set to $0$, the second with the RandomSeed option set to $1$, and the third run is with the starting points $(0, 0)$, $(-0.5, 0)$, and $(0, -0.5)$.
The third run is to guarantee that the point $f_Y = 0$ is visited at least once,
    since this point corresponds to the $M/L_*$ expected from SPS models.

To further reduce the needed computation time we rounded $f_Y$ and $f_{\varepsilon_*}$ to $0.01$ before any calculation.
For (un-rounded) $f_Y$ and $f_{\varepsilon_*}$ that give the same rounded values as a previous calculation, we reused the previous results without a new computation.

This fitting method is much simpler than the MCMC method used in \cite{Li2018}.
Still, as we will see in Appendix~\ref{sec:rar},
    we found similar results for the stellar $M/L_*$ as \cite{Li2018} for a standard MOND model.
In addition, for SFDM we could not set up informative priors on $f_{\varepsilon_*}$ anyway since there is so far no cosmology from which to infer such a prior.

In the SPARC data, the Newtonian acceleration due to gas sometimes points outward from the galactic center, not toward it.
Usually, such a negative gas contribution is countered by the positive contributions from the stellar disk and bulge such that the total $a_b$ points to the galactic center.
But sometimes this is not the case, especially for small $f_Y$.
When this happened, we simply ignored the data points where $a_b$ is negative when calculating $\chi^2$.

As a cross-check and as a comparison for SFDM,
    we also fitted the RAR to the SPARC data, that is, we fitted the SPARC data with MOND assuming no curl term and the exponential interpolation function $\nu_e$ \citep{Lelli2017b}.
We call this the ``MOND'' model.
In this case, we have only one free fit parameter, $f_Y$.
Thus, when calculating $\chi^2$, we set $f=1$.
Also, we used the ``SimulatedAnnealing'' method of Mathematica's NMinimize function with one run instead of the NelderMead method with three runs.
We did not round $f_Y$ to $0.01$ for these MOND fits.

Below we consider modifications of both the SFDM model and the MOND model.
The SFDM-based models will be fitted as the ``SFDM'' model.
The MOND-based models will be fitted as the MOND model.
We will discuss the details of these modifications below.

For the SFDM models, we parametrize the total dark matter within the last rotation curve data point $R_{\mathrm{max}}$ by a parameter, $\fMDM$,
\begin{align}
 \label{eq:fMDM}
 \fMDM \equiv \log_{10}\left(\frac{M_{\mathrm{DM}}(R_{\mathrm{max}}) }{ M^{MP}_{\mathrm{DM}}(R_{\mathrm{max}}) } \right) \,,
\end{align}
where $M^{MP}_{\mathrm{DM}}(R)$ is defined by
\begin{align}
 \sqrt{a_0^{\mathrm{SFDM}} a_b(R,z=0)} + a_{\mathrm{SF}}(R,z=0) \equiv \sqrt{a_0^{\mathrm{RAR}} a_b(R,z=0)} \,,
\end{align}
with $a_{\mathrm{SF}} \equiv G M_{\mathrm{DM}}^{MP}/R^2$ and with the SPS $M/L_*$ values for $a_b$ (i.e. $M/L_* = 0.5$ for the disk and $M/L_* = 0.7$ for the bulge).
The parameter $\fMDM$ measures how far the dark matter mass at $R_{\mathrm{max}}$ is away from the reference value $M_{\mathrm{DM}}^{MP}$.
This reference value $M_{\mathrm{DM}}^{MP}$ is defined such that the associated dark matter acceleration $a_{\mathrm{SF}}$ counters the difference in $\sqrt{a_0 a_b}$ between MOND and SFDM due to the different choice for $a_0$ (see Appendix~\ref{sec:sfdm:less}).
Thus, $\fMDM$ parametrizes how large the dark matter mass is relative to the mass that cancels this $a_0$ difference.

\subsection{Two-field SFDM calculation}
\label{sec:twofieldcalc}

We can use the same calculation and fitting procedure as for standard SFDM.
We simply have to adjust the expression for the superfluid energy density and the algebraic no-curl solution of the phonon force.
Apart from that, we adjusted the calculation only in two ways that we now explain.

The superfluid energy density, $\rho_{\mathrm{SF}}$, of two-field SFDM is linear in $\hat{\mu}/m$ and depends on no other fields.
This allows the calculation to be sped up.
For a given galaxy, we first calculated one particular solution, $\hat{\mu}_{\mathrm{SF}}$, of the full, inhomogeneous equation as previously described,
\begin{align}
 \Delta \left(-\frac{\hat{\mu}_{\mathrm{SF}}}{m}\right) = \frac{1}{r_0^2} \frac{\hat{\mu}_b + \hat{\mu}_{\mathrm{SF}}}{m} \,.
\end{align}
To get solutions for the same galaxy with different boundary conditions,
    we can add solutions of the homogeneous equation to the so-obtained $\hat{\mu}_{\mathrm{SF}}$.
Since we assume spherical symmetry, the solutions to the homogeneous equation are $A \sin(r/r_0)/r$ for arbitrary $A$.
To get a solution for some desired boundary condition, we just needed to choose an appropriate $A$.

For standard SFDM, we used $\varepsilon_*(R_{\mathrm{mid}})$ in the range $\varepsilon_{*\mathrm{min}} + 10^{-2}$ to $\varepsilon_{*\mathrm{min}} + 10^4$ as the boundary condition for $\hat{\mu}$.
For two-field SFDM, we must adjust this range.
This is because, in two-field SFDM, the phonon force can more easily reach the MOND limit $|\varepsilon_*| \ll 1$ (i.e., typical values of $|\varepsilon_*|$ are much smaller).
Thus, we changed the range of $\varepsilon_*(R_{\mathrm{mid}})$ values scanned by our fit code to be
\begin{align}
 10^{-6} \leq \varepsilon_*(R_{\mathrm{mid}}) \leq 1 \,.
\end{align}
We note that $\varepsilon_{*\mathrm{min}} = 0$ in two-field SFDM.
We will later see that no galaxies end up at the boundaries of this range, so it seems to be reasonable.

\section{Results}

\begin{table*}
\caption{Median $0.5 \times Q_*$ for the best fit for different models and galaxy cuts.}
\label{tab:fYmedian}
\centering
\begin{tabular}{l|c|c|c|c}
Name & all & $q=1$ & $\mathrm{MOND}\,\log_{10}Q_\ast > -1.5$ & thermal ok \\ 
\hline 
$\mathrm{MOND}$ & 0.394 & 0.469 & 0.443 & 0.395\\ 
 \hline 
$\mathrm{MOND}\ \nu_\theta$ & $\times 0.74$ & $\times 0.74$ & $\times 0.75$ & $\times 0.73$\\ 
 \hline 
$\mathrm{MOND}\ \nu_\theta + a_0^{\mathrm{SFDM}}$ & $\times 0.97$ & $\times 0.89$ & $\times 0.93$ & $\times 0.97$\\ 
 \hline 
$\mathrm{SFDM}$ & $\times 1.18$ & $\times 1.04$ & $\times 1.10$ & $\times 1.25$\\ 
 \hline 
$\mathrm{SFDM}\ a_\theta = \sqrt{a_0 a_b}$ & $\times 0.92$ & $\times 0.87$ & $\times 0.88$ & $\times 0.92$\\ 
 \hline 
$\mathrm{SFDM}\ \beta=1.55$ & $\times 0.90$ & $\times 0.87$ & $\times 0.90$ & $\times 0.91$\\ 
 \hline 
$\mathrm{SFDM}\ |\varepsilon| < 5$ & $\times 0.96$ & $\times 0.89$ & $\times 0.92$ & $\times 0.96$\\ 
 \hline 
$\mathrm{SFDM}\ |\varepsilon| < 0.4$ & $\times 0.96$ & $\times 0.90$ & $\times 0.94$ & $\times 0.95$\\ 
 \hline 
$\mathrm{SFDM}\ |\varepsilon| < 0.4$$\;\mathrm{(no\;bad\;fits)}$ & $\times 0.94$ & $\times 0.89$ & $\times 0.92$ & $\times 0.94$\\ 
 \hline 
two-field$\,$ & $\times 0.90$ & $\times 0.87$ & $\times 0.88$ & $\times 0.91$\\ 
 \hline 
two-field $a_{\mathrm{min}}^{\mathrm{small}}$ & $\times 0.90$ & $\times 0.87$ & $\times 0.88$ & $\times 0.91$\\ 
 \hline 
two-field $a_b > a_{\mathrm{min}}^{\mathrm{small}}$ & $\times 0.90$ & $\times 0.87$ & $\times 0.88$ & $\times 0.91$
\end{tabular}
\tablefoot{
    Each row corresponds to a different model, and each column corresponds to a different cut on the included galaxies.
    The first row shows the values for the MOND fits, and the other rows show the factor relative to the MOND fits.
}
\end{table*}

\begin{table*}
\caption{Same as Table~\ref{tab:fYmedian} but for the mean instead of the median $0.5 \times Q_*$.}
\label{tab:fYmean}
\centering
\begin{tabular}{l|c|c|c|c}
Name & all & $q=1$ & $\mathrm{MOND}\,\log_{10}Q_\ast > -1.5$ & thermal ok \\ 
\hline 
$\mathrm{MOND}$ & 0.534 & 0.614 & 0.613 & 0.558\\ 
 \hline 
$\mathrm{MOND}\ \nu_\theta$ & $\times 0.74$ & $\times 0.75$ & $\times 0.74$ & $\times 0.74$\\ 
 \hline 
$\mathrm{MOND}\ \nu_\theta + a_0^{\mathrm{SFDM}}$ & $\times 0.96$ & $\times 0.96$ & $\times 0.96$ & $\times 0.96$\\ 
 \hline 
$\mathrm{SFDM}$ & $\times 1.46$ & $\times 1.29$ & $\times 1.32$ & $\times 1.55$\\ 
 \hline 
$\mathrm{SFDM}\ a_\theta = \sqrt{a_0 a_b}$ & $\times 0.85$ & $\times 0.87$ & $\times 0.85$ & $\times 0.84$\\ 
 \hline 
$\mathrm{SFDM}\ \beta=1.55$ & $\times 0.87$ & $\times 0.84$ & $\times 0.84$ & $\times 0.86$\\ 
 \hline 
$\mathrm{SFDM}\ |\varepsilon| < 5$ & $\times 0.93$ & $\times 0.92$ & $\times 0.93$ & $\times 0.93$\\ 
 \hline 
$\mathrm{SFDM}\ |\varepsilon| < 0.4$ & $\times 1.07$ & $\times 1.14$ & $\times 1.07$ & $\times 0.97$\\ 
 \hline 
$\mathrm{SFDM}\ |\varepsilon| < 0.4$$\;\mathrm{(no\;bad\;fits)}$ & $\times 0.95$ & $\times 0.96$ & $\times 0.95$ & $\times 0.94$\\ 
 \hline 
two-field$\,$ & $\times 0.84$ & $\times 0.88$ & $\times 0.84$ & $\times 0.82$\\ 
 \hline 
two-field $a_{\mathrm{min}}^{\mathrm{small}}$ & $\times 0.84$ & $\times 0.88$ & $\times 0.84$ & $\times 0.82$\\ 
 \hline 
two-field $a_b > a_{\mathrm{min}}^{\mathrm{small}}$ & $\times 0.83$ & $\times 0.87$ & $\times 0.83$ & $\times 0.81$
\end{tabular}
\end{table*}

\begin{figure*}
  \centering
  \includegraphics[width=\textwidth]{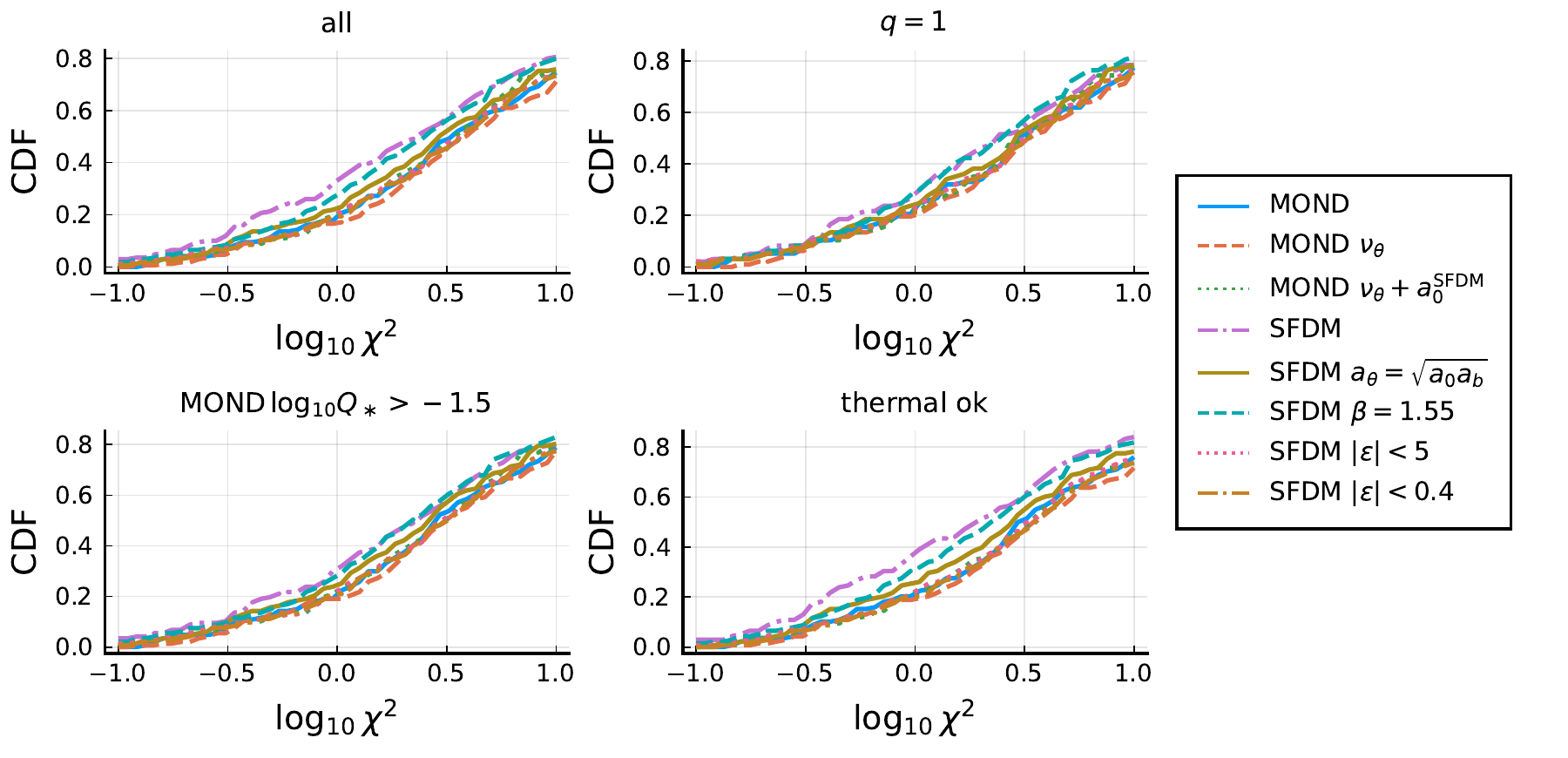}
  \caption{
      $\chi^2$ CDFs for the different MOND and SFDM models and for different galaxy cuts.
  }
  \label{fig:chi2cdf}
\end{figure*}

\begin{figure*}
 \centering
 \includegraphics[width=\textwidth]{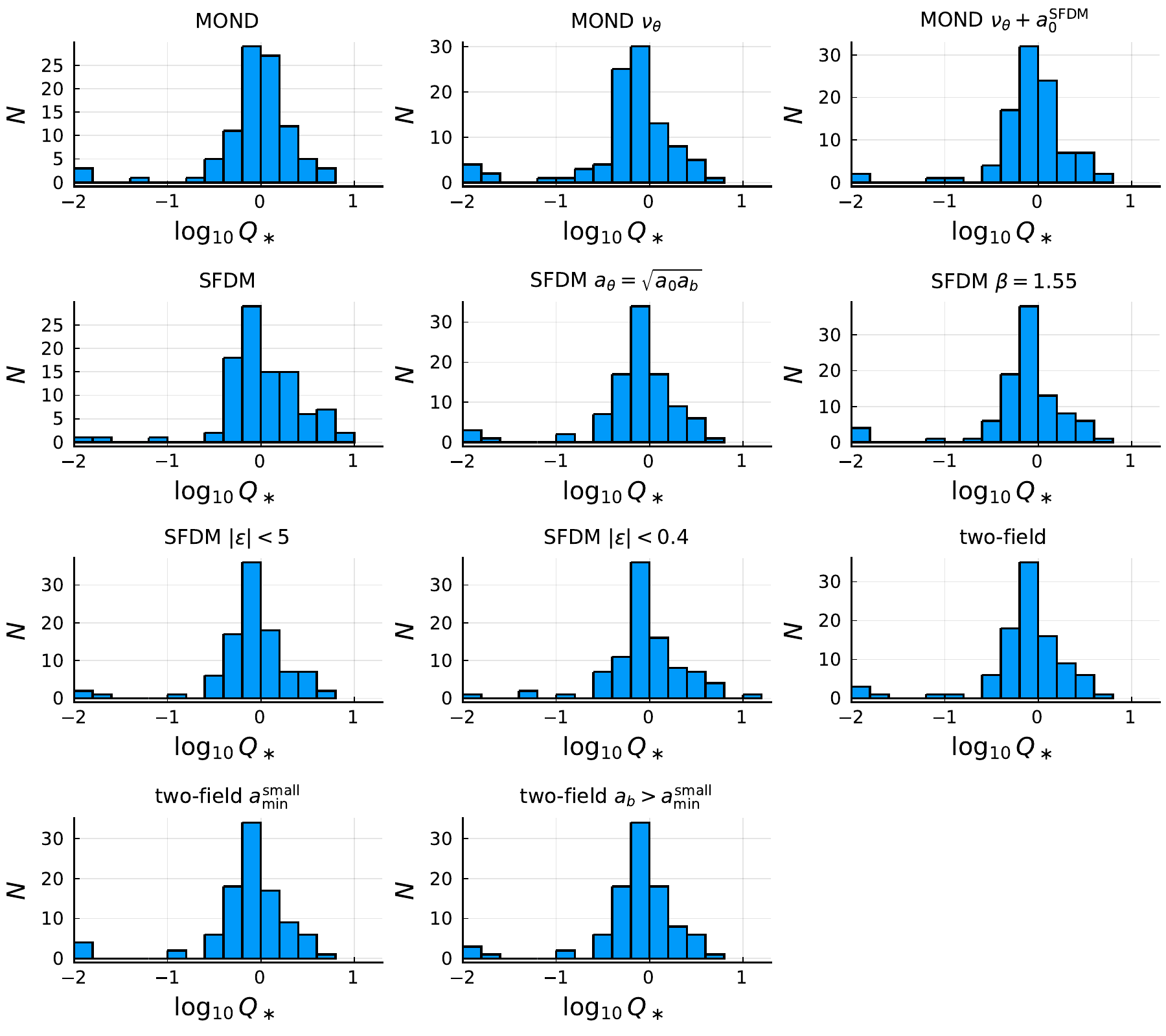}
 \caption{
     Histograms of the best-fit $f_Y$ values for the different models restricted to the $\Qual=1$ galaxies.
    }
 \label{fig:fYhist}
\end{figure*}

\subsection{$M/L_*$ in MOND}
\label{sec:rar}

Our MOND fit should give results roughly comparable to \cite{Li2018},
    which also fitted the RAR to SPARC galaxies.
A difference is that \cite{Li2018} used an MCMC procedure with Gaussian priors, while we used a simple parameter scan to minimize $\chi^2$.
We also did not vary distance and inclination and did not separately vary the mass-to-light ratio of the stellar disk and the bulge.
As a consequence of this simplified fitting procedure,
    the distribution of best-fit $M/L_*$ has more outliers and looks less like a Gaussian in our case compared to \cite{Li2018}.
This can be seen for example in Fig.~\ref{fig:fYhist}, which shows the histograms for the best-fit $f_Y$ for the galaxies with the SPARC quality flag $\Qual=1$.

Still, the median best-fit stellar mass-to-light ratios and the best-fit $\chi^2$ values are similar to those from \cite{Li2018}.
The median stellar disk $M/L$ is $0.39$.
When we restrict the ourselves to $\Qual=1$ galaxies, this becomes $0.47$.
This is shown in Table~\ref{tab:fYmedian}.
This is in reasonable agreement with \cite{Li2018}, who obtained $0.50$.
We show the cumulative $\chi^2$ distribution in Fig.~\ref{fig:chi2cdf}, which is also in reasonable agreement with \cite{Li2018}.

In Fig.~\ref{fig:fYhist}, one sees that some galaxies end up at the minimum stellar mass-to-light ratio allowed in our fitting method, corresponding to $Q_* \approx 0.01$.
If we do not restrict ourselves to $\Qual=1$, this peak at $Q_* \approx 0.01$ is even more pronounced.
These galaxies with $Q_* \approx 0.01$ come about as follows.
Consider a galaxy where the observed $V_{\mathrm{obs}}$ is smaller than that computed in MOND.
The computed rotation curve can be brought closer to $V_{\mathrm{obs}}$ by decreasing $M/L_*$.
It can happen that $M/L_*$ must be reduced so much that the gas component, which is unaffected by $M/L_*$, dominates.
When this happens and when $V_{\mathrm{obs}}$ is still smaller than the computed rotation velocity,
    the fitting code will continue to decrease $M/L_*$ to improve $\chi^2$.
But of course this will only barely change $\chi^2$ since the gas component dominates, and the galaxy ends up at the minimum allowed mass-to-light ratio corresponding to $Q_* = 0.01$.
This does not happen in \cite{Li2018}, since varying the distance and inclination can avoid such situations and also because the Gaussian priors discourage going to the minimum allowed $M/L_*$.

Thus, the best-fit mass-to-light ratios of the galaxies at $Q_* = 0.01$ should not be taken seriously.
They are an artifact of our simplified fitting procedure.
They have a comparably good fit also with larger $Q_*$.
We verified that the galaxies at $Q_* = 0.01$ are gas-dominated at their best-fit $Q_*$.
To check that our results do not depend on these outlier galaxies, we also include a column in Table~\ref{tab:fYmedian} that averages only over the galaxies where the MOND fit gives $\log_{10} \, Q_* > -1.5$.
This gives a median stellar disk $M/L$ of $0.44$.
This lies between the result for the $\Qual=1$ galaxies and the one we got when not restricting the galaxies.

In Table~\ref{tab:fYmedian} and Fig.~\ref{fig:chi2cdf}, we also show the results for a fourth quality cut we call ``thermal ok.''
This refers to the galaxies where, in our SFDM fit discussed below, a simple estimate shows that all SPARC data points lie within the superfluid core of the galaxy (see Appendix~\ref{sec:results:sfdm:thermal} for more details).
Here, we just note that this quality cut does not qualitatively change our results.

We show the mean stellar disk mass-to-light ratios in Table~\ref{tab:fYmean}.
These differ from the median values for all quality cuts because the resulting $f_Y$ distributions are not Gaussian as already discussed.

\subsection{$M/L_*$ in SFDM}
\label{sec:results:sfdm}

\begin{figure*}
 \centering
 \includegraphics[width=\textwidth]{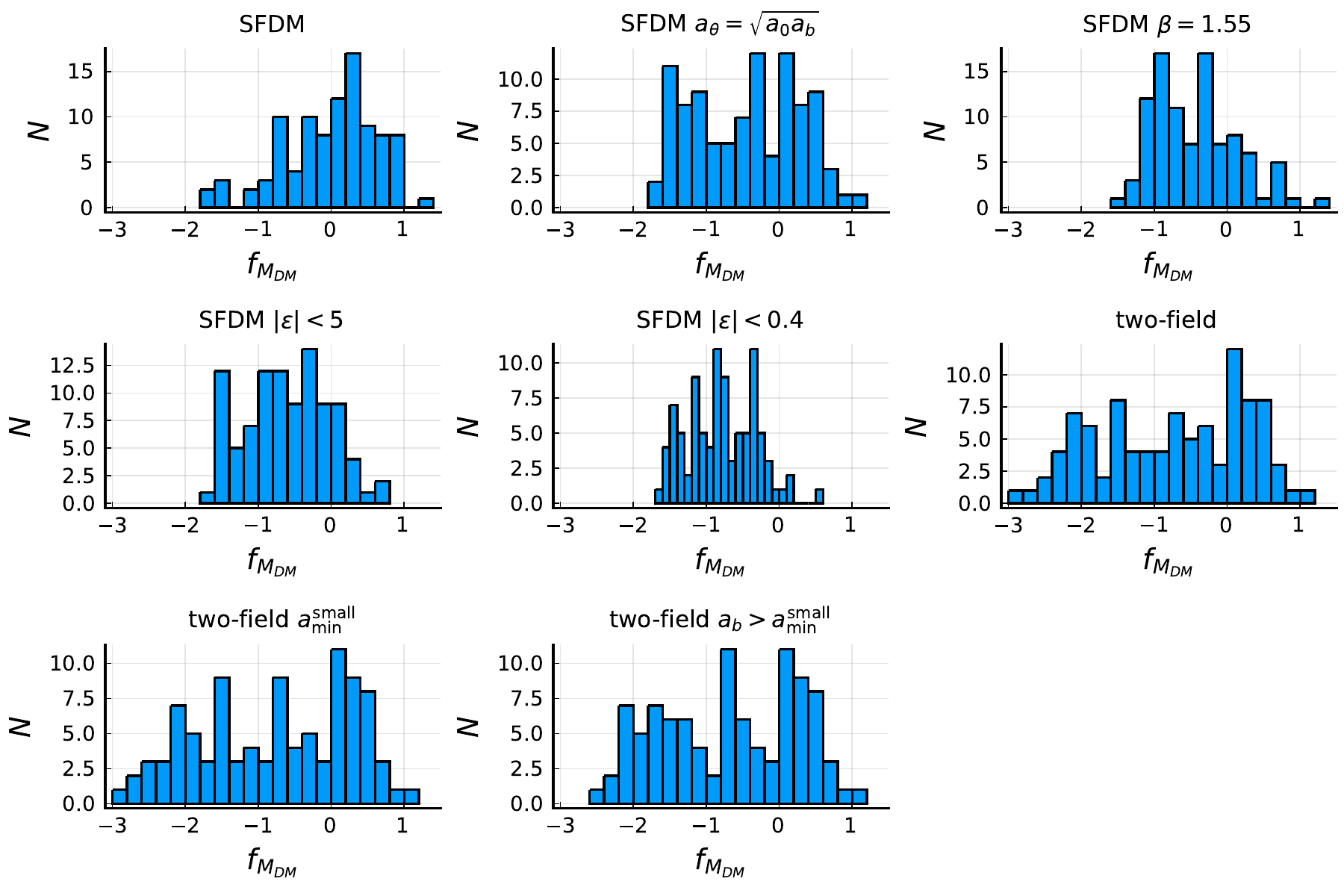}
 \caption{
     Histograms of the best-fit $\fMDM$ values for the different models restricted to the $\Qual=1$ galaxies.
    }
 \label{fig:fMDMhist}
\end{figure*}

For SFDM, we show the $\chi^2$ CDF in Fig.~\ref{fig:chi2cdf} and the $Q_*$ and $\fMDM$ histograms for the $\Qual=1$ galaxies in Fig.~\ref{fig:fYhist} and Fig.~\ref{fig:fMDMhist}.
The $\chi^2$ CDF and the $Q_*$ histogram look qualitatively similar to those from the MOND fit,
    just with some numerical differences.
For example, as for the MOND fit, there are some galaxies at the minimum value $Q_* = 0.01$.
These are the galaxies that become gas-dominated during the fitting procedure as explained in Appendix~\ref{sec:rar}.
The precise distribution of best-fit $\fMDM$ values should not be taken too seriously, especially at smaller values.
This is because the superfluid halo's Newtonian gravitational pull is often subdominant in SFDM,
    so that our fitting method cannot really distinguish different $\fMDM$ values, as long as $a_{\mathrm{SF}}$ stays subdominant.

\subsubsection{Stellar mass-to-light ratio}
\label{sec:results:sfdm:ML}

Our initial question was whether or not SFDM needs a smaller $M/L_*$ than standard MOND models.
We find that this is not necessarily the case.
In SFDM, the median stellar disk $M/L$ is $0.49$ for the $\Qual=1$ galaxies.
This is not much smaller than MOND.
The numerical details depend on whether one considers the mean or the median and on the chosen galaxy cuts (see Tables~\ref{tab:fYmedian} and \ref{tab:fYmean}).
Still, a robust finding across all of these choices is that SFDM does not give a significantly smaller $Q_*$ than MOND.

There are two reasons for this.
First, contrary to what one would hope for in SFDM, many galaxies do not end up in the MOND limit $|\varepsilon_*| \ll 1$.
If the phonon force $a_\theta$ is close to its MOND-limit value $\sqrt{a_0 a_b}$, SFDM does give smaller averaged $Q_*$ than MOND.
Second, even in the MOND limit, the best-fit $Q_*$ values in SFDM are systematically smaller than in  MOND only for certain galaxy types.
Such trends are not expected from SPS models.
It also means we can never say that SFDM universally requires a smaller or larger $Q_*$ than MOND.
We can make such statements only for a given galaxy sample.
We now discuss these two points in more detail.

\subsubsection{Effect of going outside the MOND limit on $M/L_*$ }
\label{sec:largeML}

As discussed in Appendix~\ref{sec:sfdm:mond}, going to $\varepsilon_* \gg 1$ allows us to make $a_\theta$ smaller so that a larger $M/L_*$ is possible.
This could be one reason why the averaged $M/L_*$ is relatively large in SFDM.

\begin{figure*}
 \centering
 \includegraphics[width=\textwidth]{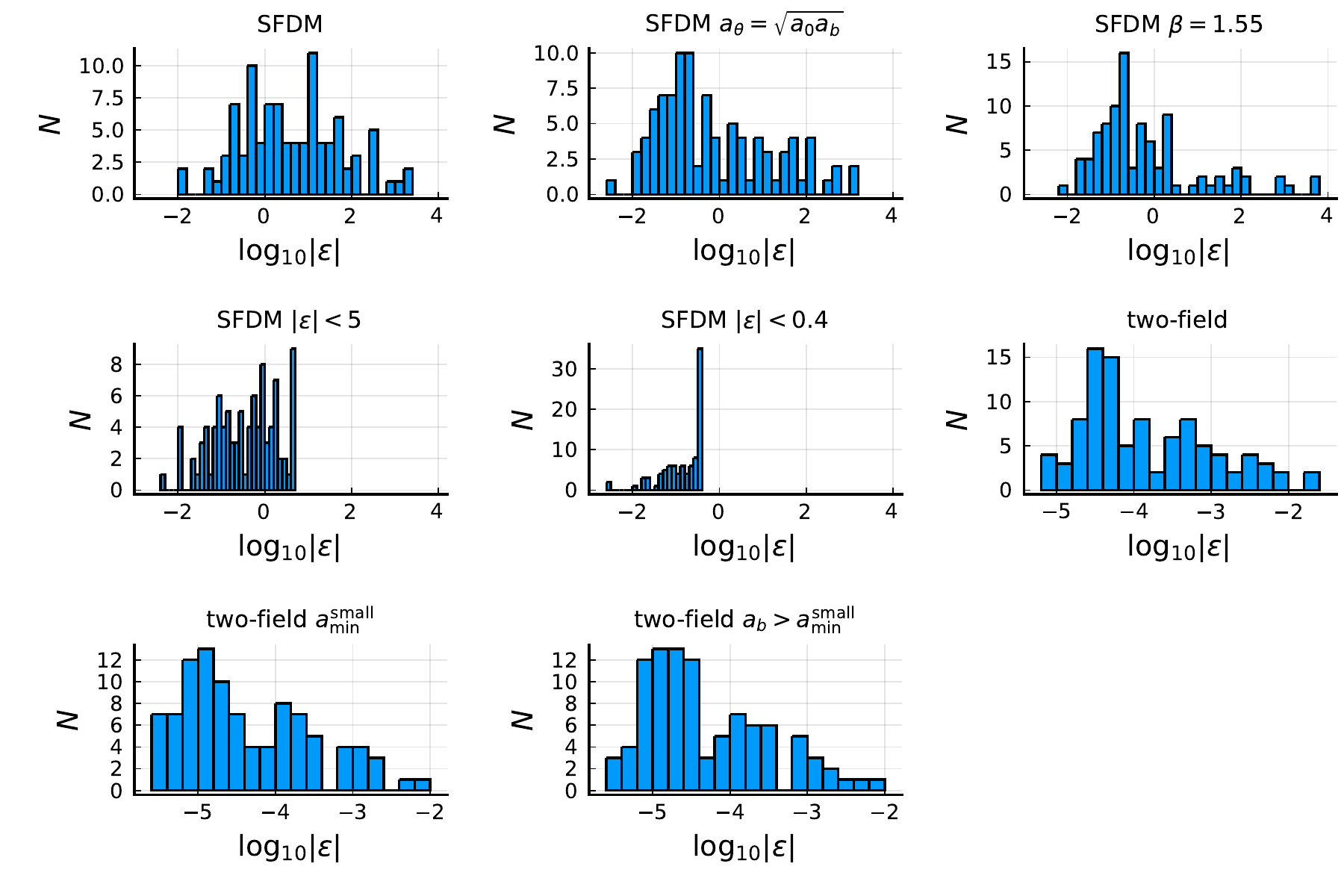}
 \caption{
     Histograms of the best-fit values of $\varepsilon = \varepsilon_*(R_{\mathrm{mid}})$ for the different models restricted to the $\Qual=1$ galaxies.
    }
 \label{fig:estarhist}
\end{figure*}

As a first check, we show a scatter plot of $Q_*$ versus $\varepsilon_*(R_{\mathrm{mid}})$ (see Fig.~\ref{fig:estarscatter}).
Indeed, many galaxies have $\varepsilon_* \gg 1$.
In addition, Fig.~\ref{fig:estarscatter} shows a correlation between $\varepsilon_*$ and $Q_*$.
Galaxies with $\varepsilon_* \gg 1$ tend to have a larger $Q_*$.
This fits with the idea that we do not find a smaller $M/L_*$ for SFDM because many galaxies are not in the MOND limit.

To confirm this, we redid the SFDM fit, but with the phonon acceleration $a_\theta$ replaced by its MOND limit value $\sqrt{a_0 a_b}$ when calculating rotation curves.
The calculation of $\hat{\mu}$ was left untouched.
This is the model shown as ``SFDM $a_\theta = \sqrt{a_0 a_b}$'' in, for example, Table~\ref{tab:fYmedian} and Fig.~\ref{fig:fYhist}.
With this model, the trick of going to $\varepsilon_* \gg 1$ to make the phonon acceleration $a_\theta$ small does not work.
Indeed, the averaged $M/L_*$ is now significantly smaller than for the original SFDM fit.
This result is again robust against different choices for the galaxies we consider and different choices for the averaging function.
We can also see explicitly in Fig.~\ref{fig:estarhist} that the distribution of best-fit $\varepsilon$ values has migrated to smaller values compared to the original SFDM fit.

As a third check, we redid the SFDM fit, but with the model parameter $\beta$ set to $1.55$ instead of $2$.
This choice makes it much harder to make the phonon acceleration $a_\theta$ small by going to large $\varepsilon_* \gg 1$,
    as can be seen from Fig.~\ref{fig:atheta-estar}.
This is the model shown as ``SFDM $\beta=1.55$'' in, for example, Table~\ref{tab:fYmedian} and Fig.~\ref{fig:fYhist}.
If our explanation for the large $M/L_*$ in SFDM is correct, this modified model should again have significantly smaller $M/L_*$.
Indeed, this SFDM $\beta=1.55$ model gives results that are comparable to those from the SFDM $a_\theta=\sqrt{a_0 a_b}$ model.
That is, $M/L_*$ is now significantly smaller than for the SFDM fit.
Similarly, the resulting $\varepsilon$ values are much smaller than in the original SFDM fit (see Fig.~\ref{fig:estarhist}).

Thus, one reason for the relatively large $M/L_*$ in SFDM is indeed that many galaxies are not actually in the MOND limit.

\subsubsection{Enforcing the MOND limit}
\label{sec:nomondrequired}

Its MOND limit is one of the main motivations of SFDM, because then rotation curves are automatically MOND-like.
This is not the case outside the MOND limit (i.e., when $\varepsilon_*$ is not small).
Then MOND-like rotation curves are not possible without adjusting the boundary condition separately for each galaxy.
Thus, our fit results for SFDM go against the original motivation behind SFDM.

We may wonder if large $\varepsilon_*$ values are really necessary for SFDM to get reasonable fits of the SPARC data.
It is possible that our fit code went to $\varepsilon_* \gg 1$ for little gain in $\chi^2$.
To check this,
    we redid the SFDM fit,
    but with $\varepsilon = \varepsilon_*(R_{\mathrm{mid}})$ restricted to $|\varepsilon| < 0.1$.
Whenever we solved the SFDM equations and found $|\varepsilon| \geq 0.1$, we manually set $\chi^2 = 10^{7}(1 + |\varepsilon|)$ so that the fitting code went elsewhere.
The $|\varepsilon|$ in $10^7(1+|\varepsilon|)$ is to help the $\chi^2$-minimizing fit algorithm to find small $|\varepsilon|$.
In this fit, all galaxies are restricted to stay in the proper MOND limit $|\varepsilon| \ll 1$ of SFDM.

\begin{figure}
 \centering
 \includegraphics[width=.48\textwidth]{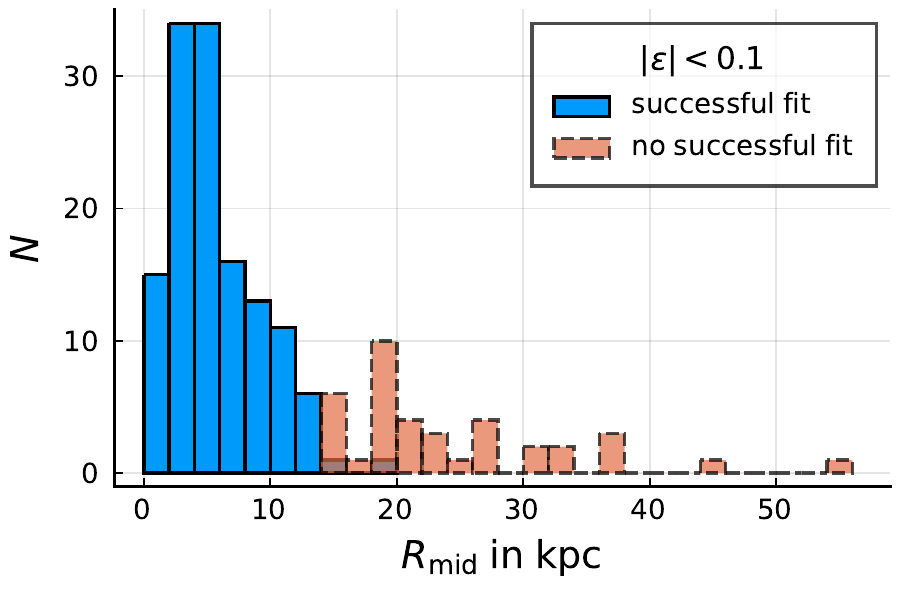}
 \includegraphics[width=.48\textwidth]{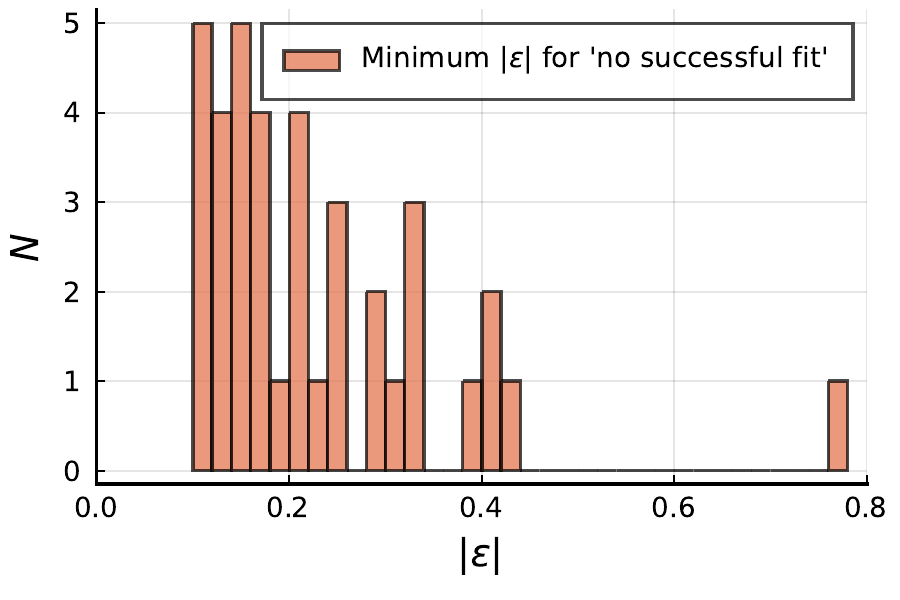}
 \caption{
     Fit results illustrating the ability of galaxies to satisfy the MOND limit condition $|\varepsilon| \ll 1$.
     Top:
     Histograms of the radius $R_{\mathrm{mid}}$ for galaxies where we did (blue) and did not (red) find a fit with the restriction $|\varepsilon| < 0.1$.
     We see that only smaller galaxies tend to be able to satisfy the condition $|\varepsilon| < 0.1$,
        consistent with the simple estimate $\varepsilon_* \gtrsim 0.1 + 0.4 \cdot (r/18\,\mathrm{kpc} - 1)$ from Eq.~\ref{eq:estarlowerbound}.
    Bottom:
    Minimum possible $|\varepsilon|$ for the galaxies where we could find no fit with $|\varepsilon| < 0.1$.
    We see that many only barely fail to satisfy $|\varepsilon| < 0.1,$ and we can get a fit for almost all galaxies with $|\varepsilon| < 0.4$.
    }
 \label{fig:smallestar01fails}
\end{figure}

This works, but only for galaxies that are not too large.
The restriction $|\varepsilon| < 0.1$ is impossible to satisfy for many larger galaxies.
Our fitting procedure did not find any fit for $38$ out of the $169$ SPARC galaxies (i.e., $38$ galaxies end up with $\chi^2 > 10^7$).
This is not unexpected since our estimate $\varepsilon_* \gtrsim 0.1 + 0.4 \cdot ( r/18\,\mathrm{kpc} -1)$ from Eq.~\ref{eq:estarlowerbound} rules out $|\varepsilon| < 0.1$ for many larger galaxies.
Indeed, Fig.~\ref{fig:smallestar01fails}, top, shows that the galaxies that cannot be fit with $|\varepsilon| < 0.1$ tend to be those with $R_{\mathrm{mid}} \gtrsim 15\,\mathrm{kpc}$.

Since we set $\chi^2 = 10^7(1 + |\varepsilon|)$ when the condition $|\varepsilon| < 0.1$ was not satisfied,
    our fitting algorithm actually minimized $|\varepsilon|$ until it satisfied $|\varepsilon| < 0.1$.
Thus, we can get the minimum possible $|\varepsilon|$ for each galaxy where $|\varepsilon| < 0.1$ could not be satisfied.
The results are shown in Fig.~\ref{fig:smallestar01fails}, bottom.
We see that many galaxies only barely fail to satisfy the condition $|\varepsilon| < 0.1$.
Indeed, if we allowed $|\varepsilon|$ up to $0.4$, almost all galaxies could be fit.

Of course, $0.4$ is not that small, so it is debatable whether or not a value $|\varepsilon| = 0.4$ still counts as the proper MOND limit $|\varepsilon_*| \ll 1$.
Here, we do not dwell on this point.
However, we did redo our fit with the condition $|\varepsilon| < 0.1$ replaced by the condition $|\varepsilon| < 0.4$.
As expected, we then obtained a fit for almost all galaxies.
We did not find a fit for only four galaxies.
Thus, if by ``proper MOND limit'' we mean $|\varepsilon| < 0.1$,
    SFDM's proper MOND limit does not work for lager galaxies.
But if we allow $|\varepsilon|$ up to $0.4$, it might.

\begin{figure}
 \centering
 \includegraphics[width=.48\textwidth]{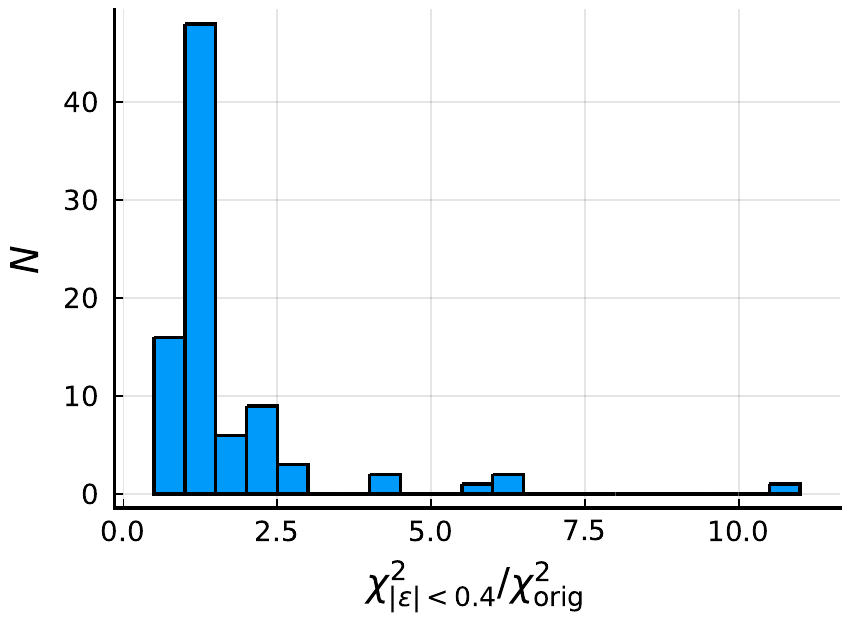}
 \caption{
     Histogram of the change in best-fit $\chi^2$ values for the $\Qual=1$ galaxies when switching from the unrestricted SFDM fit to that with the restriction $|\varepsilon| < 0.4$.
     This is excluding the galaxies that only barely satisfy $|\varepsilon| < 0.4$ and therefore have a bad $\chi^2$,
        i.e., excluding galaxies with $\varepsilon > 0.38$ and $\chi^2 > 100$.
    }
 \label{fig:smallestar04-chi2-changes}
\end{figure}

The resulting best-fit $\chi^2$ CDF is shown in Fig.~\ref{fig:chi2cdf}.
In Fig.~\ref{fig:smallestar04-chi2-changes},
    we show the changes in $\chi^2$ between the ``SFDM $|\varepsilon| < 0.4$'' fit and the unrestricted SFDM fit.
The resulting $\chi^2$ values tend to be worse than for the unrestricted SFDM fit, but generally still acceptable.
Indeed, they are quite similar to those of the MOND fit (see the CDF in Fig.~\ref{fig:chi2cdf}).

For some galaxies, the condition $|\varepsilon| < 0.4$ can barely be satisfied.
After satisfying this condition they have basically no freedom left to actually fit the observed rotation curve data and they end up with very bad $\chi^2$.
Specifically, there are seven galaxies with $\varepsilon > 0.38$ and $\chi^2 > 100$.
Since these are hardly useful in assessing the $M/L_*$ required in SFDM, we separately list the $M/L_*$ results with these galaxies excluded in Tables~\ref{tab:fYmean} and~\ref{tab:fYmedian}.
We also exclude them in Fig.~\ref{fig:smallestar04-chi2-changes}.

Somewhat surprisingly, some galaxies even have a better best-fit $\chi^2$ with the $|\varepsilon| < 0.4$ restriction than without (see Fig.~\ref{fig:smallestar04-chi2-changes}).
Some of these are just very slightly better than the previous best fit.
For two galaxies it improves by more than $15\%$,
    specifically by $46\%$ for NGC1090 and by $19\%$ for NGC2683.
For all galaxies with an improved $\chi^2$, the corresponding best-fit $M/L_*$ changes by less than $10\%$.
These differences are insignificant for our purposes.
They just show that our fitting algorithm is not perfect and does not always find the very best $\chi^2$.

If we exclude the galaxies with a bad $\chi^2$ because they can only barely satisfy $|\varepsilon| < 0.4$ as described above,
    the resulting averaged stellar mass-to-light ratios are between $4\%$ and $11\%$  smaller than for the MOND fit.
The numerical details depend on the averaging procedure and the cut of galaxies.
We discuss the resulting $M/L_*$ values in more detail in Appendix~\ref{sec:results:sfdm:MLtrends}.

An alternative to the proper MOND limit $|\varepsilon_*| \ll 1$ is the pseudo-MOND limit discussed in Appendix~\ref{sec:sfdm:mond}.
At $\varepsilon_* = \mathcal{O}(1)$ (for $\beta = 2$), the phonon acceleration $a_\theta$ is still numerically close to its MOND limit value $\sqrt{a_0 a_b}$ although it does not satisfy a MOND-like equation (see Fig.~\ref{fig:atheta-estar}, left).
To test this regime, we redid the SFDM fit with $\varepsilon$ restricted to $|\varepsilon| < 5$.
This is the model shown as ``SFDM $|\varepsilon| < 5$'' in, for example, Table~\ref{tab:fYmedian} and Fig.~\ref{fig:fYhist}.
Allowing values of $|\varepsilon|$ up to $5$ allows the phonon acceleration $a_\theta$ to deviate by up to about $5\%$ from its MOND limit value $\sqrt{a_0 a_b}$ (at $R = R_{\mathrm{mid}}$) (see Fig.~\ref{fig:atheta-estar}).
The resulting $\chi^2$ values and stellar mass-to-light ratios are roughly comparable to those of the $|\varepsilon| < 0.4$ fits.
As always, the numerical details depend on the choice of galaxies and on whether we average using the median or the mean.

Thus, fitting the SPARC data does not require $\varepsilon_* \gg 1$ (i.e., it does not require going outside the MOND limit).
Both the proper MOND limit and the pseudo-MOND limit also give reasonable $\chi^2$ values.
In this case, the averaged $M/L_*$ is a bit smaller than in standard MOND models.
In Appendix~\ref{sec:results:sfdm:MLtrends}, we discuss the $M/L_*$ of these fits in more detail.

\subsubsection{Trends of $M/L_*$ with galaxy type}
\label{sec:results:sfdm:MLtrends}

We now come back to the question of why SFDM does not necessarily need a smaller averaged $M/L_*$ compared to MOND.
Above, we already identified one reason, namely that many galaxies are not in the MOND limit $|\varepsilon_*| \ll 1$.
But this is not the whole story, as we will now explain.
To this end, we consider the fits with the MOND limit enforced as introduced in the previous subsection.
This excludes effects from going outside the MOND limit.

\begin{figure}
 \centering
 \includegraphics[width=.48\textwidth]{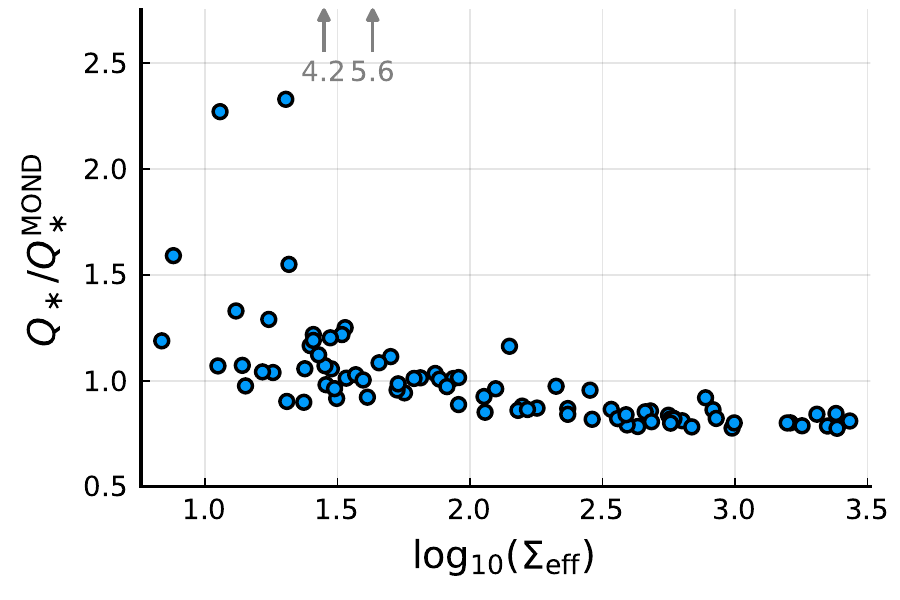}
 \includegraphics[width=.48\textwidth]{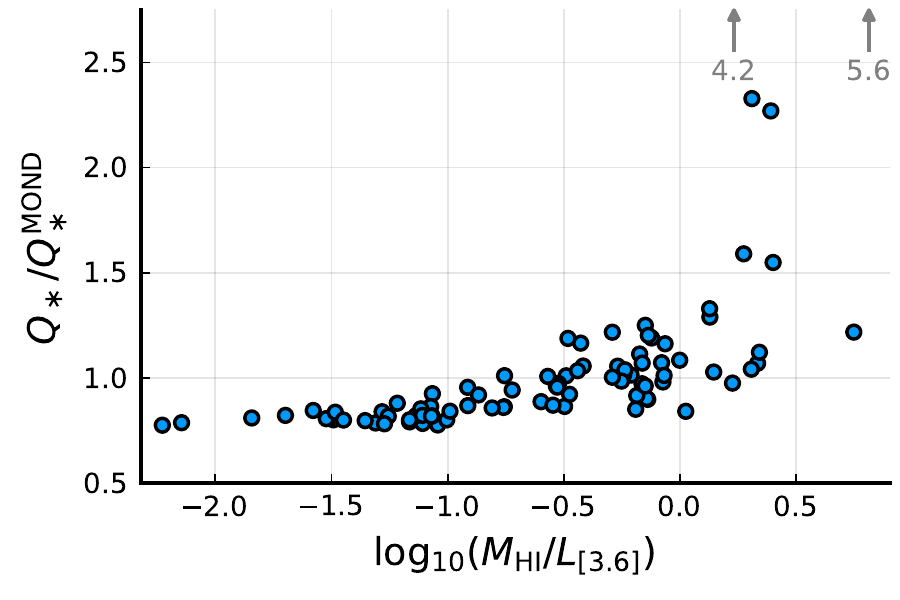}
  \caption{
        Same as Fig.~\ref{fig:this} but for galaxy properties other than the flat rotation curve velocity, $V_{\mathrm{flat}}$.
        Top: For the effective surface brightness, $\Sigma_{\mathrm{eff}}$.
        Bottom: For $M_{HI}/L_{[3.6]}$.
    }
 \label{fig:this23}
\end{figure}

In Appendix~\ref{sec:sfdm:less},
    we argued that the MOND limit of SFDM likely requires a systematically smaller $M/L_*$ than MOND only for high-acceleration galaxies.
Galaxies with smaller accelerations may even require a larger $M/L_*$ than MOND.
The main reason is that SFDM has a smaller value of $a_0$ than MOND,
    which becomes important at small accelerations.
If this is true, having a smaller or larger $Q_*$ in SFDM is not just a property of the model but also a property of the galaxy sample.
An example of this is Fig.~\ref{fig:this},
which shows the best-fit $Q_*$ of each galaxy in the SFDM $|\varepsilon| < 0.4$ model relative to the best-fit $Q_*$ for MOND as a function of the observed flat rotation velocity $V_{\mathrm{flat}}$.
A larger $V_{\mathrm{flat}}$ is associated with larger accelerations.
Indeed, large $V_{\mathrm{flat}}$ values are where SFDM systematically gives smaller $Q_*$ than MOND.
Similarly, a smaller gas fraction and a larger surface brightness are associated with larger accelerations.
The effective surface brightness $\Sigma_{\mathrm{eff}}$ and the ratio $M_{HI}/L_{[3.6]}$, a proxy for gas fraction, of each galaxy are also part of SPARC.
And indeed, Fig.~\ref{fig:this23} shows that SFDM has a systematically smaller $M/L_*$ than MOND for galaxies with a large $\Sigma_{\mathrm{eff}}$ and a small $M_{HI}/L_{[3.6]}$.

\begin{figure}
 \centering
 \includegraphics[width=.48\textwidth]{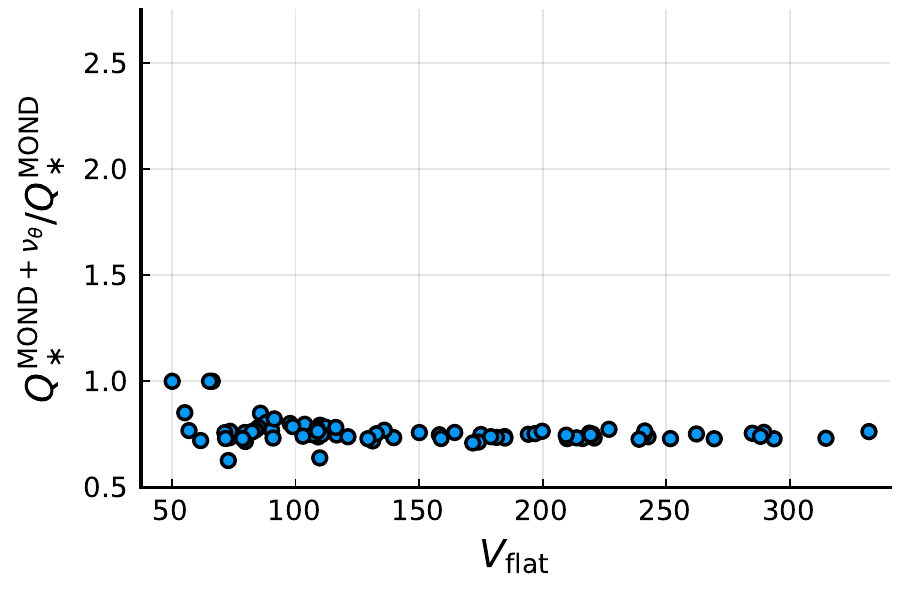}
  \caption{
      Same as Fig.~\ref{fig:this} but for MOND fits that use the SFDM-like interpolation function, $\nu_\theta$, instead of $\nu_e$ and keeping the larger value $a_0^{\mathrm{MOND}}$ of $a_0$ (rather than using the smaller value, $a_0^{\mathrm{SFDM}}$).
    }
 \label{fig:thisRARsimp}
\end{figure}

To further test our understanding,
    we redid the MOND fit but using both the interpolation function $\nu_\theta$ instead of $\nu_e$ (see Appendix~\ref{sec:sfdm:less}) and the smaller value $a_0^{\mathrm{SFDM}}$ of $a_0$.
This should give fits qualitatively similar to those of the MOND limit of SFDM.
Indeed, we verified that the resulting best-fit $M/L_*$ show similar trends with, for example, $V_{\mathrm{flat}}$ as SFDM.
As we explain in Appendix~\ref{sec:sfdm:less},
    the different shape of the interpolation function is responsible for the systematically smaller $M/L_*$ at large accelerations.
The smaller $a_0$ value is responsible for the fact that this is not true at smaller accelerations.
Thus, in a MOND model with the SFDM-like interpolation function $\nu_\theta$ but with the larger $a_0$ value $a_0^{\mathrm{MOND}}$,
    we would expect to see a smaller $M/L_*$ consistently across all accelerations.
To test this, we redid the MOND fit with the interpolation function $\nu_\theta$ but keeping the larger value $a_0^{\mathrm{MOND}}$ of $a_0$.
And indeed, this gives a consistently smaller $M/L_*$ than in MOND.
There is no clear trend with, for example, $V_{\mathrm{flat}}$ (see Fig.~\ref{fig:thisRARsimp}).

Trends of the stellar $M/L$ in the [3.6] micron band with galaxy properties are not expected from SPS models
\citep{Schombert2019}.
This disfavors SFDM, especially compared to MOND, which does not show such trends.
If anything, one expects the opposite trend: more massive galaxies should have higher mass-to-light ratios than dwarfs, especially in optical bands.

More precisely, when we do not normalize $Q_*$ to $Q_*^{\mathrm{MOND}}$,
    our fitting procedure reproduces the SPS expectations for neither MOND nor SFDM.
But this is a consequence of our simplistic fitting procedure.
More sophisticated fits do reproduce the SPS expectations for MOND \citep{McGaugh2004}.
Moreover, since our simple fitting procedure is suited to identify relative differences between MOND and SFDM, we expect the trends in $Q_*/Q_*^{\mathrm{MOND}}$ to be robust. It is unlikely that these would be mitigated by
a more sophisticated fitting procedure.
That is, we expect that this conflict between SFDM and SPS expectations is real.

Figures \ref{fig:this} and \ref{fig:this23} not only show that SFDM has a systematically small $Q_*$ for large accelerations.
They also show that there is more scatter in $Q_*/Q_*^{\mathrm{MOND}}$ at smaller accelerations.
One reason for this scatter is that many of the small-acceleration galaxies are gas-dominated.
In gas-dominated galaxies, the formal best-fit value for the stellar mass-to-light ratio (i.e., $Q_*$)
    may not mean much, since $a_b$ is relatively independent of $Q_*$.
This allows for more scatter in $Q_*$.

\subsubsection{The Milky Way}

\cite{Hossenfelder2020} find that SFDM requires about $20\%$ less baryonic mass than standard MOND models to fit the Milky Way rotation curve at $R \lesssim 25\,\mathrm{kpc}$.
Specifically, it requires about $20\%$ less baryonic mass than the MOND model from \cite{McGaugh2019b}.
This is a significantly larger difference than what we find, on average, for the SPARC galaxies (see Tables~\ref{tab:fYmedian} and \ref{tab:fYmean}).
Here, we confirm and discuss this result.

To confirm the result of \cite{Hossenfelder2020},
    we fitted the Milky Way with the same method we used for the SPARC galaxies.
For this, we used $\rho_b$ and $V_{\mathrm{obs}}$ as in \cite{Hossenfelder2020}.
The $V_{\mathrm{obs}}$ data based on \cite{Portail2017} that is used in \cite{Hossenfelder2020} for $R \lesssim 2\,\mathrm{kpc}$ does not come with error bars.
As a simple way to still get a result, we assumed an error of $5\,\mathrm{km}/\mathrm{s}$ for these $V_{\mathrm{obs}}$ data points.
For easier comparison to the SPARC fits,
    we rescaled the stellar disk and bulge densities such that stellar mass-to-light ratios of $0.5$ (for the stellar disk) and $0.7$ (for the bulge) correspond to the baryonic mass model used in \cite{McGaugh2019b}.
That is, the factor $10^{f_Y}$ tells us how much less stellar mass SFDM uses compared to the standard MOND model from \cite{McGaugh2019b}.
We found a best-fit $\chi^2$ of $2.69$, a best-fit $Q_*$ of $0.79$, and a best-fit $\varepsilon$ of $4.37$.
This confirms the estimate from \cite{Hossenfelder2020} of about $20\%$ less baryonic mass compared to standard MOND.
This fit stays roughly within the pseudo-MOND limit $\varepsilon_* = \mathcal{O}(1)$.
With the best-fit parameters, $\varepsilon_*$ stays below $15$ at $R < 25\,\mathrm{kpc}$.
Thus, the phonon acceleration cannot be much suppressed and cannot allow for an increased $M/L_*$,
    in contrast to galaxies at $\varepsilon_* \gg 1$ (see Appendix~\ref{sec:largeML}).

The Milky Way's $a_b$ ranges from about $10^{-9}\,\mathrm{m}/\mathrm{s}^2$ to about $10^{-10.8}\,\mathrm{m}/\mathrm{s}^2$ at $R < 25\,\mathrm{kpc}$ for the best-fit parameters.
These are relatively large.
So, from the discussion in Appendix~\ref{sec:results:sfdm:MLtrends}, a smaller $M/L_*$ than in MOND is what one would expect.

\subsubsection{SFDM model parameters}
\label{sec:sfdm:params}

Above, we used the fiducial parameter values from \cite{Berezhiani2018} and kept them fixed during the fitting procedure.
Here, we discuss whether our conclusions could be changed by adjusting these parameters.

Our calculation does not depend on each of the four parameters, $\alpha$, $\Lambda$, $m$, and $\beta$, separately.
We need only the combinations
\begin{align}
 a_0 = \frac{\alpha^3 \Lambda^2}{M_{\mathrm{Pl}}}\,, \qquad \beta \,, \qquad \frac{m^2}{\alpha} \,.
\end{align}
This can be seen directly from the SFDM Lagrangian \citep{Berezhiani2015} by rescaling the phonon field, $\theta \to \theta (\alpha \Lambda/M_{\mathrm{Pl}})^{-1}$.
It also follows from $(\Lambda m^3)^2 = a_0 (m^2/\alpha)^3 M_{\mathrm{Pl}}$.

First, the acceleration scale $a_0$.
To reproduce MOND, $a_0$ must be close to $10^{-10}\,\mathrm{m}/\mathrm{s}^2$.
Still, we could choose the same value as in standard MOND rather than the somewhat smaller value that \cite{Berezhiani2018} chose.
This would give $M/L_*$ values that are smaller than those for MOND for all galaxies,
    not just the small-acceleration ones (see Appendix~\ref{sec:results:sfdm:MLtrends}),
    at least as long as the superfluid's gravitational pull stays negligible.
That is, SFDM would give $M/L_*$ values similar to our ``MOND $\nu_\theta$'' fit (see Appendix~\ref{sec:results:sfdm:MLtrends}).
These are relatively small.
To get closer to $M/L_*$ values as expected from SPS modeling, one would have to change not only the value of $a_0$ but also the form of the interpolation function $\nu_\theta$.
This might be possible by adjusting the Lagrangian, that is, by changing what is usually called the function $P(X)$ in superfluid low-energy effective field theories.
Exploring this is beyond the scope of this paper.

One effect of the parameter $\beta$ is that it controls the phonon force outside the proper MOND limit $|\varepsilon_*| \ll 1$ (see Appendix~\ref{sec:sfdm:mond}).
For one example ($\beta = 1.55$),
    we explicitly explored the effect of a different value of $\beta$ (see Appendix~\ref{sec:largeML}).
Still, it is better not to tune this parameter for better fits to the data.
This is because the value of $\beta$ and the form of the finite-temperature corrections it is supposed to represent are completely ad hoc.
So they might turn out to be unphysical.
It is better to not rely too sensitively on any specific value of $\beta$ for the fits.

The combination $m^2/\alpha$ multiplies both $\varepsilon_*$ and the superfluid's energy density $\rho_{\mathrm{SF}}$.
One motivation to change $m^2/\alpha$ is to make $|\varepsilon_*|$ small in order to allow  more galaxies to reach SFDM's proper MOND limit $|\varepsilon_*| \ll 1$ (see Appendix~\ref{sec:sfdm:reachingmond}).
The problem with this is  that then $\rho_{\mathrm{SF}} \propto (m^2/\alpha) \cdot f_\beta(\varepsilon_*)$ becomes small as well.
Indeed, in this case the problem regarding strong lensing described in Appendix~\ref{sec:M200estar04} becomes even worse.
Conversely, making $m^2/\alpha$ larger in order to solve the strong lensing tension means even fewer galaxies can reach the MOND limit $|\varepsilon_*| \ll 1$.
For example, increasing $m^2/\alpha$ by a factor of $10$ would give $\varepsilon_* \gtrsim 0.1 + 0.41 (r/1.8\,\mathrm{kpc} - 1)$ from Eq.~\eqref{eq:estarlowerbound}.
Then, only the smallest galaxies could reach the proper MOND limit.

An adjusted function $P(X)$ might invalidate this argument since this function determines not only the phonon force but also the superfluid's energy density in SFDM.
But, again, this is beyond the scope of the present work.

To sum up, adjusting the SFDM model parameters might change some of our conclusions regarding the best-fit $M/L_*$ values, but probably not in a way that is completely satisfactory from the perspective of SPS models.
Moreover, we expect that the tradeoff between having MOND-like rotation curves and producing sufficient strong lensing described in Appendix~\ref{sec:M200estar04} remains.

\subsubsection{Thermal radius check}
\label{sec:results:sfdm:thermal}

We assumed that all SPARC rotation curve data points of a given galaxy lie within this galaxy's superfluid core.
This is necessary for one of the main motivations behind SFDM, namely to automatically reproduce MOND-like rotation curves without having to adjust the boundary condition separately for each galaxy.
The superfluid phase ends at the very latest when $\rho_{\mathrm{SF}}$ reaches zero.
Thus, we discarded solutions where $\rho_{\mathrm{SF}}$ vanishes inside the $V_{\mathrm{obs}}$ data points,
    as discussed in Appendix~\ref{sec:sfdm:calcSF}.

But this may not be sufficient since the superfluid phase may end even before $\rho_{\mathrm{SF}}$ reaches zero.
Consider, for example, the simplest estimate for the radius where the superfluid phase transitions to the non-superfluid phase, the so-called thermal radius $R_T$ \citep{Berezhiani2018}.
According to this estimate, the superfluid phase corresponds to
\begin{align}
 \Gamma > t_{\mathrm{dyn}}^{-1} \,,
\end{align}
where $\Gamma$ is the local self-interaction rate and $t_{\mathrm{dyn}}$ is the dynamical time.
Here, $\Gamma = (\sigma / m) \, \mathcal{N} \, v \, \rho$, where $\sigma$ is the self-interaction rate, $\mathcal{N} = (\rho/m) (2\pi/mv)^3$
is the Bose-degeneracy factor, and $v$ is the average velocity of the particles.
Following \cite{Berezhiani2018}, we take $ \sigma/m = 0.01\,\mathrm{cm}^2/\mathrm{g} $ and $t_{\mathrm{dyn}} = R/v$.

As a simple check, we evaluated the quantity $\Gamma/t_{\mathrm{dyn}}^{-1}$ for each galaxy at the last $V_{\mathrm{obs}}$ data point at $R_{\mathrm{max}}$ for the SFDM best fits.
We found that 31 of 169 galaxies violate the condition $\Gamma > t_{\mathrm{dyn}}^{-1}$ at $R_{\mathrm{max}}$.
In principle, we should discard these solutions, just as we discard solutions where $\rho_{\mathrm{SF}}$ reaches zero before $R_{\mathrm{max}}$.
Here, we do not do this.
The reason is that the condition $\Gamma > t_{\mathrm{dyn}}^{-1}$ is quite ad hoc.
For example, the value of $\sigma/m$ is chosen ad hoc and not derived from an underlying Lagrangian.
Also, the transition radius derived from $\Gamma = t_{\mathrm{dyn}}^{-1}$ can, in general, differ wildly from the transition radius derived from the so-called NFW matching procedure \citep{Berezhiani2018, Hossenfelder2020} where the density and pressure are matched to those of an NFW halo for a fixed NFW concentration parameter, as discussed in \cite{Mistele2020}.
This makes any particular choice for discarding solutions based on the thermal or NFW matching somewhat arbitrary.

We avoid this arbitrariness by discarding solutions only based on the criterion that $\rho_{\mathrm{SF}}$ must be positive.
Still, this means we do not discard some solutions that we maybe should discard.
This might affect our $M/L_*$ results.
To get a rough idea of possible effects due to this, we also show our results for $M/L_*$ in Table~\ref{tab:fYmedian} and Table~\ref{tab:fYmean} with the 31 galaxies violating $\Gamma > t_{\mathrm{dyn}}^{-1}$ excluded, labeled as ``thermal ok.''
Often, this does not significantly change the $M/L_*$ fit results, though it does give a larger $M/L_*$ for example for the SFDM model.

Overall, we expect that actually enforcing the rotation curve data to lie within the superfluid phase  does not significantly change our $M/L_*$ results.
But keep in mind that there is a considerable theoretical uncertainty around the transition from the superfluid to the non-superfluid phase.
We stress that the tension with strong lensing (see Appendix~\ref{sec:M200estar04}) is not affected by this uncertainty,
    because there we anyway choose the transition radius to maximize the resulting total dark matter mass.

\subsection{The RAR implied by the SFDM best fits}
\label{sec:RARplots}

\begin{figure*}
\centering
\includegraphics[width=.33\textwidth]{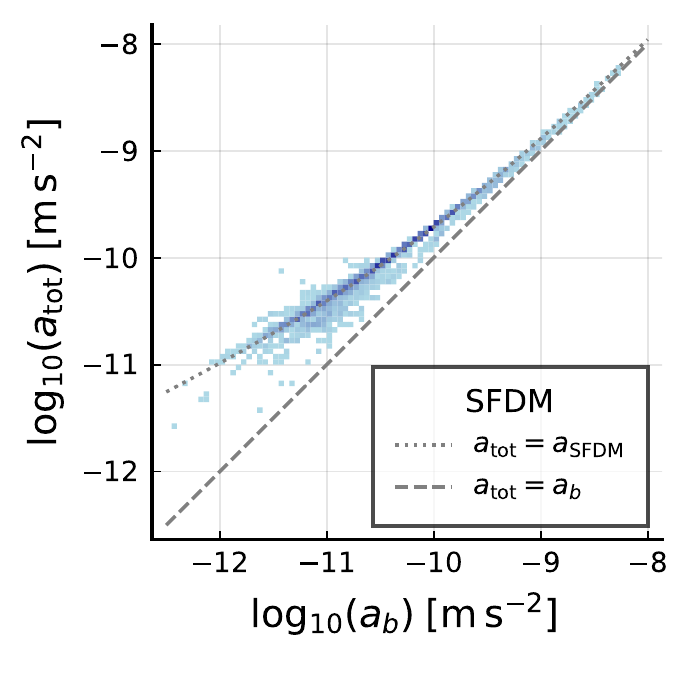}
\includegraphics[width=.33\textwidth]{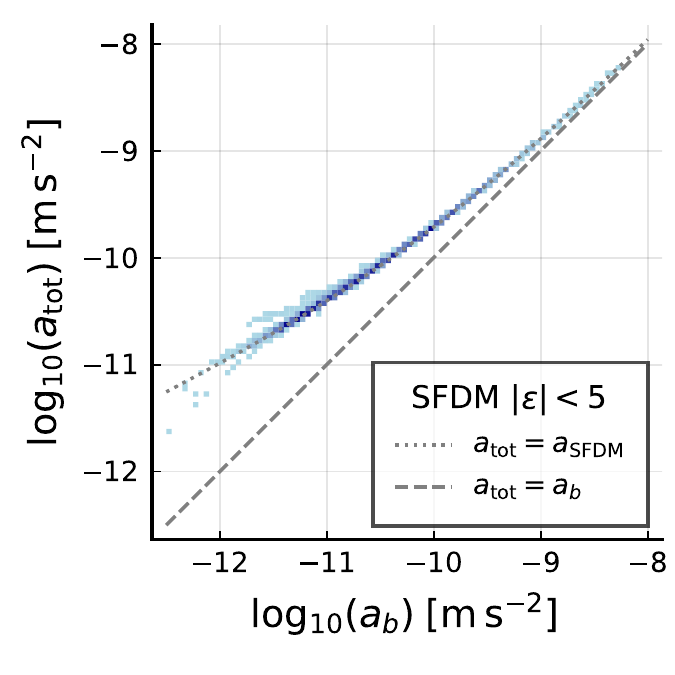}
\includegraphics[width=.33\textwidth]{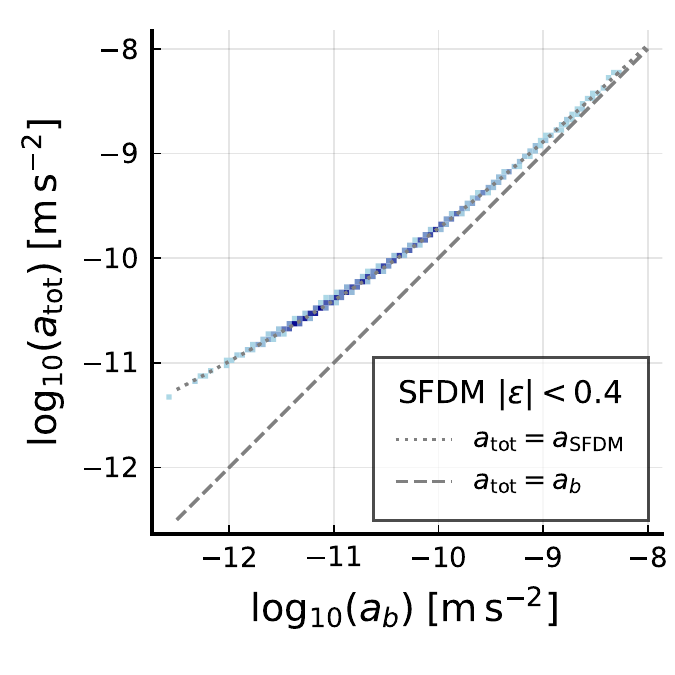}
\caption{
    RAR of the $\Qual=1$ galaxies implied by our SFDM fits for the unrestricted fit (left), the fit restricted to the pseudo-MOND limit ($|\varepsilon| < 5$, middle), and the fit restricted to the proper MOND limit ($|\varepsilon| < 0.4$, right).
    Each panel also shows the RAR implied by MOND with the SFDM-like interpolation function $\nu_\theta$ and $a_0 = 0.87 \cdot 10^{-10}\,\mathrm{m}/\mathrm{s}^2$, i.e., $a_{\mathrm{SFDM}} = a_b \, \nu_\theta(a_b/a_0)$ (dotted gray line)
        and by Newtonian gravity with no dark matter (dashed gray line).
    For the $|\varepsilon| < 0.4$ fit we exclude the galaxies that only barely satisfy $|\varepsilon| < 0.4$ and therefore have a bad $\chi^2$,
    i.e., we exclude galaxies with $\varepsilon > 0.38$ and $\chi^2 > 100$.
}
\label{fig:RAR-SFDM}
\end{figure*}

We show the RAR (i.e., the relation between $a_b$ and $a_{\mathrm{tot}}$) implied by our SFDM fits for the $\Qual = 1$ galaxies in Fig.~\ref{fig:RAR-SFDM}.
For each galaxy, we show one data point for each radius with a SPARC rotation curve data point $V_{\mathrm{obs}}$.
As expected, the fit restricted to the proper MOND limit ($|\varepsilon| < 0.4$) shows a tight relation with almost no scatter.
The unrestricted fit deviates from this in two ways.
First, it produces significantly more scatter.
Second, it systematically puts more data points below the relation implied by the proper MOND limit fit (i.e., it puts more data points at larger $a_b$, smaller $a_{\mathrm{tot}}$, or both).
Of course, outside the MOND limit of SFDM, there is no guarantee that one ends up with a MOND-like RAR.
Still, we can understand in more detail why our unrestricted fits look the way they do.

One contribution to the increased scatter is as follows.
The observed rotation curves contain some scatter from observational uncertainties.
Our unrestricted SFDM fits match these observed rotation curves more closely than the fits restricted to the MOND limit (see, e.g., Fig.~\ref{fig:smallestar04-chi2-changes}).
Thus, overfitting the noise in the observational data may, at least partly, explain the increased scatter in the RAR implied by our unrestricted fits.

We can also identify a possible cause for the systematic deviation from a MOND-like RAR.
Namely, some observed rotation curves can be matched more closely outside the MOND limit (i.e., with a large $|\varepsilon_*|$).
As discussed in Appendices~\ref{sec:sfdm:mond} and~\ref{sec:largeML}, going outside the MOND limit allows for larger $M/L_*$:
a large $|\varepsilon_*|$ makes $a_{\mathrm{tot}}$ smaller, which must be countered by a larger $M/L_*$.
These larger $M/L_*$ values explain why the RAR implied by our unrestricted SFDM fit puts significantly more data points below the tight RAR from the MOND limit than above it.
Since this mechanism affects only some galaxies and since it affects different galaxies differently, this likely also contributes to the increased scatter.

When comparing the RAR implied by our fits to the actually observed RAR, one should keep in mind that,
    as mentioned above, observations always add scatter on top of what an underlying theory predicts.
So, for example, if our unrestricted SFDM best fits are the ground truth, the observed RAR should contain even more scatter than what Fig.~\ref{fig:RAR-SFDM}, left, shows.
At least if one ignores that some of the scatter in Fig.~\ref{fig:RAR-SFDM}, left, already reflects the scatter in the observed rotation curve data due to overfitting.
The result may be even more scatter than what one obtains in actual observations with the simple prescription $(M/L_*)_{\mathrm{disk}} = 0.5$ and $(M/L_*)_{\mathrm{bulge}} = 0.7$ \citep{Lelli2017b}.
Since SFDM aims to explain this observational fact, one might reject our unrestricted fit results purely on theses grounds.

Still, the objective of our fits was to match the observed rotation curve data,
    not to get a tight RAR.
So our fit results do not directly imply that a non-tight RAR is an intrinsic property of going outside the MOND limit of SFDM.
It may still be possible to get a tight RAR even outside the MOND limit.
However, as discussed in Sect.~\ref{sec:models}, for this one would have to carefully adjust the boundary condition $\varepsilon$ for each galaxy.
Then, one must rely on galaxy formation to always pick the right values.
This is contrary to the aim of SFDM to explain scaling relations such as the RAR without having to resort to the details of galaxy formation.

\subsection{Tension with strong lensing in SFDM}
\label{sec:M200estar04}

The proper MOND limit $|\varepsilon_*| \ll 1$ of SFDM is useful for fitting rotation curves due to its MOND-like phonon force.
Above, we saw that most SPARC rotation curves can be reasonably fit with the proper MOND limit $|\varepsilon_*| \ll 1$.
At least if we count values of $|\varepsilon_*|$ as large as $0.4$ as still satisfying $|\varepsilon_*| \ll 1$.
But for certain other observables, such as strong lensing, the phonon force plays no role.
This is because the observation of GW170817 requires that the phonon force does not affect photons \citep{Hossenfelder2019, Sanders2018, Boran2018}.
Thus, the strong lensing signal is produced only by the standard gravitational pull of the mass of the baryons and the superfluid,
not by the phonon force.

As a consequence, the MOND limit of SFDM has a serious problem,
because the superfluid's energy density is too small to produce significant lensing.
Specifically, assuming the whole superfluid core to be in the MOND limit $|\varepsilon_*| \ll 1$,
a rough upper bound is (see Appendix~\ref{sec:appendix:M200})
\begin{align}
\frac{M^{\mathrm{DM}}_{200}}{M_b}
 < \frac1{\sqrt{2\pi}} \frac94 \left(1 - \frac{\beta}{3}\right)^3 \left(\frac{m^2}{\alpha}\right)^3 \frac{\sqrt{a_0^3 M_b}}{\rho_{200}^2}
 = 0.9 \cdot \left(\frac{M_b}{10^{10}\,M_\odot}\right)^{1/2} \,,
\end{align}
where we assumed the numerical parameter values from \cite{Berezhiani2018}.
Producing sufficient strong lensing and a superfluid core in the MOND limit seem to be mutually exclusive.
Choosing different parameter values may help, especially a larger $m^2/\alpha$.
But this would imply that fewer galaxies can reach the MOND limit $|\varepsilon_*| \ll 1$,
as discussed in Appendix~\ref{sec:sfdm:params}.

This conclusion can, in principle, be avoided if the superfluid core is in the MOND limit only at smaller radii
(where rotation curves are measured),
but not at larger radii (where part of the lensing signal comes from).
But even in this case there are limits on how large $M^{\mathrm{DM}}_{200}/M_b$ can be.
This is because, given that we have $|\varepsilon_*| \ll 1$ at relatively small radii,
    the superfluid's energy density cannot be arbitrarily large at larger radii.
Here, we check whether or not the SPARC galaxies can possibly have a sufficiently large $M_{200}/M_b$ for strong lensing,
if we assume the proper MOND limit for the rotation curves.

Specifically, for each galaxy, we will find the largest possible value $M_{200,\mathrm{max}}^{\mathrm{DM}}$ of $M_{200}^{\mathrm{DM}}$ that is compatible with a rotation curve in the proper MOND limit.
For the requirement that the rotation curve is in the proper MOND limit we impose $|\varepsilon|  < 0.4$ as above.

\begin{figure}
 \centering
 \includegraphics[width=.48\textwidth]{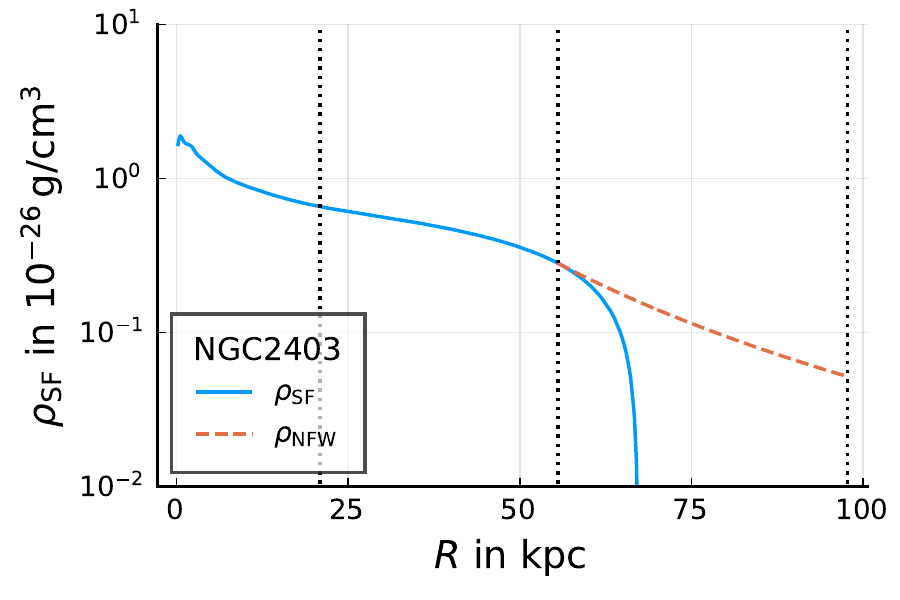}
 \caption{
     Dark matter energy density for NGC2403 giving the largest possible $M_{200}^{\mathrm{DM}}$ compatible with a rotation curve in the MOND limit ($|\varepsilon| < 0.4$).
     The energy density is that of $\rho_{\mathrm{SF}}$ (solid blue line) for $r < r_{\mathrm{NFW}}$ and that of an NFW $1/r^3$ tail at larger radii (dashed red line).
     The two contributions are matched to each other at a radius $r_{\mathrm{NFW}}$, which is chosen to maximize $M_{200}^{\mathrm{DM}}$.
     This is for $(M/L_*)_{\mathrm{disk}} = 0.5$ and $(M/L_*)_{\mathrm{bulge}} = 0.7$ and gives $M_{200}^{\mathrm{DM}} = 9.8\cdot 10^{10}\,M_\odot$.
     The dotted vertical lines denote $R_{\mathrm{max}}$, $r_{\mathrm{NFW}}$, and $r_{200}$.
    }
 \label{fig:rhoSFforM200max}
\end{figure}

In SFDM one usually assumes that the superfluid ends at a radius $r_{\mathrm{NFW}}$ beyond which the energy density is that of an NFW halo \citep{Berezhiani2015, Berezhiani2018, Hossenfelder2020}.
For simplicity, we assumed the NFW profile $\rho_{\mathrm{NFW}}(r)$ to be proportional to $1/r^3$ at the radii of interest, that is to say, we continued the superfluid density with an NFW tail instead of a full NFW profile.
We do not expect this to significantly affect our results \citep{Hossenfelder2019}.
The usual procedure for matching the superfluid density to the NFW profile is heuristic and not derived from first principles \citep{Berezhiani2018, Mistele2020}.
To be independent of the details of this matching procedure we matched the NFW density $\rho_{\mathrm{NFW}}$ to the superfluid density $\rho_{\mathrm{SF}}$ at a radius $r_{\mathrm{NFW}}$ that is chosen to maximize $M_{200}^{\mathrm{DM}}$.
This gives the most conservative upper bound for $M_{200}^{\mathrm{DM}}$.
We restricted $r_{\mathrm{NFW}}$ only in two ways.
First, we assumed a positive superfluid energy density within the superfluid core, which implies that $r_{\mathrm{NFW}}$ is smaller than the radius $r_m$ where $\rho_{\mathrm{SF}}$ vanishes.
Second, we assumed all rotation curve data points to be within the superfluid core.
Thus, we also restricted $r_{\mathrm{NFW}}$ to be larger than $R_{\mathrm{max}}$.
This is illustrated in Fig.~\ref{fig:rhoSFforM200max}.

To get the largest possible $M_{200}^{\mathrm{DM}}$ compatible with our constraints, we scanned values of $\varepsilon$ in $(\varepsilon_{*\mathrm{min}}, 0.4)$ and values of $r_{\mathrm{NFW}}$ in $(R_{\mathrm{max}}, r_m)$ and recorded the largest $M_{200}^{\mathrm{DM}}$ as an upper bound $M_{200,\mathrm{max}}^{\mathrm{DM}}$.
For a given $\varepsilon$ and $r_{\mathrm{NFW}}$ we calculated $M_{200}^{\mathrm{DM}}$ by solving the equation
\begin{align}
 \frac{4\pi}{3} \rho_{200} r_{200}^3 = M_{\mathrm{SF}}(r_{\mathrm{NFW}}) + 4 \pi r_{\mathrm{NFW}}^3 \, \rho_{\mathrm{SF}}(r_{\mathrm{NFW}}) \ln\left(\frac{r_{200}}{r_{\mathrm{NFW}}}\right)
\end{align}
for $r_{200}$ and then plugging the result into $M_{200}^{\mathrm{DM}} = (4\pi/3) \rho_{200} r_{200}^3$.
We calculated $M_{\mathrm{SF}}(r)$ as $-(\hat{\mu}_{\mathrm{SF}}'(r)/m) (r^2/G)$.
For some galaxies, it is not possible to solve the $\hat{\mu}_{\mathrm{SF}}$ equation with $|\varepsilon| < 0.4$,
    if we require a positive $\rho_{\mathrm{SF}}$ up to the last rotation curve data point.
Such galaxies are excluded in the results shown below.

For the baryonic mass distribution we used the same values $(M/L_*)_{\mathrm{disk}} = 0.5$ and $(M/L_*)_{\mathrm{bulge}} = 0.7$ for all galaxies.
These may not give the best fit to the measured rotation curves for all galaxies.
However, here we are not interested in the fit to the rotation curve data,
    but in whether or not it is possible to have, at the same time, both a rotation curve in the proper MOND limit and sufficient $M_{200}^{\mathrm{DM}}$ for strong lensing.
The precise value of $M/L_*$ is irrelevant for this.

We can simplify the scanning procedure a bit.
Namely, Fig.~\ref{fig:smallfMDM-NGC2403} and Fig.~\ref{fig:smallfMDM-DDO168} suggest that increasing the boundary condition $\varepsilon \equiv \varepsilon_*(R_{\mathrm{mid}})$ increases $\varepsilon_*(r)$ at all radii, not just at $R_{\mathrm{mid}}$.
We verified this numerically for various galaxies and boundary conditions.
Here, we assumed that this is true in general.\footnote{%
    Indeed, a violation of this would imply that the boundary value problem is nonunique, contrary to what we already implicitly assumed in our calculations above, where we assumed that a value $\varepsilon_*(R_{\mathrm{mid}})$ uniquely specifies a solution.
    The reason is the following.
    If a larger boundary condition $\varepsilon_*(R_{\mathrm{mid}})$ gives a smaller $\varepsilon_*(r)$ at some radius $r = r_l$,
        then by continuity there must be a radius $r_x$ between $R_{\mathrm{mid}}$ and $r_l$ where the solutions $\varepsilon_*(r)$ for two different boundary conditions have the same value.
    Thus, the boundary value problem with boundary conditions imposed at $r = r_x$ is nonunique.
}
Consequences of this are that a larger boundary condition $\varepsilon$ implies an everywhere larger superfluid energy density and a larger radius $r_m$ where this energy density reaches zero.
Therefore, for a fixed $r_{\mathrm{NFW}}$, the quantity $M_{200}^{\mathrm{DM}}$ is a monotonically increasing function of $\varepsilon$.
Thus, we can simplify our scanning procedure by always setting $\varepsilon = 0.4$ and scanning only values of $r_{\mathrm{NFW}}$ in $(R_{\mathrm{max}}, r_m)$.
For this, we used Mathematica's NMaximize function with its default options.

It can happen that the NFW radius $r_{\mathrm{NFW}}$ is bigger than the radius $r_\infty$ up to which we numerically solved for $\hat{\mu}_b$ (usually $100\,\mathrm{kpc}$; see Appendix~\ref{sec:method:data}).
In this case, we must continue $\hat{\mu}_b$ beyond $r_\infty$ since solving the $\hat{\mu}_{\mathrm{SF}}$ equation requires $\hat{\mu}_b$ and $a_b$ up to $r_{\mathrm{NFW}}$.
For simplicity, we continued $\hat{\mu}_b$ beyond $r_\infty$ assuming spherical symmetry and zero baryonic energy density at $r > r_\infty$,
\begin{align}
 \frac{\hat{\mu}_b(r > r_\infty)}{m} = \frac{\hat{\mu}(r_\infty)}{m} - a_{b,R}(r_\infty) r_\infty^2 \left(\frac1{r_\infty} - \frac1{r}\right) \,.
\end{align}
This implies $a_b(r > r_\infty) = a_{b,R}(r_\infty) r_\infty^2/r^2$.
We expect errors due to this to not significantly change our results.

Below we need the total baryonic mass $M_b$ of each galaxy.
We adopt
\begin{align}
 M_b \equiv 0.5 \cdot \left(L_{[3.6]} - L_{\mathrm{bulge}}\right) + 0.7 \cdot L_{\mathrm{bulge}} + 1.4 \cdot M_{\mathrm{HI}} \,,
\end{align}
where $L_{[3.6]}$, $L_{\mathrm{bulge}}$, and $M_{\mathrm{HI}}$ are taken directly from SPARC.
They denote the total [3.6] luminosity, the total bulge luminosity, and the HI mass, respectively.
The stellar luminosities are weighted by their respective mass-to-light ratio.
The factor $1.4$ in front of the HI mass is to take into account helium and molecular hydrogen \citep{McGaugh2020}.

We show the results in Fig.~\ref{fig:M200max}.
We see that large ratios $M_{200}^{\mathrm{DM}}/M_b$ are easier to reach for galaxies with relatively small baryonic masses $M_b$.
In part, this is due to the factor $1/M_b$ in $M_{200}^{\mathrm{DM}}/M_b$.
For strong lensing, relatively large baryonic masses are relevant.
To illustrate this, we also show the best-fit values from the SFDM strong lensing analysis from \cite{Hossenfelder2019} in Fig.~\ref{fig:M200max}.
These lensing galaxies tend to have $M_b \gtrsim 10^{11}\,M_\odot$ and $M_{200}^{\mathrm{DM}}/M_b \gtrsim 1000$.
In contrast, the SPARC galaxies with $M_b > 10^{11}\,M_\odot$ almost all have $M_{200}^{\mathrm{DM}}/M_b < 10$ when restricted to have rotation curves in the MOND limit (i.e., when restricted to $|\varepsilon| < 0.4$).
This is a stark contrast.
The SPARC galaxies do not reach baryonic masses quite as large as the lensing galaxies from \cite{Hossenfelder2019}.
But from Fig.~\ref{fig:M200max}, it seems clear that the trend goes into the wrong direction:
The larger the galaxy, the smaller the ratio $M_{200,\mathrm{max}}^{\mathrm{DM}}/M_b$ (assuming $|\varepsilon| < 0.4$).

Thus, it seems that a rotation curve in the MOND limit and sufficient dark matter for strong lensing are indeed mutually exclusive in standard SFDM.
A caveat is that the galaxies in the SPARC sample are not ellipticals, in contrast to the lensing sample used in \cite{Hossenfelder2019}.
One might think that, for a given $M_b$, the maximum possible $M_{200}^{\mathrm{DM}}$ is not sensitive to the details of the baryonic mass distribution since the main contributions to $M_{200}^{\mathrm{DM}}$ come from large radii where, to a first approximation, only the total $M_b$ plays a role.
However, we impose the condition $|\varepsilon| < 0.4$ at relatively small radii $R = R_{\mathrm{mid}}$ where the details of the baryonic mass distribution may still matter.
Indeed, the sensitivity of $M_{200,\mathrm{max}}^{\mathrm{DM}}$ to these details is reflected in the scatter in Fig.~\ref{fig:M200max} (see also the end of Appendix~\ref{sec:appendix:twofieldlensing}).
Still, even being generous with this scatter, it seems unlikely that the difference in galaxy type explains why the SPARC galaxies restricted to $|\varepsilon| < 0.4$ cannot reach larger values of $M_{200}^{\mathrm{DM}}/M_b$.
Also, Fig.~\ref{fig:M200max} suggests that for $M_{200}^{\mathrm{DM}}/M_b$ values closer to those required for strong lensing (when restricting to $|\varepsilon| < 5$ instead of $|\varepsilon| < 0.4$) there is less scatter (i.e., less sensitivity to the baryonic mass distribution beyond the total $M_b$).

This suggests that strong lensing galaxies cannot have their inner parts (where rotation curves or velocity dispersions are measured) be in the proper MOND limit $|\varepsilon_*| \ll 1$.
This is not in direct contradiction with measurements.
Indeed, \cite{Hossenfelder2019} successfully fitted strong lensing data in SFDM.
But one cannot easily keep the key idea of SFDM that the inner parts of galaxies are always in the proper MOND limit.
Either one has to give up this key idea or one has to postulate that it does not apply to strong lensing galaxies for some reason.

\subsection{$M/L_*$ in two-field SFDM}
\label{sec:twofield:ML}

\begin{figure}
 \centering
 \includegraphics[width=.48\textwidth]{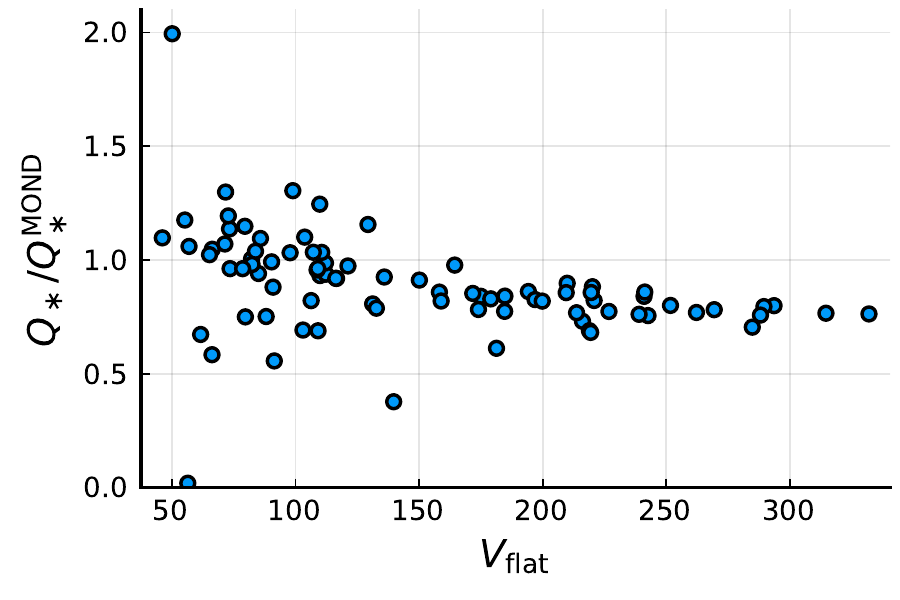}
  \caption{
      Same as Fig.~\ref{fig:this} but for two-field SFDM.
    }
 \label{fig:this-twofield}
\end{figure}

By construction, the phonon acceleration in two-field SFDM should almost always be close to $\sqrt{a_0 a_b}$ \citep{Mistele2020}.
The superfluid's Newtonian gravitational pull can be comparable to that of standard SFDM \citep{Mistele2020}.
Thus, we expect the fit results for two-field SFDM to be close to that of the SFDM $a_\theta = \sqrt{a_0 a_b}$ model discussed above.

This should be true at least for the stellar mass-to-light ratios and the best-fit $\chi^2$.
The results for $f_{M_{\mathrm{DM}}}$ may be not as close since the superfluid halo's shape is different.
We discuss the superfluid halo in more detail below.

As expected, the best fits for two-field SFDM are almost all in the $|\varepsilon_*| \ll 1$ limit so that their phonon force $a_\theta$ is close to $\sqrt{a_0 a_b}$.
This is shown in Fig.~\ref{fig:estarhist}.
Only for two galaxies (NGC6789, UGC07232) does $\varepsilon_*$ become larger than $0.1$.
Its largest values is $0.36$ for NGC6789.

\begin{figure*}
  \centering
  \includegraphics[width=\textwidth]{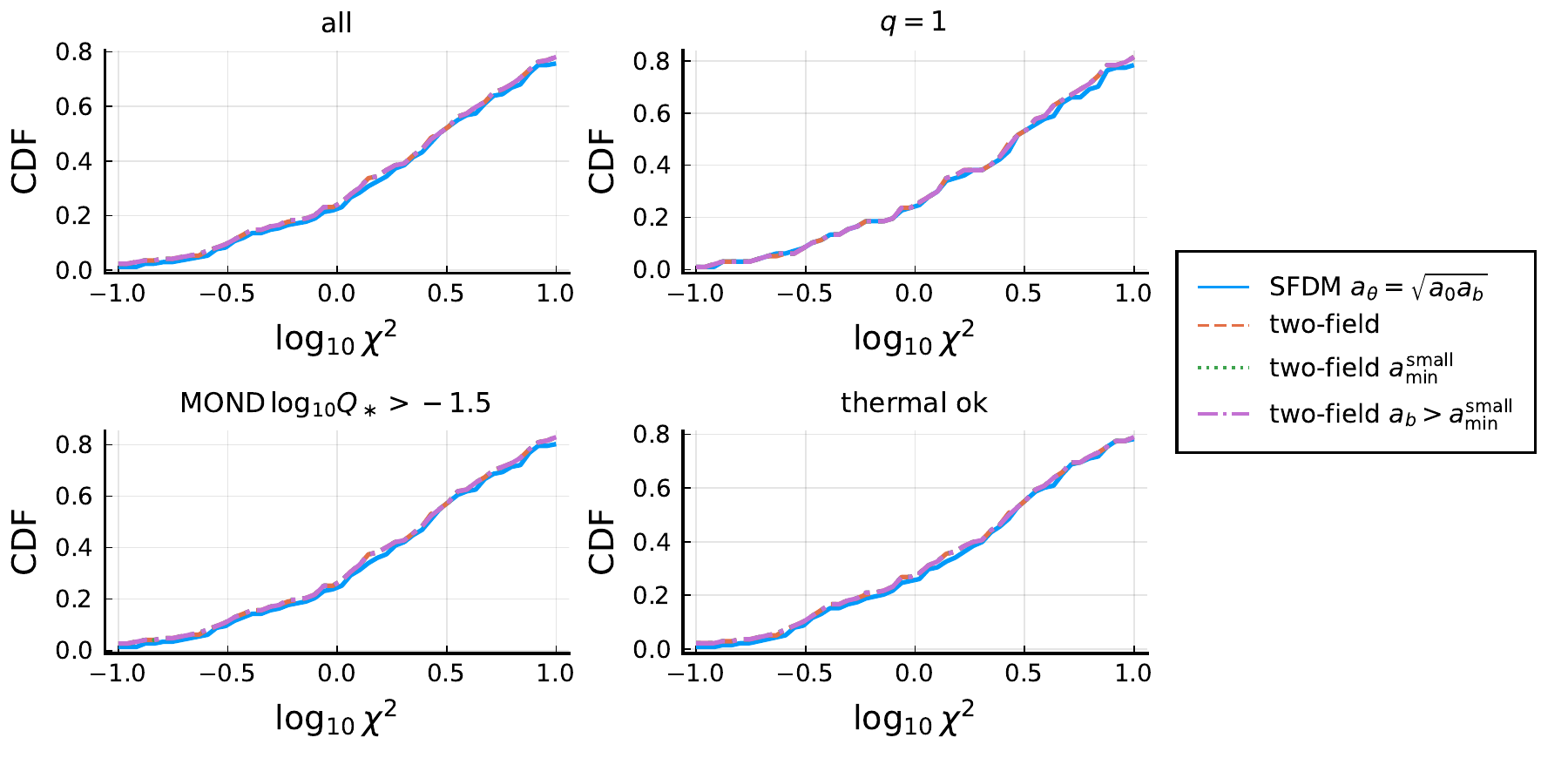}
  \caption{
      $\chi^2$ CDFs for the different two-field models and galaxy cuts.
     Also shown is the SFDM $a_\theta = \sqrt{a_0 a_b}$ fit for comparison.
  }
  \label{fig:chi2cdf-twofield}
\end{figure*}

The averaged best-fit stellar mass-to-light ratios and the best-fit $\chi^2$ can be found in Table~\ref{tab:fYmedian}, Table~\ref{tab:fYmean}, and Fig.~\ref{fig:chi2cdf-twofield}.
As expected,
    these are almost identical to those of the SFDM $a_\theta = \sqrt{a_0 a_b}$ model discussed above.

Two-field SFDM gives a systematically smaller $M/L_*$ than MOND only for high-acceleration galaxies, not for low accelerations, just as standard SFDM (see Sect.~\ref{ssec:rar} and Appendix~\ref{sec:results:sfdm:MLtrends}, and see Fig.~\ref{fig:this-twofield} for an example).

\subsubsection{The minimum acceleration}
\label{sec:twofield:amin}

In two-field SFDM, the equilibrium on galactic scales is stable only for phonon accelerations above a certain minimum acceleration.
This minimum acceleration depends on the value of the field $\hat{\mu}/m$.
Assuming $|\varepsilon_*| \ll 1$, as is usually the case, stability requires
\begin{align}
 \label{eq:abarcond}
 a_b \left(10^7 \frac{\hat{\mu}}{m} \right)^2 > \bar{a}\,,
\end{align}
where $\bar{a}$ is one of this model's parameters.
Often, $\hat{\mu}/m$ is on the order of $10^{-7}$ on galactic scales.
Thus, roughly, stability requires
\begin{align}
 a_b \gtrsim \bar{a} \,.
\end{align}
When $\hat{\mu}/m$ is smaller, the instability sets in earlier.
\cite{Mistele2020} chose $\bar{a} = 10^{-12}\,\mathrm{m}/\mathrm{s}^2$ so that this model does not predict standard MOND-like behavior for dwarf spheroidals that may start to deviate from MOND-like behavior around $a_b \sim 10^{-12}\,\mathrm{m}/\mathrm{s}^2$ \citep{Lelli2016}.
In our two-field SFDM fit we adopted this value of $\bar{a}$.

For galaxies violating Eq.~\eqref{eq:abarcond}, we should in principle model what happens beyond the minimum acceleration in two-field SFDM.
Here, we did not do this for two reasons.
First, this regime is not well-understood.
Second, we are interested in the MOND regime inside the superfluid core.
Modeling the behavior beyond the minimum acceleration will not help us understand whether or not the MOND regime of two-field SFDM requires a larger or smaller $M/L_*$ than standard MOND models.

Still, 98 of 169 SPARC galaxies violate the condition Eq.~\eqref{eq:abarcond} at $R_{\mathrm{max}}$.
For these galaxies, our fit is not meaningful since it relies on an unstable equilibrium.
This is expected for the dwarf spheroidals with $a_b$ around $10^{-12}\,\mathrm{m}/\mathrm{s}^2$.
But the 98 galaxies violating Eq.~\eqref{eq:abarcond} include many more galaxies, also at $a_b \gg \bar{a}$.
We will have to deal with this one way or another.

To further explore this, we redid the two-field SFDM fit with the much smaller value $\bar{a} = 10^{-14}\,\mathrm{m}/\mathrm{s}^2$.
This is listed as ``two-field $a_{\mathrm{min}}^{\mathrm{small}}$'' in our tables and figures.
The resulting $M/L_*$ and $\chi^2$ values are almost identical to those of the previous two-field SFDM fit.
But still 71 galaxies violate Eq.~\eqref{eq:abarcond}.

Thus, Eq.~\eqref{eq:abarcond} is often not violated because $a_b$ is smaller than $\bar{a}$ but because $\hat{\mu}/m$ is smaller than $10^{-7}$.
A small $\hat{\mu}/m$ corresponds to a small superfluid mass $M_{\mathrm{DM}}$.
Indeed, Fig.~\ref{fig:fMDMhist} shows that many galaxies have a smaller $f_{M_{\mathrm{DM}}}$ in two-field SFDM compared to standard SFDM, for both $\bar{a}$ values discussed above.

This raises two questions.
The first is whether two-field SFDM really needs to go to small $\hat{\mu}/m$ to fit the SPARC data,
    thus often violating Eq.~\eqref{eq:abarcond}.\ The second is why standard SFDM tends to end up at larger $f_{M_{\mathrm{DM}}}$ values than two-field SFDM.
\subsubsection{Origin of the small $\fMDM$ values in two-field SFDM}
\label{sec:twofield:minimumfMDM}

The best-fit $\chi^2$ and $M/L_*$ are almost identical for the two-field SFDM fits and for the standard SFDM fit with $a_\theta = \sqrt{a_0 a_b}$.
But the superfluid halos of two-field SFDM reach much smaller masses than those in standard SFDM (see Fig.~\ref{fig:fMDMhist}).
For example, there are no galaxies at $\fMDM < -2.3$ in any standard SFDM fit, but many such galaxies in two-field SFDM.
The only relevant difference between two-field SFDM and the SFDM $a_\theta = \sqrt{a_0 a_b}$ model is the difference in their $\rho_{\mathrm{SF}}$.
Thus, this different superfluid energy density must be the reason for the qualitative difference in $\fMDM$.
Here, we explain this in more detail.

As discussed above, we enforced a positive superfluid energy density at radii smaller than the last rotation curve data point (i.e., $\rho_{\mathrm{SF}} > 0$ at $R \leq R_{\mathrm{max}}$).
The difference between standard and two-field SFDM regarding small dark matter masses boils down to what this condition implies.

Consider first two-field SFDM.
In two-field SFDM, the superfluid energy density vanishes when $\hat{\mu}/m = 0$.
Equivalently, when $\varepsilon_* = 0$.
Typically, $\hat{\mu}/m$ is a decreasing function of galactocentric radius.\footnote{
    This is always the case in spherical symmetry as long as $\rho_{\mathrm{SF}}$ is positive.
    Indeed, then $\hat{\mu}'(r)/m = - G (M_b + M_{\mathrm{SF}})/r^2 < 0$.
    Here, we assumed spherical symmetry only for $\hat{\mu}_{\mathrm{SF}}$, but not for $\hat{\mu}_b$.
    Thus, only $\hat{\mu}_{\mathrm{SF}}$ is guaranteed to be a decreasing function of galactocentric radius, not the total $\hat{\mu}$.
    Still, the total $\hat{\mu}(R,z=0)$ typically decreases as a function of $R$ also in our case.
}
Thus, whenever the condition $\rho_{\mathrm{SF}} > 0$ is fulfilled at the largest radius of interest, $R = R_{\mathrm{max}}$, it is fulfilled at all radii of interest (i.e., at $R \leq R_{\mathrm{max}}$).
In a given galaxy, a smaller superfluid mass $M_{\mathrm{SF}}(R)$ at some radius, for example at $R = R_{\mathrm{max}}$, implies a smaller $\hat{\mu}/m$ everywhere in the superfluid core (not just at $R = R_{\mathrm{max}}$).
Indeed, different superfluid masses just correspond to adding a term $\sin(r/r_0)/r$ with a different prefactor to $\hat{\mu}_{\mathrm{SF}}$; see Appendix~\ref{sec:twofieldcalc}.
Thus, if we go to smaller and smaller $\fMDM$, we go to smaller and smaller $\hat{\mu}/m$.
At some point, $\hat{\mu}/m$ will become negative at $R = R_{\mathrm{max}}$.
Then, we are at the minimum possible $\fMDM$ allowed by our condition $\rho_{\mathrm{SF}} > 0$.

\begin{figure}
 \centering
 \includegraphics[width=.48\textwidth]{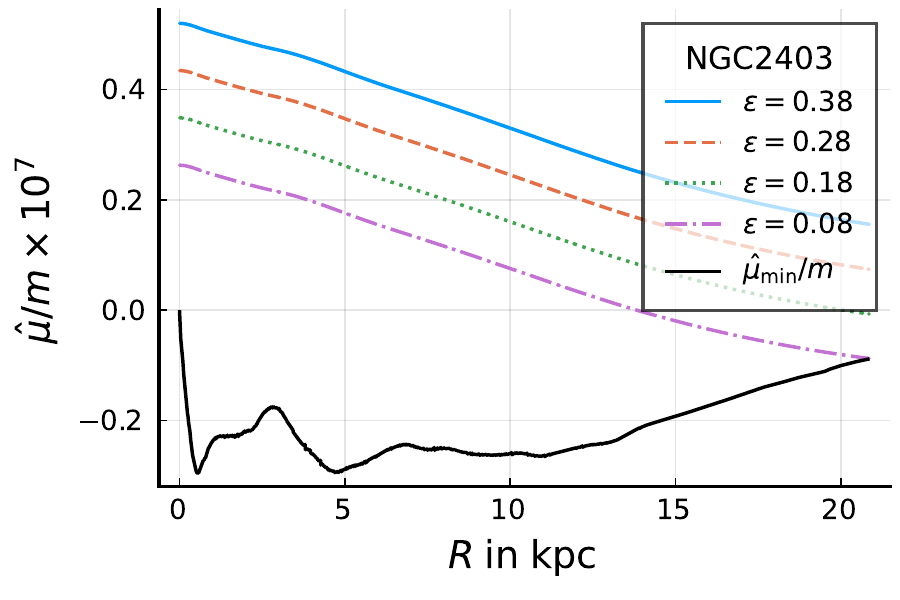}
 \includegraphics[width=.48\textwidth]{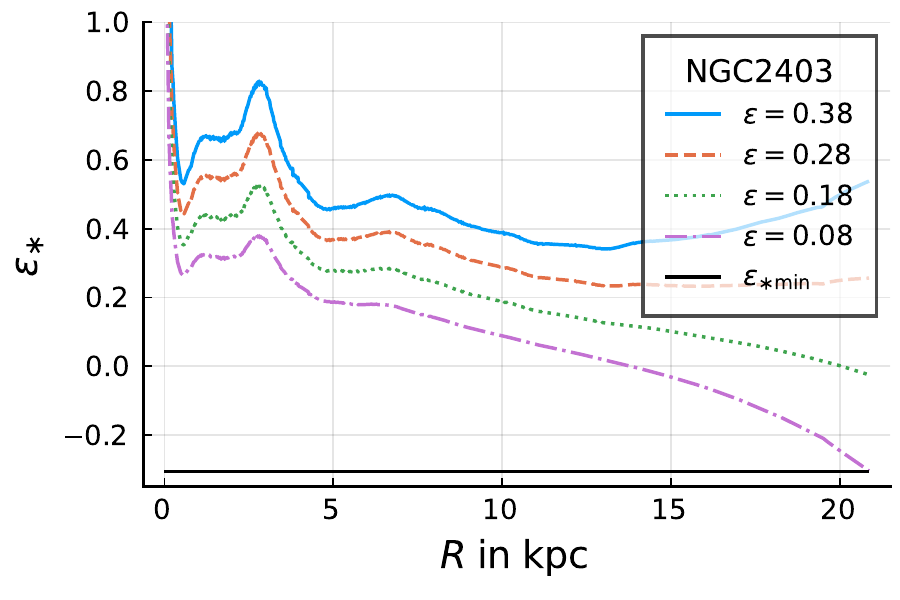}
 \caption{
     Solutions for $\hat{\mu}/m$ and $\varepsilon_*$ for various boundary conditions, $\varepsilon$, for the galaxy NGC2403 assuming the best-fit $Q_*$ for SFDM.
     Top: Quantity $\hat{\mu}/m$ at $z=0$.
     The solid black line shows the minimum possible value of $\hat{\mu}/m$ allowed by the condition $\rho_{\mathrm{SF}} > 0$ at each radius.
     The smallest boundary condition, $\varepsilon = 0.08$, corresponds to the minimum possible dark matter mass for the given baryonic mass distribution.
     Smaller masses would require that the condition $\rho_{\mathrm{SF}} > 0$ is violated before $R = R_{\mathrm{max}}$.
     Bottom: Same as the top panel but showing $\varepsilon_*$ instead of $\hat{\mu}/m$.
     The minimum possible $\varepsilon_*$ allowed by $\rho_{\mathrm{SF}} > 0$ is a constant.
    }
 \label{fig:smallfMDM-NGC2403}
\end{figure}

The condition $\rho_{\mathrm{SF}} > 0$ at $R = R_{\mathrm{max}}$ enforces a minimum possible mass also in standard SFDM.
A difference is that the superfluid energy density in standard SFDM vanishes not at $\varepsilon_* = 0$ but at a negative value $\varepsilon_{*\mathrm{min}} \approx -0.31$ (for $\beta = 2$).
Using the definition of $\varepsilon_*$ from Eq.~\eqref{eq:estar}, this constant negative lower bound on $\varepsilon_*(\vec{x})$ becomes a nonconstant lower bound $\hat{\mu}_{\mathrm{min}}(\vec{x})$ on $\hat{\mu}(\vec{x})$.
This is illustrated in Fig.~\ref{fig:smallfMDM-NGC2403} for NGC2403.
The top panel shows solutions $\hat{\mu}/m$ for various boundary conditions $\varepsilon$.
The bottom panel shows $\varepsilon_*$ for the same boundary conditions.
Both panels also show the lower bounds on $\varepsilon_*$ and $\hat{\mu}$, respectively, which ensure $\rho_{\mathrm{SF}} > 0$.
For the smallest boundary condition value $\varepsilon = 0.08$ shown in Fig.~\ref{fig:smallfMDM-NGC2403}, both $\varepsilon_*$ and $\hat{\mu}$ reach this lower bound at $R = R_{\mathrm{max}}$.
Thus, $\varepsilon = 0.08$ corresponds to the smallest possible superfluid mass that is allowed by the condition $\rho_{\mathrm{SF}} > 0$.
Lower superfluid masses would need $\rho_{\mathrm{SF}}$ to reach zero before $R = R_{\mathrm{max}}$, which we do not allow.
This is similar to as in two-field SFDM.

\begin{figure}
 \centering
 \includegraphics[width=.48\textwidth]{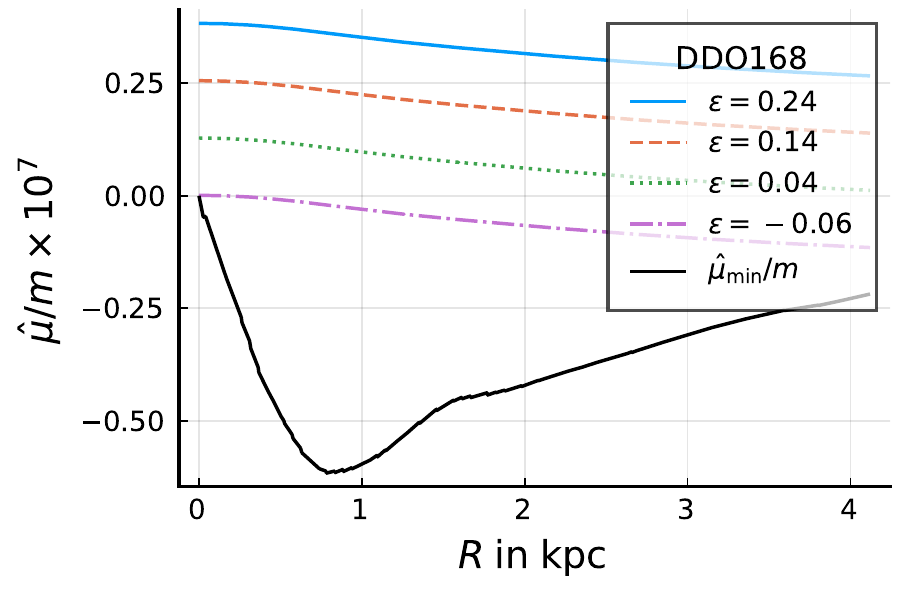}
 \includegraphics[width=.48\textwidth]{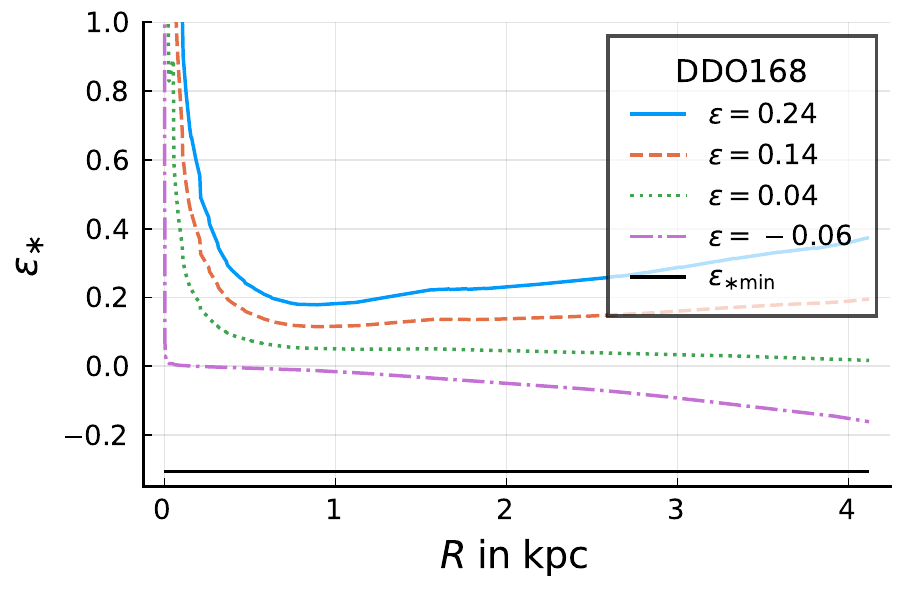}
 \caption{
     Same as Fig.~\ref{fig:smallfMDM-NGC2403} but for DDO168.
     The smallest boundary condition shown, $\varepsilon = -0.06$, again corresponds to the minimum possible mass allowed by $\rho_{\mathrm{SF}} > 0$.
     But now smaller masses would violate this condition at $R = 0$ instead of at $R = R_{\mathrm{max}}$.
    }
 \label{fig:smallfMDM-DDO168}
\end{figure}

However, in contrast to two-field SFDM, this constraint at $R = R_{\mathrm{max}}$ is not the only constraint on superfluid masses in standard SFDM.
In standard SFDM, having $\rho_{\mathrm{SF}} > 0$ at $R = R_{\mathrm{max}}$ does not imply $\rho_{\mathrm{SF}} > 0$ at smaller radii.
The main issue is at $R = 0$.
This is illustrated in Fig.~\ref{fig:smallfMDM-DDO168} for DDO168.
For the smallest boundary condition value $\varepsilon = -0.06$ shown in Fig.~\ref{fig:smallfMDM-DDO168}, at $R = R_{\mathrm{max}}$,
    both $\hat{\mu}/m$ and $\varepsilon_*$ are not close to the minimum values that $\rho_{\mathrm{SF}} > 0$ allows.
However, at $R = 0$, the minimum values are reached (i.e., $\rho_{\mathrm{SF}}$ vanishes).
Thus, the boundary condition $\varepsilon = -0.06$ corresponds to the minimum possible $\fMDM$ allowed by our condition $\rho_{\mathrm{SF}} > 0$.
But now the constraint comes from $R = 0$ rather than $R = R_{\mathrm{max}}$.

To understand this, consider small radii, $R \to 0$.
At $R \to 0$, the field $\hat{\mu}$ reaches a finite value $\hat{\mu}(0)$.
The Newtonian baryonic acceleration $|\vec{a}_b(R)|$ goes to zero, typically $a_b \propto R$ for $R \to 0$.
Thus, we have from the definition of $\varepsilon_*$ Eq.~\eqref{eq:estar}
\begin{align}
 \varepsilon_*(R\to 0) \propto \frac{\hat{\mu}(0)}{R} \to \pm \infty \,.
\end{align}
That is, $\varepsilon_*$ tends to $\pm\infty$ with the sign being that of $\hat{\mu}(0)$.
A positive $\rho_{\mathrm{SF}}$ requires $\varepsilon_*(0) \geq \varepsilon_{*\mathrm{min}}$.
Thus, we must have
\begin{align}
 \hat{\mu}(0) \geq 0 \,.
\end{align}
This condition $\hat{\mu}(0) > 0$ must also hold in two-field SFDM.
But, in two-field SFDM, $\hat{\mu}(0) > 0$ follows from $\rho_{\mathrm{SF}} > 0$ at $R = R_{\mathrm{max}}$.
This is because $\hat{\mu}$ is a decreasing function of radius and because $\rho_{\mathrm{SF}} > 0$ requires $\hat{\mu} > 0$ even at $R = R_{\mathrm{max}}$.
So this adds nothing in two-field SFDM.

This is different in standard SFDM.
A positive $\rho_{\mathrm{SF}}$ at $R = R_{\mathrm{max}}$ does not necessarily imply a positive $\hat{\mu}(0)$ and thus a positive $\rho_{\mathrm{SF}}$ at $R = 0$.
Since $\hat{\mu}$ is a decreasing function of radius,
    violating $\hat{\mu}(0) > 0$ is possible only if $\hat{\mu}$ is negative already at $R = R_{\mathrm{max}}$.
Consider this case where $\hat{\mu}$ is negative at $R = R_{\mathrm{max}}$.
Since $\hat{\mu}$ decreases with radius,
    it can happen that $\hat{\mu}$ grows sufficiently between $R = R_{\mathrm{max}}$ and $R=0$ to become positive at $R=0$.
In this case the condition $\hat{\mu}(0) > 0$, corresponding to $\rho_{\mathrm{SF}} > 0$ at $R = 0$, gives no additional constraint.
However, this is not guaranteed to happen.
When $\hat{\mu}$ does not grow sufficiently, the condition $\hat{\mu}(0) > 0$ gives an additional constraint.
This is what happens for DDO168 as illustrated in Fig.~\ref{fig:smallfMDM-DDO168}.

This, then, is the reason why two-field SFDM allows smaller superfluid masses (i.e. smaller $\fMDM$) compared to standard SFDM.
In both models, small superfluid masses correspond to $\rho_{\mathrm{SF}}$ close to zero.
And in both cases one needs to be careful not to let this density become negative (or ill-defined) at $R = R_{\mathrm{max}}$.
However, in standard SFDM, being close to $\rho_{\mathrm{SF}} = 0$ implies a negative $\hat{\mu}$,
    and in this case there can be an additional constraint at $R=0$, as just discussed.
This second constraint is absent in two-field SFDM where $\rho_{\mathrm{SF}}$ vanishes already at $\hat{\mu} = 0$.

As mentioned above, this second constraint in standard SFDM occurs only when $\hat{\mu}$ does not grow sufficiently between $R = R_{\mathrm{max}}$ and $R = 0$.
We now make this more precise.
The minimum allowed superfluid mass from the constraint at $R = R_{\mathrm{max}}$
    corresponds to $\rho_{\mathrm{SF}} = 0$ at $R = R_{\mathrm{max}}$.
This constraint is present in both standard and two-field SFDM.
Reaching this minimum mass implies
\begin{align}
 \hat{\mu}(R_{\mathrm{max}}) = \varepsilon_{*\mathrm{min}} \frac{\alpha M_{\mathrm{Pl}} |\vec{a}_b(R_{\mathrm{max}})|}{2m} \,.
\end{align}
A second constraint $\hat{\mu}(0)> 0$ from $R = 0$ is avoided whenever $\hat{\mu}$ grows between $R=R_{\mathrm{max}}$ and $R=0$ by at least
\begin{align}
 \hat{\mu}(0) - \hat{\mu}(R_{\mathrm{max}}) > -\varepsilon_{*\mathrm{min}} \frac{\alpha M_{\mathrm{Pl}} |\vec{a}_b(R_{\mathrm{max}})|}{2m}  \,.
\end{align}
As the example of DDO168 (see Fig.~\ref{fig:smallfMDM-DDO168}) shows, this is not always guaranteed in standard SFDM.
But in some cases it is.
Namely, as in Appendix~\ref{sec:sfdm:calc}, we can write $\hat{\mu} = \hat{\mu}_b + \hat{\mu}_{\mathrm{SF}}$.
Both $\hat{\mu}_b$ and $\hat{\mu}_{\mathrm{SF}}$ typically decrease with radius, with the amount they decrease being determined by the baryonic and superfluid mass, respectively.
Once we know the baryonic mass distribution, we know a lower bound on how much the total $\hat{\mu}$ grows, independently of the superfluid energy density and the boundary condition $\varepsilon$.
As a result, if the baryonic mass alone makes $\hat{\mu}$ grow sufficiently, we never get an additional constraint from $\hat{\mu}(0) > 0$.
This is the case if
\begin{align}
 \label{eq:mubavoidsecondconstraint}
 \hat{\mu}_b(0) - \hat{\mu}_b(R_{\mathrm{max}}) > -\varepsilon_{*\mathrm{min}} \frac{\alpha M_{\mathrm{Pl}} |\vec{a}_b(R_{\mathrm{max}})|}{2m} \,.
\end{align}

Thus, more quantitatively, we claim that standard SFDM cannot get masses as low as two-field SFDM because Eq.~\eqref{eq:mubavoidsecondconstraint} is often not satisfied.
That is, there is an additional constraint from $R = 0$ in standard SFDM that is not present in two-field SFDM.
We expect that, without this constraint, galaxies would end up at smaller $\fMDM$ also in standard SFDM.

As a consequence, we expect that the galaxies that end up at very small $\fMDM$ in two-field SFDM but not in standard SFDM violate Eq.~\eqref{eq:mubavoidsecondconstraint}.
To check this, we selected the 30 galaxies that have $\fMDM < -2.3$ in two-field SFDM.
As mentioned above, no galaxies have such small $\fMDM$ in standard SFDM.
For the most direct comparison against two-field SFDM we use the best-fit stellar $M/L_*$ from the SFDM $a_\theta = \sqrt{a_0 a_b}$ fits.
We find that these 30 galaxies all violate Eq.~\eqref{eq:mubavoidsecondconstraint} so that they face an additional constraint in standard SFDM.
This confirms our explanation why galaxies reach smaller $\fMDM$ values in two-field SFDM compared to standard SFDM.

\subsubsection{Enforcing the minimum acceleration in two-field SFDM}

In Appendix~\ref{sec:twofield:amin}, we saw that many best fits in two-field SFDM violate the minimum acceleration condition Eq.~\eqref{eq:abarcond}, even with a reduced $\bar{a}$ value.
The reason is that many galaxies end up at small $\hat{\mu}/m$, or, equivalently, at small superfluid masses corresponding to small $\fMDM$ (see Appendix~\ref{sec:twofield:minimumfMDM}).

For small superfluid masses, the precise mass is often not important for fitting rotation curves since the corresponding $a_{\mathrm{SF}}$ is subdominant.
Thus, it may be possible that we can find fits with sufficiently large $\hat{\mu}/m$ so that all galaxies satisfy the minimum acceleration condition without getting significantly worse fits.
To check this, we redid the two-field SFDM fit with $\bar{a} = 10^{-14}\,\mathrm{m}/\mathrm{s}^2$ but with the minimum acceleration condition Eq.~\eqref{eq:abarcond} enforced.
That is, whenever this condition is violated we set $\chi^2 = 10^{10}$ so that our fit code goes elsewhere.
We label this model ``two-field $a_b > a_{\mathrm{min}}^{\mathrm{small}}$.''

As we can see from Table~\ref{tab:fYmedian}, Table~\ref{tab:fYmean}, and Fig.~\ref{fig:chi2cdf-twofield},
    this gives almost identical results for the best-fit $\chi^2$ and $M/L_*$ values as the previous two-field SFDM fits.
This shows that it is not necessary for two-field SFDM  to violate the minimum acceleration condition.
The small $\hat{\mu}/m$ values are not required to get a reasonable fit (see also Fig.~\ref{fig:fMDMhist}).

\subsection{Tension with strong lensing in two-field SFDM}
\label{sec:appendix:twofieldlensing}

In two-field SFDM, the $|\varepsilon_*| \ll 1$ condition is almost always fulfilled so that the phonon force is almost always close to the MOND-like value $\sqrt{a_0 a_b}$.
So, in contrast to standard SFDM, simultaneously being in the $|\varepsilon_*| \ll 1$ limit and producing a sufficient strong lensing signal is not a problem.
However, in two-field SFDM, a small $\varepsilon_*$ does not imply that $a_{\mathrm{SF}}$ is negligible.
So two-field SFDM may still not be able to produce MOND-like rotation curves and sufficient strong lensing at the same time.

To check this, we calculated a maximum total dark matter mass $M_{200}^{\mathrm{DM}}$ in the same way as we did for standard SFDM in Appendix~\ref{sec:M200estar04}.
But instead of imposing $|\varepsilon| < 0.4$ we imposed
\begin{align}
 \label{eq:twofieldlensingaSFcondition}
 a_{\mathrm{SF}} < 0.3 \left(a_b + a_\theta \right) \,
\end{align}
at $R = R_{\mathrm{max}}$.
For simplicity, we assumed $a_\theta = \sqrt{a_0 a_b}$, which is usually a good approximation.
Then, Eq.~\eqref{eq:twofieldlensingaSFcondition} becomes
\begin{align}
 \label{eq:twofieldlensingfMDMcondition}
 \frac{G M_{\mathrm{DM}}(R_{\mathrm{max}})}{R_{\mathrm{max}}^2} < 0.3 \cdot (a_b + \sqrt{a_0 a_b})\,.
\end{align}
To implement this numerically, we used $M_{\mathrm{DM}}(R_{\mathrm{max}})$ instead of $\varepsilon_*(R_{\mathrm{mid}})$ as a boundary condition.
This translates to a boundary condition on $\hat{\mu}_{\mathrm{SF}}'(R_{\mathrm{max}})/m$ instead of on $\hat{\mu}(R_{\mathrm{mid}})/m$.
We can still solve the two-field equations as described in Appendix~\ref{sec:twofieldcalc} with this modified boundary condition.

The value $0.3$ is somewhat arbitrary.
It is chosen to give a non-negligible but still subdominant $a_{\mathrm{SF}}$.
There is a clear trade-off here.
Larger values allow for larger total dark matter masses $M_{200}^{\mathrm{DM}}$ but also larger deviations from MOND-like rotation curves.
Smaller values give more MOND-like rotation curves but also smaller total dark matter masses.

One way to find the largest possible $M_{200}^{\mathrm{DM}}$ compatible with the condition Eq.~\eqref{eq:twofieldlensingfMDMcondition} is to just scan over all possible $M_{\mathrm{DM}}(R_{\mathrm{max}})$ satisfying the condition.
But we do not have to do this here for the same reason we did not have to do it for standard SFDM (see Appendix~\ref{sec:M200estar04}).
Basically, since a larger $M_{\mathrm{DM}}(R_{\mathrm{max}})$ always gives a larger $M_{200}^{\mathrm{DM}}$.

Above, we considered different values of the parameter $\bar{a}$, which determines the minimum acceleration.
Here, we use $\bar{a} = 10^{-12}\,\mathrm{m}/\mathrm{s}^2$.
But the choice of $\bar{a}$ does not actually affect the maximum $M_{200}^{\mathrm{DM}}$ we calculate since we keep $r_0$ and $a_0$ fixed.
It only determines how small $\varepsilon_*$ is since $m^2/\alpha \propto (\bar{a}/a_0)^{1/4}$.
That is, it determines how well our approximation $a_\theta = \sqrt{a_0 a_b}$ works.
Since, by construction, $|\varepsilon_*| \ll 1$ for any reasonable value of $\bar{a}$, this approximation works well for any reasonable $\bar{a}$.

The result is shown in Fig.~\ref{fig:M200maxtf} and discussed in Sect.~\ref{sec:twofield}.
Comparing Fig.~\ref{fig:M200maxtf} for two-field SFDM and Fig.~\ref{fig:M200max} for standard SFDM,
    there is less scatter in the derived relation between $M^{\mathrm{DM}}_{200,\mathrm{max}}$ and $M_b$ for two-field SFDM.
This is likely because this relation in two-field SFDM is less sensitive to the details of the baryonic mass distribution at a given total baryonic mass $M_b$.
There are a few reasons for this.
First, the superfluid energy density $\rho_{\mathrm{SF}}$ scales as $\sqrt{a_b}$ in standard SFDM (at least in the MOND limit $|\varepsilon_*| \ll 1$) but not in two-field SFDM.
Second, we impose the condition $|\varepsilon| < 0.4$ in standard SFDM at smaller radii $R = R_{\mathrm{mid}}$ than $a_{\mathrm{SF}}/(a_b + a_\theta) < 0.3$ in two-field SFDM, which we impose at $R = R_{\mathrm{max}}$.
There is less variation in the baryonic mass distribution at larger radii.
Third, these conditions depend on different quantities.
The left-hand side of $|\varepsilon| < 0.4$ in standard SFDM scales as $\hat{\mu}/a_b$ while the left-hand side of $a_{\mathrm{SF}}/(a_b + a_\theta) < 0.3$ in two-field SFDM scales as $M_{\mathrm{SF}}/\sqrt{a_b}$, at least at larger radii where $a_\theta \propto \sqrt{a_b}$ dominates.
The acceleration $a_{\mathrm{SF}}$ depends only on $\hat{\mu}_{\mathrm{SF}}$, while $\hat{\mu}$ also depends on $\hat{\mu}_b$, which is much more sensitive to the details of the baryonic mass distribution.
Similarly, $\sqrt{a_b}$ is less sensitive to these details than $a_b$.

\end{appendix}

\end{document}